\documentclass[11pt,twoside]{article} 
\usepackage{times,fancyhdr}
\usepackage{cite}
\usepackage{url}
\usepackage{graphicx}
\usepackage{color}
\usepackage{subfigure}
\usepackage{multicol}
\usepackage{latexsym}
\usepackage{amsmath}
\usepackage{amsfonts}
\usepackage{amssymb}
\usepackage{harvard}
\citationstyle{dcu}
\usepackage[amsmath]{ntheorem}
\theoremheaderfont{\normalfont \bfseries}
\theorembodyfont{\normalfont}

\def\sech{\mathop{\rm sech}\nolimits}

\usepackage{lastpage}

\date{\footnotesize}
\title{\bf Mathematical Physics Properties of\\ Waves on Finite Background} 
\author{Natanael Karjanto{\small \rm $^{1,}$}\thanks{\setlength{\parindent}{0pt} E-mail address: {\footnotesize \protect\url{natanael.karjanto@nottingham.edu.my}}. (Corresponding author.)}\; and E. van Groesen{\small $^{2,3}$}\\
{\footnotesize {$\rm ^1$}School of Applied Mathematics, The University of Nottingham Malaysia Campus, Malaysia}\\
{\footnotesize {$\rm ^2$}Department of Applied Mathematics, University of Twente, The Netherlands}\\
{\footnotesize {$\rm ^3$}LabMath-Indonesia, Bandung, Indonesia}} 

\begin{document} 
\pagestyle{fancy}
\fancyhead{} 
\fancyhead[EC]{Natanael Karjanto and E. van Groesen}
\fancyhead[EL,OR]{\thepage}
\fancyhead[OC]{Mathematical Physics Properties of Waves on Finite Background}
\fancyfoot{} 
\renewcommand\headrulewidth{0.5pt}
\addtolength{\headheight}{2pt} 
\maketitle 

\fancypagestyle{title}{%
  \setlength{\headheight}{0pt}%
  \fancyhf{}
  \renewcommand{\headrulewidth}{0pt}
  \renewcommand{\footrulewidth}{0pt}
  \fancyhead[L]{In: Handbook of Solitons\\ Editors: S.P. Lang and S.H. Bedore, pp.~\thepage--\pageref{LastPage} \vspace{0.5cm}\\ \textsl{\textbf{Chapter 14}}}
  \fancyhead[R]{ISBN 978-1-60692-596-6\\ \copyright \ 2009 Nova Science Publishers, Inc.\\ \vspace{0.5cm}}
}%
\thispagestyle{title}
\tableofcontents
\hrulefill

\begin{abstract}
Several mathematical and physical aspects of waves on a finite background are reported in this article. \addcontentsline{toc}{section}{Abstract} The evolution of the complex wave packet envelope of these type of waves is governed by the focussing-type of the nonlinear Schr\"{o}dinger (NLS) equation. The NLS equation admits a number of exact solutions; in this article, we only discuss waves on finite background type of solutions that have been proposed as theoretical models for freak wave events. Three types of waves on finite background considered in this article are known as the Soliton on Finite Background (SFB), the Ma solution and the rational solution. In particular, two families of the SFB solutions deserve our special attention. These are SFB$_{1}$ and SFB$_{2}$, where the latter one belongs to higher order waves on finite background type of solution. These families of solutions describe the Benjamin-Feir modulational instability phenomenon, which has been verified theoretically, numerically and experimentally as the phenomenon that a uniform continuous wave train is unstable under a very long modulational perturbations of its envelope.

A distinct difference between the two families of solutions can be observed in the spectral domain, where SFB$_{1}$ has one pair of initial side-bands and SFB$_{2}$ has two pairs of initial side-bands within the interval of instability. The relationship between SFB$_{1}$ and SFB$_{2}$ are explained and some important physical characteristics of the two solutions are discussed. These include the amplitude amplification factor, the spatial evolution of complex-valued envelopes, their corresponding physical wave fields and the evolution of the corresponding wave signals. Interestingly, wavefront dislocation and phase singularity are observed in both families of the solution with different patterns, depending on the value of the modulation wavelength and on the choice of parameters in SFB$_{2}$.
\end{abstract}

{\small
{\setlength{\parindent}{0pt}
{\bf 2008 PACS}: 02.30.Jr; 05.45.Yv; 46.40.Cd; 52.35.Mw}
}

\section{Introduction}

The theory of waves has a long history since the time of Issac Newton (1643-1727) when he worked in optics to explain the diffraction of light. The content of this article is mainly concerned with the applications in water waves, although relevant applications in optics and other fields can be considered as well. An excellent overview of the origins and the development of water wave theory can be found in \cite{Craik04}. Furthermore, \citeasnoun{Craik05} provides a thorough examination of Stokes' papers and letters concerning water waves as well as on how Stokes (1819-1903) built on the earlier foundations to establish a definite theory of linear and weakly nonlinear waves.

Particularly we are interested in the motion of nonlinear and dispersive waves governed by an evolution equation. In this context, dispersion generally refers to frequency dispersion, which means that waves of different wavelength travel at different phase velocities. Nonlinear means that the combination of waves does not satisfy the superposition principle. A governing equation that deserves our attention is the nonlinear Schr\"{o}dinger (NLS) equation. It is a partial differential equation which has applications in various areas of physical, biological and engineering sciences. The NLS equation is a nonlinear version of the Schr\"{o}dinger equation, an equation that describes how quantum state of a physical system varies. The equation is named after an Austrian-Irish physicist Erwin Schr\"{o}dinger (1887-1961) who proposed the equation in 1926 \cite{Schrodinger}.

The NLS equation is also known as the cubic Schr\"{o}dinger equation since the nonlinearity is of order three. Based on nonlinearity classification, others  called it as the NLS equation with Kerr law nonlinearity \cite{Biswas07}, and the name is based on a Scottish physicist John Kerr (1824-1907). Together with the Korteweg-de Vries equation and the sine-Gordon equation, the NLS equation belongs to the classical soliton equations. Each one of these three and many other completely integrable equations possesses solutions with soliton properties. A brief history on solitons can be found in \cite{Scott05}. An extensive discussion on the classical soliton equations and their applications are given in \cite{Scott03}.

\section{Nonlinear Dispersive Wave Equation}

This section discusses the NLS equation as a governing equation for the evolution of our waves on finite background. An overview of the Benjamin-Feir modulational instability phenomenon and the stability analysis of the plane-wave solution of the NLS equation are also explained in this section.

\subsection{The Nonlinear Schr\"{o}dinger Equation}

The NLS equation describes the evolution of the envelope of nonlinear and dispersive wave packets. As mentioned above, it plays a vital role in many applied fields, including fluid dynamics, nonlinear optics, plasma physics and protein chemistry. For applications in hydrodynamics of nonlinear envelope waves, see e.g. \cite{Benney67,Newell74,Whitham74,Yuen82,Johnson97}, in nonlinear optics, see e.g. \cite{Kelley65,Talanov65,Karpman69,Hasegawa73,Akhmediev97,Agrawal06}, the propagation of a heat pulse in a solid \cite{Tappert70}, in plasma physics, see e.g. \cite{Taniuti68,Ichikawa72,Shimizu72,Zakharov72,Kakutani74,Ichikawa79}, nonlinear instability problems \cite{Stewartson71,Nayfeh71}, the propagation of solitary waves in piezoelectric semiconductors \cite{Pawlik75}, protein chemistry \cite{Daniel02}, see also \cite{Davydov73} for the quantum theory of contraction of $\alpha$-helical proteins under their excitation.

The NLS equation was first discussed and pioneered by \citeasnoun{Benney67} for nonlinear dispersive waves in general. Later on, it was re-derived independently by \citeasnoun{Zakharov68} using a spectral method in the context of gravity waves on deep water. Further, \citeasnoun{Hasimoto72} and \citeasnoun{Davey72} derived the NLS equation for finite depth independently using multiple scale methods. In addition, \citeasnoun{Yuen75} derived it using the averaged Lagrangian formulation from Whitham's theory \cite{Whitham65a}. \citeasnoun{ZakharovShabat72} developed an ingenious inverse scattering method to show that the NLS equation is completely integrable. Moreover, a heuristic derivation of the NLS equation has been given by several authors \cite{Kadomtsev71,Karpman75,Jeffrey82,Dingemans97,Dingemans01}. A similar derivation of the spatial NLS equation can be found in \cite{Djordjevic78}, but under an assumption of slowly varying bottom.

The NLS equation describing the slow modulation of a harmonic wave moving over a surface of a two dimensional channel is derived using the multiple scale method by \citeasnoun{Johnson76} and \citeasnoun{Johnson97}. The derivation through perturbation theory, including a discussion on the Korteweg-de Vries (KdV)-induced long wave pole in the nonlinear coefficient of the NLS equation, resonance effects and numerical illustrations can be found in \cite{Boyd01}. Furthermore, the derivation from the KdV type of equation with exact dispersion relation is given by \citeasnoun{vanGroesen98} and \citeasnoun{Cahyono02}. In this article, we will consider, different from most of the other cited papers, the NLS equation as the focusing and spatial type of equation since this it is more suitable for freak wave generation in a hydrodynamic laboratory \cite{Karjanto06}.

The NLS equation that will be discussed in this article is given by
\begin{equation}
\partial_{\xi}A + i \beta \partial_{\tau}^{2} A + i\gamma |A|^{2}A = 0.
\label{spatialNLS}
\end{equation}
That it is of focusing type is a consequence if the coefficients of the dispersive term and the nonlinear term have the same sign, i.e., $\beta \gamma > 0$. The spatial type of equation is specified by a partial derivative in space of the evolution term, i.e. $\partial_{\xi}A$ and the dispersive term has the second partial derivative in time. i.e., $\partial_{\tau}^{2}A$.\footnote{If the coefficients of the dispersive term and the nonlinear term have the opposite sign, i.e., $\beta \gamma < 0$, then we have the defocusing type of NLS equation. Literature generally refers to the NLS equation as the temporal type, given as $\partial_{\tau}A + i \beta \partial_{\xi}^{2} A + i\gamma |A|^{2}A = 0$. Opposite to the spatial type of NLS equation, this type of equation has a time derivative of the evolution term, i.e. $\partial_{\tau}A$ and the second derivative in space of the dispersive term, i.e., $\partial_{\xi}^{2}A$.} In \eqref{spatialNLS}, $A(\xi,\tau)$ is a complex-valued function describing the amplitude of the corresponding wave packet $\eta(x,t)$, where $\xi$ and $\tau$ are variables related to the physical variables space $x$ and time $t$. The wave packet $\eta$ describes a wave surface elevation from an equilibrium and describes a physical wave field in the $xt$-plane, related to the complex-valued amplitude as follows:
\begin{equation}
  \eta(x,t) = \epsilon A(\xi,\tau) e^{i(k_{0}x - \omega_{0}t)} + \textmd{higher order terms} + \textmd{complex conjugate}. \label{eta}
\end{equation}
In this context, $\eta$ satisfies the KdV type of equation with exact dispersion relation, given by \citeasnoun{vanGroesen98}:
\begin{equation}
  \partial_{t}\eta + i \Omega(-i\partial_{x})\eta + \alpha\,\eta \partial_{x}\eta = 0, \qquad \alpha \in \mathbb{R}.
  \label{KdVexactdispersion}
\end{equation}
Applying the multiple scale method with $\xi = \epsilon^{2}x$ and $\tau = \epsilon(t - x/\Omega'(k_{0}))$, substituting \eqref{eta} into \eqref{KdVexactdispersion} will give us the NLS equation \eqref{spatialNLS}.
In the following subsection, we will see that the simplest nontrivial solution of the NLS equation is unstable under a very long modulational perturbation.

\subsection{Benjamin-Feir Modulational Instability}

The simplest nontrivial solution of the NLS equation is a uniform continuous wave train, also known as the `plane-wave' solution. Explicitly it is given by $A_{0}(\xi) = r_{0} e^{-i\gamma r_{0}^{2}\xi}$, where $r_{0}$ is a constant. This essentially represents the fundamental component of a Stokes wave. It has been verified theoretically, numerically, experimentally that this solution is unstable under a very long modulational perturbation of its envelope. This phenomenon is known as `modulational instability' or `sideband instability' or `Benjamin-Feir instability' in the context of water waves \cite{Benjamin67}. They established analytically using a perturbation approach that progressive waves of finite amplitude on deep water, also known as Stokes waves, are unstable. \citeasnoun{Benjamin67} also reported experimental data in fairly good agreement with the predictions regarding the wavenumber and growth rate of the instability. A fine overview of the history of T. Brooke Benjamin (1929-1995) and his contributions to nonlinear wave theory is presented by \citeasnoun{Hunt05}.

A similar independent result on modulational instability is observed in nonlinear liquids by \citeasnoun{Bespalov66}. The instability
of weakly nonlinear waves in dispersive media has been investigated \cite{Lighthill65,Whitham65b,Lighthill67,2Benjamin67,Ostrovsky67,Whitham67,Zakharov67}.
Much other research, later on, shows that modulational instability is observed in almost any field of wave propagation in nonlinear
media \cite{Dodd82,Newell85,Lvov94,Remoissenet99}, including plasma physics \cite{Taniuti68,Hasegawa70}, nonlinear optics \cite{Karpman69,Hasegawa80,Tai86,Hasegawa95,Agrawal06} and quite recently in Bose-Einstein condensates \cite{Wu01,Konotop02,Smerzi02,Baizakov02,Salasnich03}. Modulational instability is also observed in even more complex settings such as spinor Bose-Einstein condensates \cite{Robin01} and in the presence of traps and time-dependent potentials \cite{Theocharis03}. It is interesting to note that although there has been a continued research activity of modulational instability in many other fields since the 1960s and the 1970s down to this date, the phenomenon in Bose-Einstein condensates started being a topic of major interest only in the 2000s.

The references given in this paragraph refer to the temporal NLS equation instead. The long time behaviour of the modulational instability of the NLS equation is investigated in \cite{Janssen81}. The results are in qualitative agreement with experimental findings of \cite{Lake77} and the numerical computation of \cite{Yuen78a,Yuen78b}. Their works show that unstable modulations grow to a maximum limit and then subside. The energy is transferred from the primary wave to the sidebands for a certain period of time and is then recollected back into the primary wave mode. The long time evolution of an unstable wave train leads to a series of modulation-demodulation cycles in the absence of viscosity, known as the `Fermi-Pasta-Ulam recurrence' phenomenon \cite{Fermi55}. An alternative treatment of the Benjamin-Feir instability mechanism of the two-dimensional Stokes waves on deep water was given by \cite{Stuart78}. Regarding the effect of dissipation, \citeasnoun{Segur05} show that any amount of a certain type of dissipation stabilises the Benjamin-Feir instability for waves with narrow bandwidth and moderate amplitude. On the other hand, \citeasnoun{Bridges07} recently show that there is an overlooked mechanism whereby the addition of dissipation leads to an enhancement of the Benjamin-Feir instability.

To investigate the stability of the plane-wave solution of the NLS equation, we substitute $A(\xi,\tau) = A_{0}(\xi) [B_{0} + \epsilon B(\xi,\tau)]$ into the NLS equation~\eqref{spatialNLS}, where $B_{0} = e^{i\phi_{0}} \in \mathbb{C}$ is the phase of the plane-wave for $\xi \rightarrow -\infty$ and $B(\xi,\tau)$ is a perturbation function.\footnote{A similar analysis can be done if one chooses $B_{0} = 1$. In our case, we choose $B_{0}$ as a unit complex number in order to match the phase with the analysis obtained from the SFB.} The corresponding linearized equation is obtained by neglecting higher order terms and it reads
\begin{equation}
  B_{0}^{\ast} \partial_{\xi}B + i \beta B_{0}^{\ast} \partial_{\tau}^{2}B + i\gamma r_{0}^{2}(B_{0}^{\ast} B + B_{0} B^{\ast}) = 0, \label{linearNLS}
\end{equation}
where $B^{\ast}(\xi,\tau)$ denotes the complex conjugate of the perturbation function $B(\xi,\tau)$. We seek an Ansatz for the perturbation function in the form $B(\xi,\tau) = B_{1} e^{(\sigma \xi + i \nu \tau)} + B_{2} e^{(\sigma^{\ast} \xi - i \nu \tau)}$, where $B_{1}$, $B_{2} \in \mathbb{C}$, $\sigma$ is a growth rate and $\nu$ is the modulation frequency. Substituting this Ansatz into the linear equation~\eqref{linearNLS}, we obtain a set of two equations:
\begin{equation}
  \left(%
\begin{array}{cc}
  [\sigma - i(\beta \nu^{2} - \gamma r_{0}^{2})]e^{-i\phi_{0}} & i\gamma r_{0}^{2} e^{i\phi_{0}}  \\
  -i\gamma r_{0}^{2} e^{-i\phi_{0}} & [\sigma + i(\beta \nu^{2} - \gamma r_{0}^{2})] e^{i\phi_{0}} \\
\end{array}%
\right)\left(%
\begin{array}{c}
  B_{1} \\
  B_{2}^{\ast} \\
\end{array}%
\right) = \left(
\begin{array}{c}
  0 \\
  0 \\
\end{array}
\right). \label{B1B2}
\end{equation}
Requiring the determinant of the matrix above to be zero, we have
the condition $\sigma^{2} = 2\beta \gamma r_{0}^{2} \nu^{2} -
\nu^{4}$. The coefficient $\sigma$ is real for sufficiently small
$\nu$, and is then the growth rate of the instability. The growth rate $\sigma > 0$ is given by $\sigma = \nu \sqrt{2 \beta \gamma r_{0}^{2} - \beta^{2}\nu^{2}}$. Defining a normalized modulation frequency
$\tilde{\nu}$ by $\nu = r_{0} \sqrt{\gamma/\beta} \tilde{\nu}$, the growth rate $\sigma$ becomes $\sigma = \gamma r_{0}^{2} \tilde{\sigma}$, with
$\tilde{\sigma} = \tilde{\nu} \sqrt{2 - \tilde{\nu}^{2}}$. Note
that for instability, the normalized modulation
frequency has to be in the instability interval $0 < \tilde{\nu} < \sqrt{2}$. The maximum instability occurs at $\tilde{\nu} = 1$ with a maximum growth rate of $\tilde{\sigma}_\textmd{max} = 1$.
\begin{figure}[h]
  \begin{center}
  \includegraphics[width=0.5\textwidth]{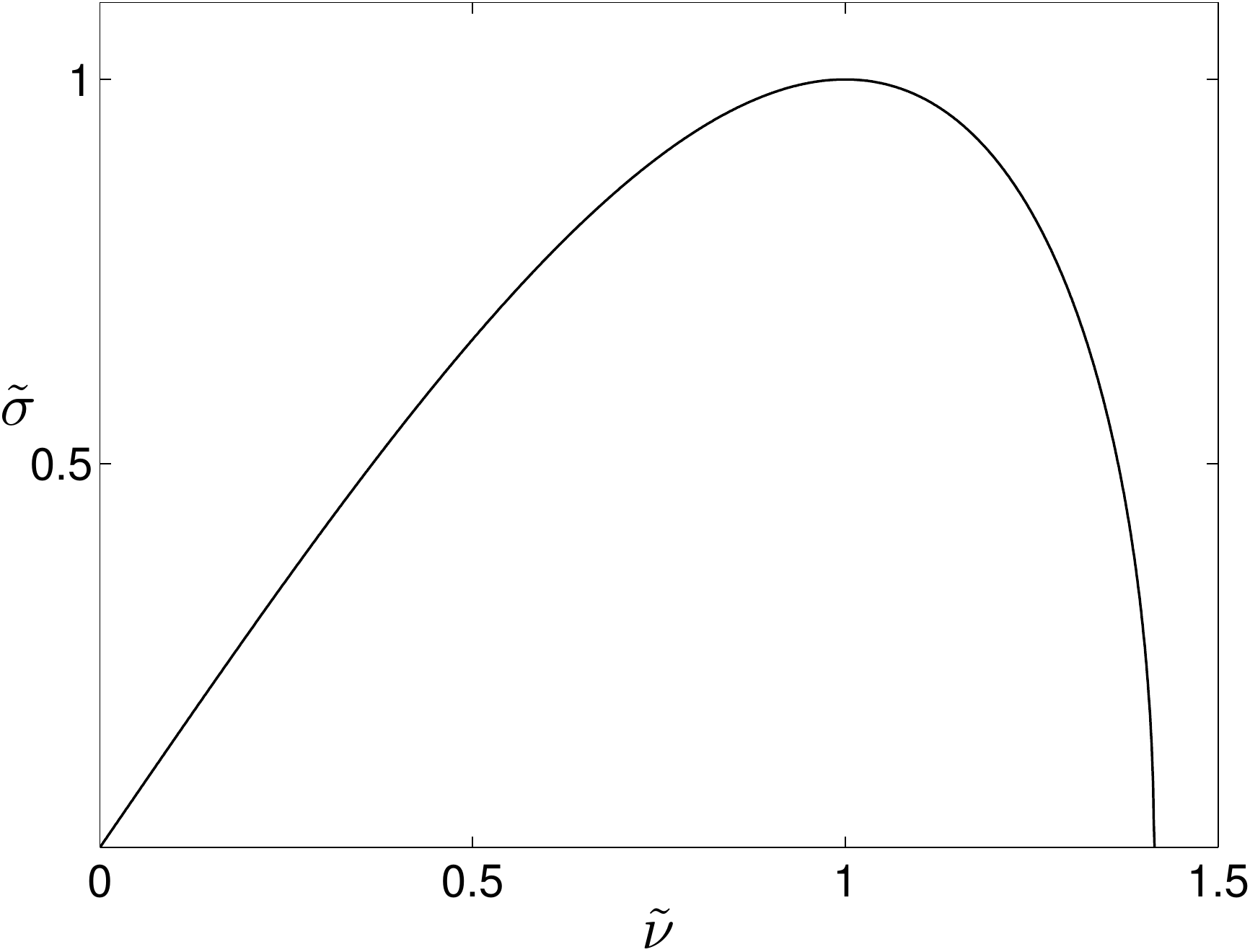}
  \caption[]{The growth rate $\tilde{\sigma}$ as a function of the modulation frequency $\tilde{\nu}$, where $0 < \tilde{\nu} < \sqrt{2}$. The maximum growth rate is reached when $\tilde{\nu} = 1$.}
  \end{center}
\end{figure}

From the above two equations for $B_{1}$ and $B_{2}$~\eqref{B1B2}, we obtain $(B_{1} e^{-i\phi_{0}})/(B_{2}^{\ast} e^{i\phi_{0}}) = (\tilde{\nu}^{2} - 1) - i\tilde{\sigma}$. Taking the modulus of this ratio, we have $\left|B_{1}/ B_{2}^{\ast}\right| = 1$. Since $|B_{2}| = |B_{2}^{\ast}|$, we have also $|B_{1}| = |B_{2}|$. Therefore we
write $B_{j} = |B_{j}|e^{i\phi_{j}}$, $j = 1, 2$, where $\phi_{1}$ and $\phi_{2}$ are the phases corresponding to the lower and upper sideband frequencies, respectively. The phase relationship is given as follows:
\begin{equation}
  e^{i(\phi_{1} + \phi_{2} - 2\phi_{0})} = (\tilde{\nu}^{2} - 1) - i\tilde{\sigma}.
  \label{anglephases}
\end{equation}
From the following section, we discover that the phase of the sideband frequencies is equal, $\phi_{1} = \phi_{2}$ and the phase of the plane-wave is given by  $e^{i\phi_{0}} = \tilde{\nu}^{2} - 1 + i \tilde{\sigma}$. This implies
\begin{equation}
  e^{i(\phi_{1} - \phi_{0})} = \tilde{\nu}^{2} - i \tilde{\sigma} \qquad \textmd{and} \qquad
  e^{i \phi_{1}} = \tilde{\nu}^{2} + i \tilde{\sigma}.
\end{equation}
The solution $A$ which describes the instability of the plane-wave solution now reads
\begin{equation}
  A(\xi,\tau) = A_{0}(\xi) \left[e^{i\phi_{0}} + 2 \epsilon |B_{1}| e^{i\phi_{1}} e^{\sigma \xi} \cos(\nu \tau) \right].
\end{equation}
We see that for $\xi \rightarrow \infty$, this solution grows exponentially in space. This is the linear instability of the Benjamin-Feir modulated wave signal. Because of nonlinear effects from the cubic term that have been ignored in the analysis above, this growth is unbounded. In the following section, we will see that a fully nonlinear extension of the Benjamin-Feir instability can be found, it is bounded and is given by the SFB. For a very long modulation wavelength, namely when the modulation frequency $\nu \rightarrow 0$, it is given by a rational solution.

\section{Waves on Finite Background}

The type of waves on finite background in the context of this paper refers to exact solutions of the NLS equation which asymptotically become the plane-wave solution for either $\xi \rightarrow \pm \infty$ or $\tau \rightarrow \pm \infty$. Since the amplitude of the plane-wave solution remains finite in space and time and forms the background for these exact solutions, these solutions are called `waves on finite background'. However, several authors refer to this type of solutions as `breather solutions' \cite{Tajiri98,Dysthe99,Grimshaw01}. The term `breather' was introduced by \cite{Ablowitz73,Ablowitz74} in the context of the sine-Gordon equation since its solution has the characteristic property of being localised in space and oscillate (breath) in time. The opposite situation, oscillations in space and localised in time, is also denoted as a breather. It is interesting to note that only the focusing type of the NLS equation possesses the breather solutions, while the defocusing type does not possess such type of solutions.

There are a number of methods to find the exact solutions of the NLS equation that are waves on finite background type solution. For instance, \cite{Akhmediev97} propose a method of finding all exact solutions using a phase-amplitude transformation, constructing a system of ordinary differential equations and implementing a certain integration procedure. A derivation using Hirota's method and associating it with the dark-hole soliton solutions of the defocusing NLS equation can be found in \cite{Ablowitz90,Calini02}. This method is invented by Ryogo Hirota to find multi-soliton solutions of the KdV equation and the exponential lattice equation \cite{Hirota71}. Remarkably, it has also provided a method for clarifying the relationship between nonlinear equations and how to find entirely new equations which also possess multi-soliton solutions. Finding waves on finite background using the technique of inverse scattering transform can be found in \cite{Osborne00,Osborne01}. In the following subsection, we derive the waves on finite background type of solutions using a displaced phase-amplitude representation.

\subsection{Displaced Phase-Amplitude Representation}

One of a very recent method in finding waves on finite background solutions of the NLS equation is the `displaced phase-amplitude' method \cite{vanGroesen06}. The authors introduced displaced phase and displaced amplitude variables, apply the variational formulation approach to explain the method and present an interesting analogy of a nonlinear oscillator equation with the corresponding potential equation that depends on the position. In the following paragraphs, we will derive the waves on finite background solutions using the displaced phase-amplitude method.

We seek the type of solution of the NLS equation in the form of displaced phase-amplitude variables, given as follows:
\begin{equation}
  A(\xi,\tau) = A_{0}(\xi) [G(\xi,\tau) e^{i\phi(\xi)} - 1],
  \label{NLSsolution}
\end{equation}
where a real-valued function $G(\xi,\tau)$ is referred to as the displaced amplitude and a real-valued function $\phi(\xi)$ which depends only on the spatial variable is referred to as the displaced phase. Since we are interested to waves on finite background type of solution, the asymptotic behavior of the solution must be a constant. This constant eventually plays the role as background for our waves. Thus, with $|A_{0}(\xi)| = r_{0}$, it will hold that:
\begin{equation}
  \lim_{\xi  \rightarrow \pm \infty} |A(\xi,\tau)| = r_{0} =
  \lim_{\tau \rightarrow \pm \infty} |A(\xi,\tau)|.
\end{equation}

Substituting (\ref{NLSsolution}) into the NLS equation (\ref{spatialNLS}), we obtain two equations that correspond to the real and the imaginary parts, respectively given by:
\begin{eqnarray*}
  \partial_{\xi}G \cos \phi + \gamma r_{0}^{2} G^{2} \sin 2\phi - [G\phi'(\xi) + \beta \partial_{\tau}^{2}G + \gamma r_{0}^{2} G^{3}] \sin \phi &=& 0 \\%
  \partial_{\xi}G \sin \phi - \gamma r_{0}^{2} G^{2} (2\cos^{2}\phi + 1) + [G\phi'(\xi) + \beta \partial_{\tau}^{2}G + \gamma r_{0}^{2} G (G^{2} + 2)] \cos \phi &=& 0.%
\end{eqnarray*}
Multiplying the real part by $\cos \phi$ and the imaginary part by
$\sin \phi$ and adding both equations, will give a Riccati-like
equation for the displaced amplitude $G$:
\begin{equation}
  \partial_{\xi} G + \gamma r_{0}^{2} \sin 2\phi G - \gamma
  r_{0}^{2} \sin \phi G^{2} = 0. \label{riccati}
\end{equation}
Solving this equation leads us to the conclusion that $G$ can be written in a special form. By letting $G = 1/H$, equation \eqref{riccati} becomes a first order linear differential equation in $H$:
\begin{equation}
  \partial_{\xi} H - \gamma r_{0}^{2} \sin 2\phi\, H + \gamma r_{0}^{2} \sin \phi = 0. \label{1linearode}
\end{equation}
Let us take $\tilde{P}(\xi) = \textmd{exp}
\left(-\gamma r_{0}^{2} \int \sin 2 \phi(\xi)\, d\xi \right)$ as the integrating factor, multiply it to \eqref{1linearode} and integrate the result with respect to $\xi$, we obtain the solution for $H$:
\begin{equation}
   H(\xi,\tau) = \frac{- \gamma r_{0}^{2} \int \tilde{P}(\xi) \sin \phi(\xi) \, d\xi - \zeta(\tau)}{\tilde{P}(\xi)} = \frac{\tilde{Q}(\xi) - \zeta(\tau)}{\tilde{P}(\xi)},
\end{equation}
where $-\zeta(\tau)$ is a constant of integration that depends on $\tau$ and $\tilde{Q}(\xi)$ denotes the term with the integral sign. Assuming that the displaced phase $\phi(\xi)$ is an invertible function, we can write $\xi = \xi(\phi)$ and drop the tilde signs to indicate that $P = P(\phi)$ and $Q = Q(\phi)$. Consequently, the displaced amplitude $G$ is now written as a function of the displaced phase $\phi$ and the temporal variable $\tau$ as follows:
\begin{equation}
  G(\phi,\tau) = \frac{P(\phi)}{Q(\phi) - \zeta(\tau)}. \label{AnsatzG}
\end{equation}
We will observe in the following subsections that by choosing three different functions of the constant of integration $\zeta(\tau)$ will lead to three different waves on finite background solutions of the NLS equation.

On the other hand, multiplying the real part by $\sin
\phi$ and the imaginary part by $\cos \phi$, subtracting one from
the other we obtain a nonlinear oscillator equation for the displaced amplitude $G$:
\begin{equation}
  \beta \partial_{\tau}^{2} G + (\phi'(\xi) + 2\gamma r_{0}^{2}\cos^{2}\phi)G - 3 \gamma r_{0}^{2} \cos \phi G^{2} + \gamma r_{0}^{2} G^{3} = 0.%
  \label{oscillator}
\end{equation}
In deriving the waves on finite background type of solutions, we will compare this equation with the second order differential equation for $G$ after choosing a particular function of $\zeta(\tau)$. Choosing $\zeta(\tau) = \cos(\nu \tau)$ yields the SFB, choosing $\zeta(\tau) = \cosh(\mu \tau)$ yields the Ma solution and choosing $\zeta(\tau) = 1 - \frac{1}{2} \nu^{2} \tau^{2}$ or $\zeta(\tau) = 1 + \frac{1}{2} \mu^{2} \tau^{2}$ yields the rational solution.

The nonlinear oscillator equation \eqref{oscillator} can also be written as follows:
\begin{equation}
  \beta \frac{\partial^{2} G}{\partial \tau^{2}} + \frac{\partial V}{\partial G} = 0 \label{oscillator2}
\end{equation}
where $V$ denotes the potential energy. It is a quartic function of the displaced amplitude and is expressed as:
\begin{equation}
  V(G,\phi) = \frac{1}{2} \phi'(\xi) G^{2} + \frac{1}{4} \gamma r_{0}^{2} G^{2} (G - 2\cos \phi)^{2}. \label{potential}
\end{equation}
The total energy has a constant value and is given by:
\begin{equation}
  E(G,\phi) = \frac{1}{2} \beta \left(\frac{\partial G}{\partial \tau} \right)^{2} + V(G,\phi).
\end{equation}
Interestingly, for all three types of the waves on finite background solutions considered in this article, the total energy for the displaced phase-amplitude representation is zero, i.e. $E(G,\phi) = 0$ for all $(\xi,\tau) \in \mathbb{R}^{2}$ for the SFB, the Ma solution and the rational solution.

\subsection{Soliton on Finite Background}

For the choice of $\zeta(\tau) = \cos(\nu \tau)$, the corresponding family of waves on finite background is known as the SFB \cite{Akhmediev87}. The variable $\nu$ denotes the modulation frequency, and for the normalized quantity $\tilde{\nu} = \nu/\left(r_{0}\sqrt{\frac{\gamma}{\beta}}\right)$, the SFB is well defined for $0 < \tilde{\nu} < \sqrt{2}$. With this choice of $\zeta$, the corresponding differential equation for $G$ reads:
\begin{equation}
  \partial_{\tau}^{2}G = -\nu^{2} G + 3\nu^{2} \frac{Q}{P} G^{2} + 2\nu^{2}\frac{1 - Q^{2}}{P^{2}} G^{3}. \nonumber%
\end{equation}
By comparing with (\ref{oscillator}), the solution (\ref{AnsatzG})
is obtained with $P(\phi) = \tilde{\nu}^{2} Q(\phi)/\cos \phi$,
$Q^{2}(\phi) = 2 \cos^{2}\phi /(2 \cos^{2}\phi - \tilde{\nu}^{2})$
and the displaced phase satisfies
\begin{equation}
  \tan \phi(\xi) = -\frac{\tilde{\sigma}} {\tilde{\nu}^{2}} \tanh(\sigma \xi) \label{disphiSFB}
\end{equation}
where $\tilde{\sigma} = \tilde{\nu} \sqrt{2 - \tilde{\nu}^{2}}$ and
$\sigma = \gamma r_{0}^{2} \tilde{\sigma}$ is a positive quantity corresponding to the growth rate of the Benjamin-Feir modulational instability. The displaced amplitude $G$ is given by:
\begin{eqnarray}
  G(\phi(\xi),\tau) &=& \frac{\tilde{\nu}^{2} \sqrt{2}}{\sqrt{2} \cos \phi - \sqrt{2 \cos^{2}\phi - \tilde{\nu}^{2}} \cos (\nu \tau)} \\
  &=& \frac{\tilde{\nu} \sqrt{2} \sqrt{2\tilde{\nu}^{2} \cosh^{2}(\sigma \xi) - \tilde{\sigma}^{2}}}{\tilde{\nu} \sqrt{2} \cosh(\sigma \xi) - \tilde{\sigma} \cos(\nu \tau)}. \label{disampSFB}
\end{eqnarray}

\begin{figure}[h]
  \begin{center}
  \subfigure[]{\includegraphics[width=0.45\textwidth]{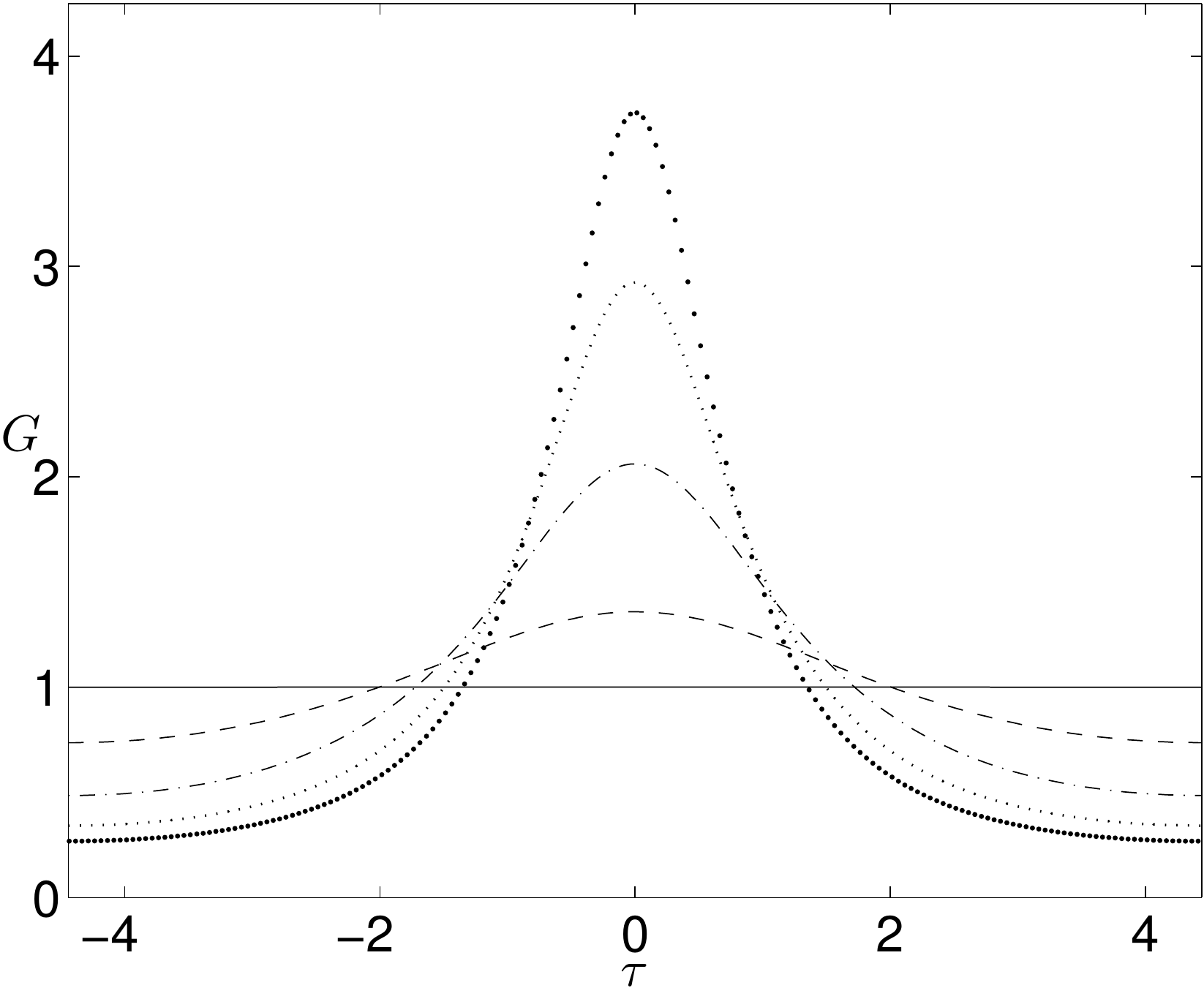}} \hspace{0.5cm}
  \subfigure[]{\includegraphics[width=0.45\textwidth]{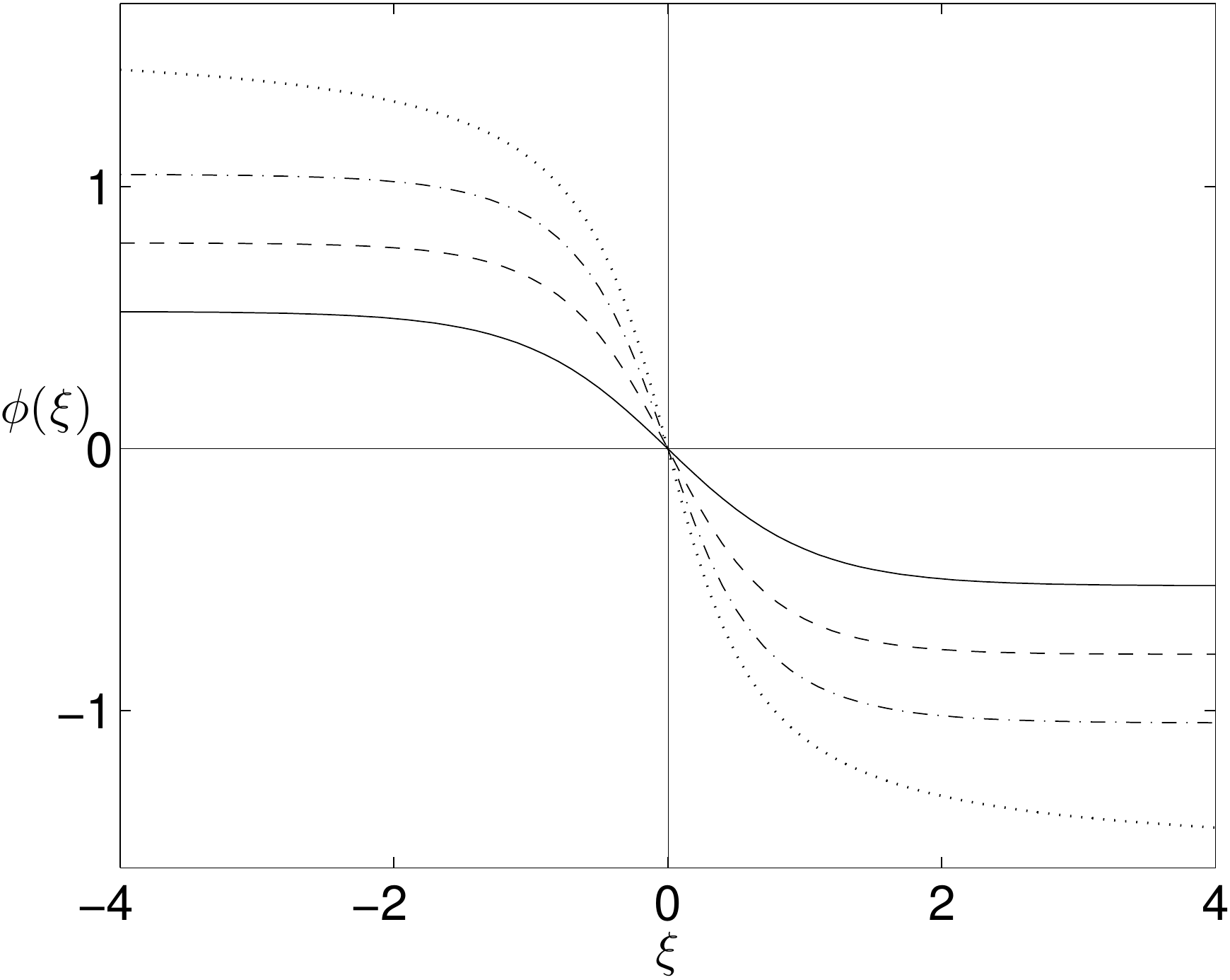}}
  \caption[]{(a). Sketch of the corresponding displaced amplitude of the SFB solution $G$ for $\tilde{\nu} = \sqrt{1/2}$ at different positions: $\xi = 0$ (heavy dotted), $\xi = \pm 1/2$ (light dotted), $\xi = \pm 1$ (dashed-dot), $\xi = \pm 2$ (dashed) and $\xi \rightarrow \pm \infty$ (solid).
  (b). Sketch of the displaced phase $\phi$ for different values of $\tilde{\nu}$: $\tilde{\nu} = \sqrt{3/2}$ (solid), $\tilde{\nu} = 1$ (dashed), $\tilde{\nu} = \sqrt{1/2}$ (dashed-dot) and $\tilde{\nu} \rightarrow 0$ (dotted).} \label{disamphiSFB}
  \end{center}
\end{figure}

Figure \ref{disamphiSFB} shows plots of the corresponding displaced amplitude  \eqref{disampSFB} for $\tilde{\nu} = \sqrt{1/2}$ and the corresponding displaced phase \eqref{disphiSFB} for several values of $\tilde{\nu}$. The displaced amplitude is simply $G = 1$ for $\xi \rightarrow \pm \infty$. Then it becomes modulated as the position progresses and eventually the middle part grows at a faster rate than the side parts decrease. It reaches its maximum at $\xi = 0$. For the displaced phase, we observe that the phase difference is larger between $\xi = -\infty$ and $\xi = \infty$ for smaller values of modulation frequency $\tilde{\nu}$. It is interesting to note that for all values of the modulation frequency where it is defined, the displaced phase at $\xi = 0$ vanishes. As a consequence, the complex-valued amplitude now reduces into a real-valued amplitude at $\xi = 0$.

\subsubsection{Physical Characteristics}

The corresponding solution of the NLS equation (\ref{NLSsolution}) with maxima at $(\xi,\tau) = (0,2n\pi/\nu)$, $n \in \mathbb{Z}$ can then be written after some manipulations as:
\begin{equation}
  A(\xi,\tau) = A_{0}(\xi) \left(\frac{\tilde{\nu}^{2} \cosh(\sigma \xi) - i \tilde{\sigma} \sinh(\sigma \xi)}{\cosh(\sigma \xi) - \sqrt{1 - \frac{1}{2} \tilde{\nu}^{2}} \cos(\nu \tau)} - 1 \right). \label{SFB}
\end{equation}
In particular, \citeasnoun{Ablowitz90} called this solution as a `homoclinic orbit' solution of the NLS equation. In particular, for $\tilde{\nu} = 1$, \citeasnoun{Osborne00} called it a `rogue wave' solution. To honor the author who discovered this solution, \citeasnoun{Dysthe99} called it as the `Akhmediev solution'. In this article, we call it the SFB.

The name SFB comes from the characteristics of its complex-valued amplitude plot. At a fixed time, particularly when it reaches its maxima, $\tau = 2n\pi/\nu$, $n \in \mathbb{Z}$, this SFB has a soliton shape but for $\xi \rightarrow \pm \infty$, it goes to a finite background of the plane-wave solution $A_{0}$. At a fixed position, the SFB is periodic in time and its periodicity depends on the modulation frequency $\nu$, which is given by $2\pi/\nu$. Thus, the smaller the modulation frequency, or the longer the modulation wavelength for the Benjamin-Feir initial signal, the longer its period will be. For $\nu \rightarrow 0$, the SFB reduces into the rational solution, which we will discuss in Subsection \ref{subsecrational}.
\begin{figure}[h]
  \begin{center}
    \subfigure[]{\includegraphics[width=0.45\textwidth]{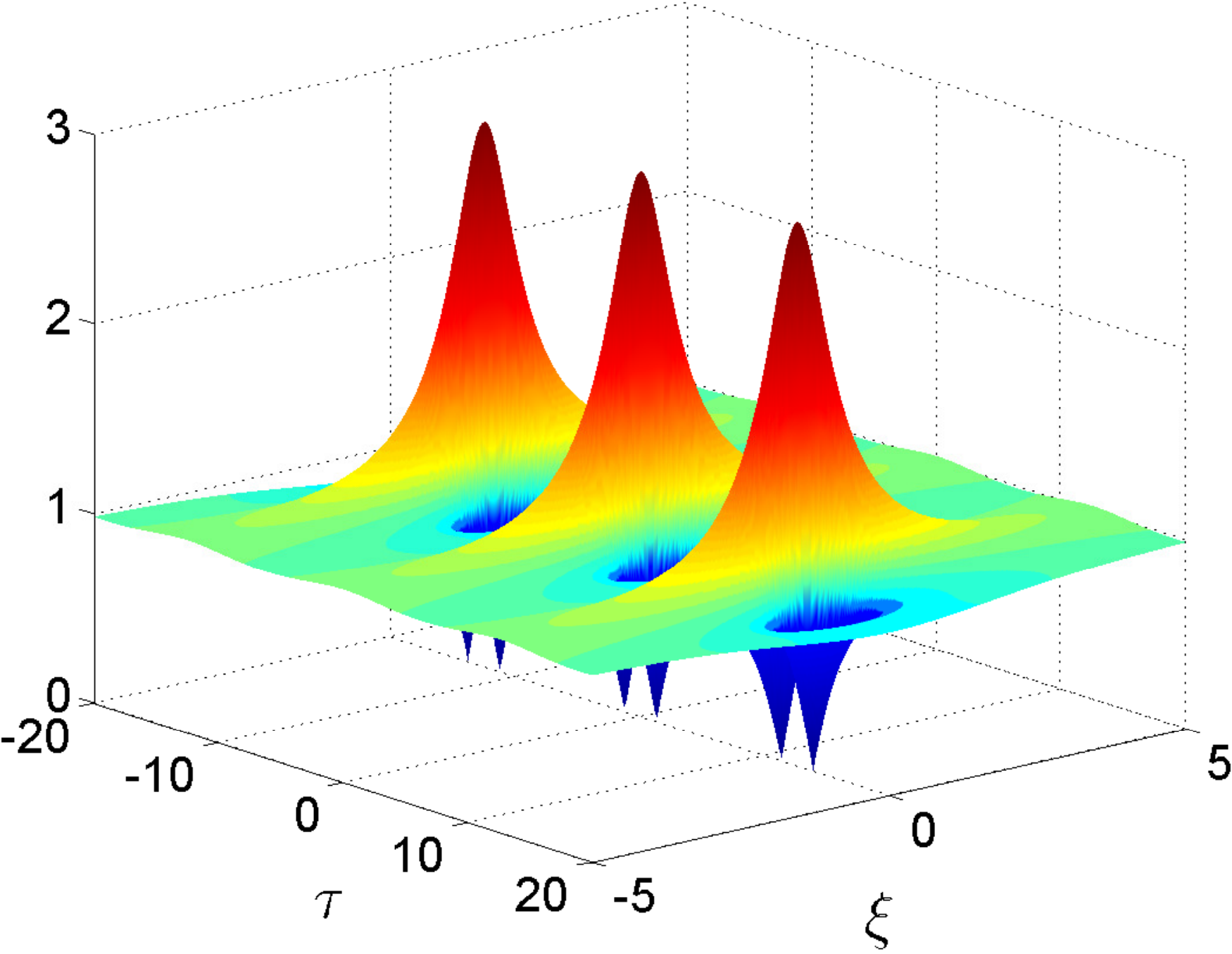}} \hspace{1.0cm}
    \subfigure[]{\includegraphics[width=0.4\textwidth]{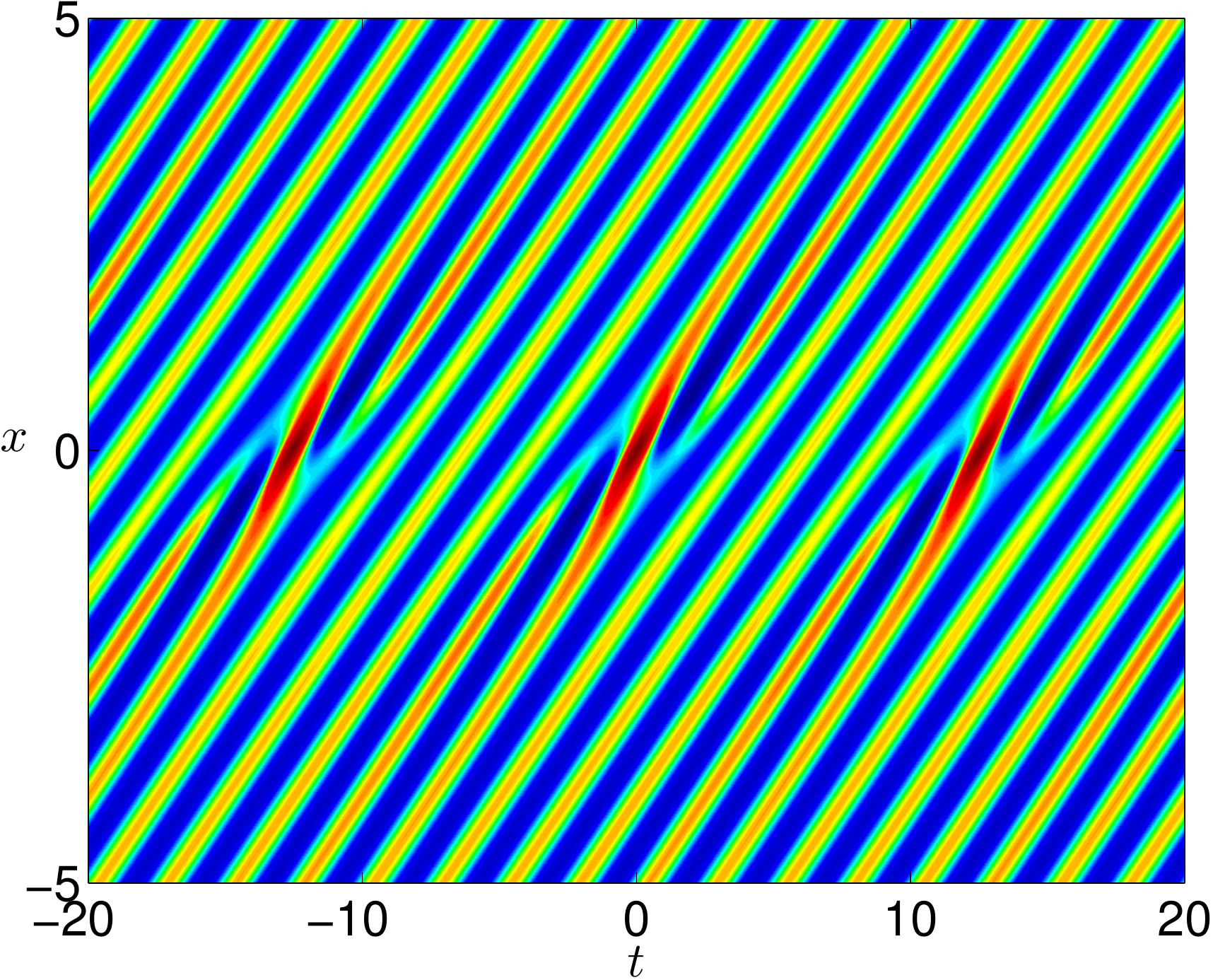}}
    \caption{(a). A three-dimensional plot of the absolute value of the SFB for the modulation frequency $\tilde{\nu} = 1/2$, the dispersion and the nonlinear coefficients of the NLS equation are taken to be 1, i.e. $\beta = 1 = \gamma$. (b). A density plot of the corresponding physical wave field.} \label{PlotSFB1}
  \end{center}
\end{figure}

As an illustration, Figure \ref{PlotSFB1}(a) shows a three-dimensional plot of $|A(\xi,\tau)|$ for $\tilde{\nu} = 1/2$. This plot represents the envelope of the wave signal that travels in the space variable $\xi$. We can observe that if it is viewed from the $\xi$-axis, this envelope has a soliton-like form with a non-vanishing background for $\xi \rightarrow \pm \infty$. If it is viewed from the $\tau$-axis, the envelope is periodic with `mountain-like' profile and `valley-like' profile come after one another for all $\tau \in \mathbb{R}$. \citeasnoun{Osborne00} call these `rogue waves' envelope and `holes', respectively. The corresponding physical wave field of any complex-valued amplitude of the NLS equation is given by \eqref{eta}. Figure \ref{PlotSFB1}(b) illustrates a density plot of the SFB physical wave field $\eta(x,t)$ excluding higher order terms for wavenumber $k_{0} = 2\pi$. A suitably chosen velocity is used to show it in a moving frame of reference. It is interesting to note that the wave field shows `wavefront dislocation' where waves are merging and splitting as they travel along the $x$-axis. Although SFB is defined for $0 < \tilde{\nu} < \sqrt{2}$, wavefront dislocation only occurs for $0 < \tilde{\nu} \leq \sqrt{3/2}$. This means that there is no wavefront dislocation for modulation frequency $\sqrt{3/2} < \tilde{\nu} < \sqrt{2}$. Thus, wavefront dislocation occurs in more than 86\% of the modulation frequency interval. More explanation for wavefront dislocation in surface water waves can be found in \cite{Karjanto07}. The concept of wavefront dislocation is originally introduced by \citeasnoun{Nye74} to explain the experimentally observed appearance and disappearance of crest and trough pairs in a wave field.
\begin{figure}[h]
  \begin{center}
    \subfigure[]{\includegraphics[width=0.45\textwidth]{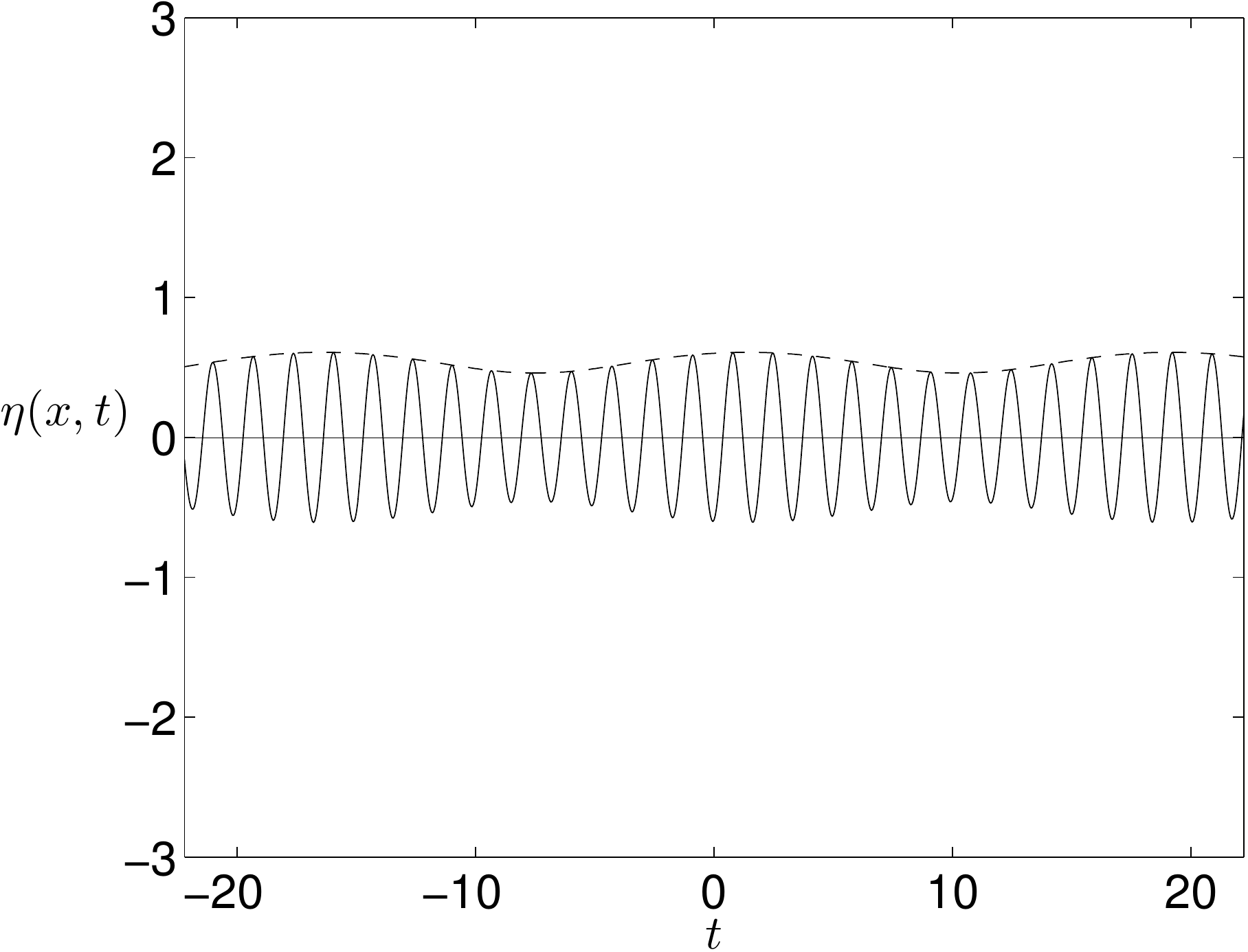}}    \hspace{1.0cm}
    \subfigure[]{\includegraphics[width=0.45\textwidth]{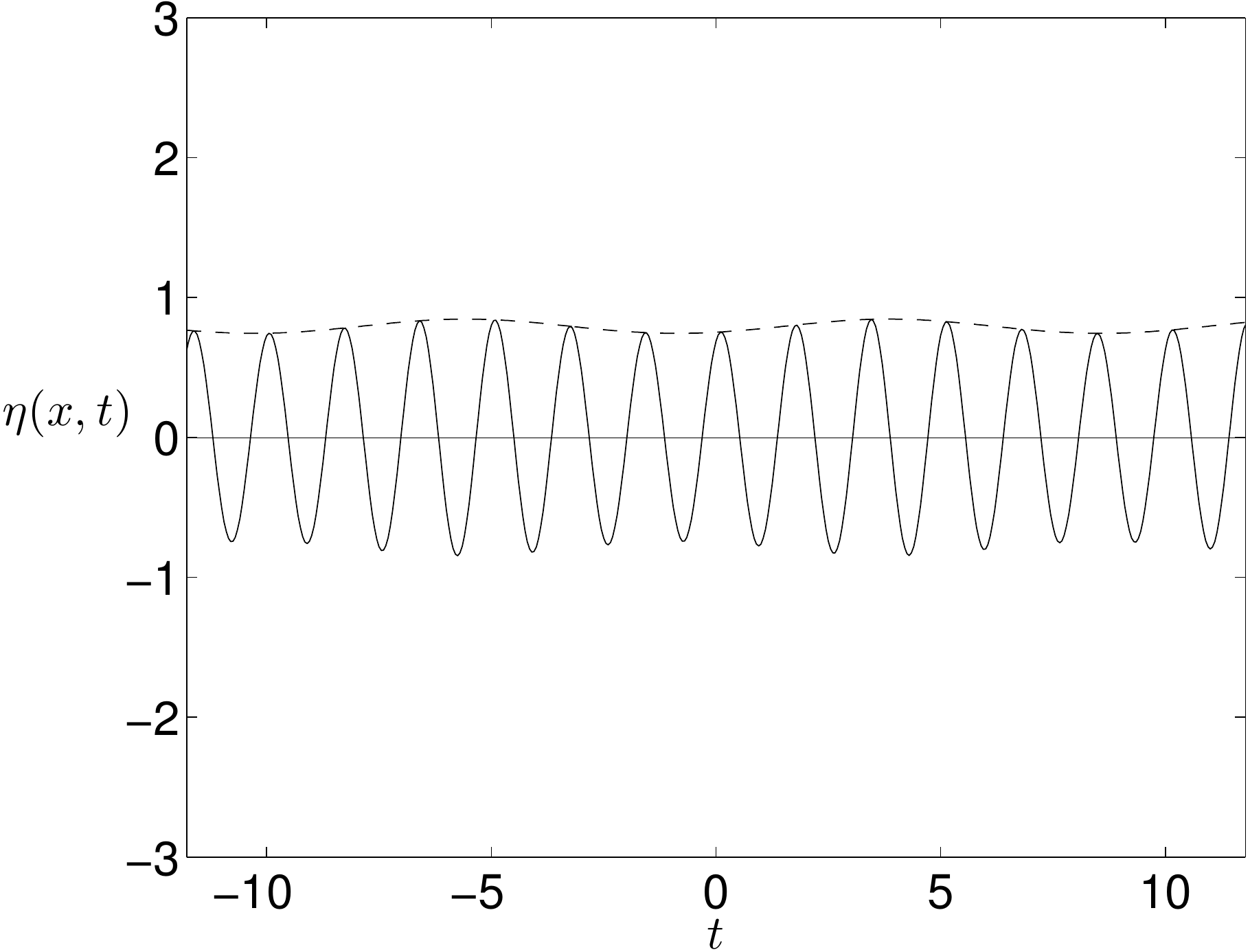}}        \vspace*{0.25cm} \\
    \subfigure[]{\includegraphics[width=0.45\textwidth]{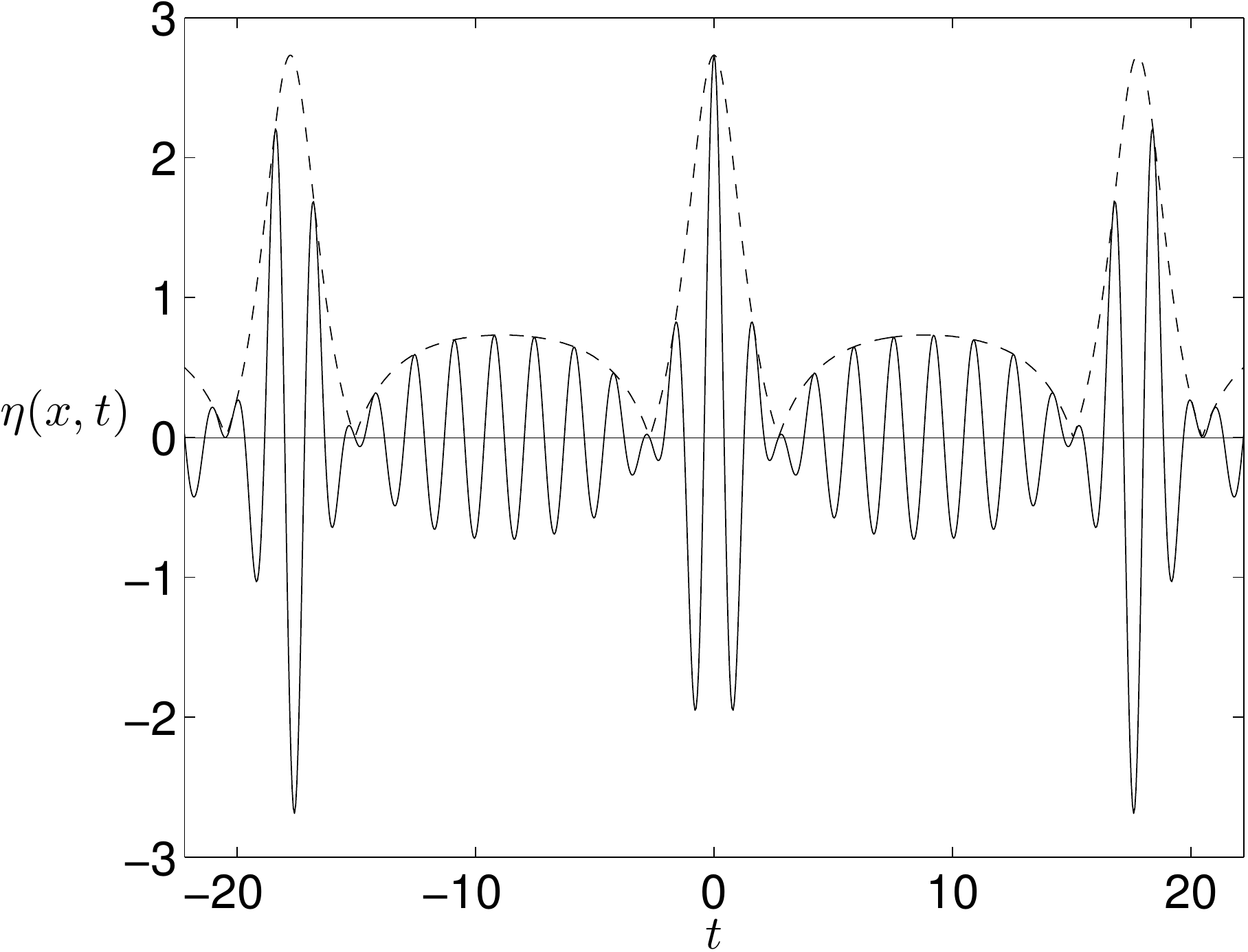}}     \hspace{1.0cm}
    \subfigure[]{\includegraphics[width=0.45\textwidth]{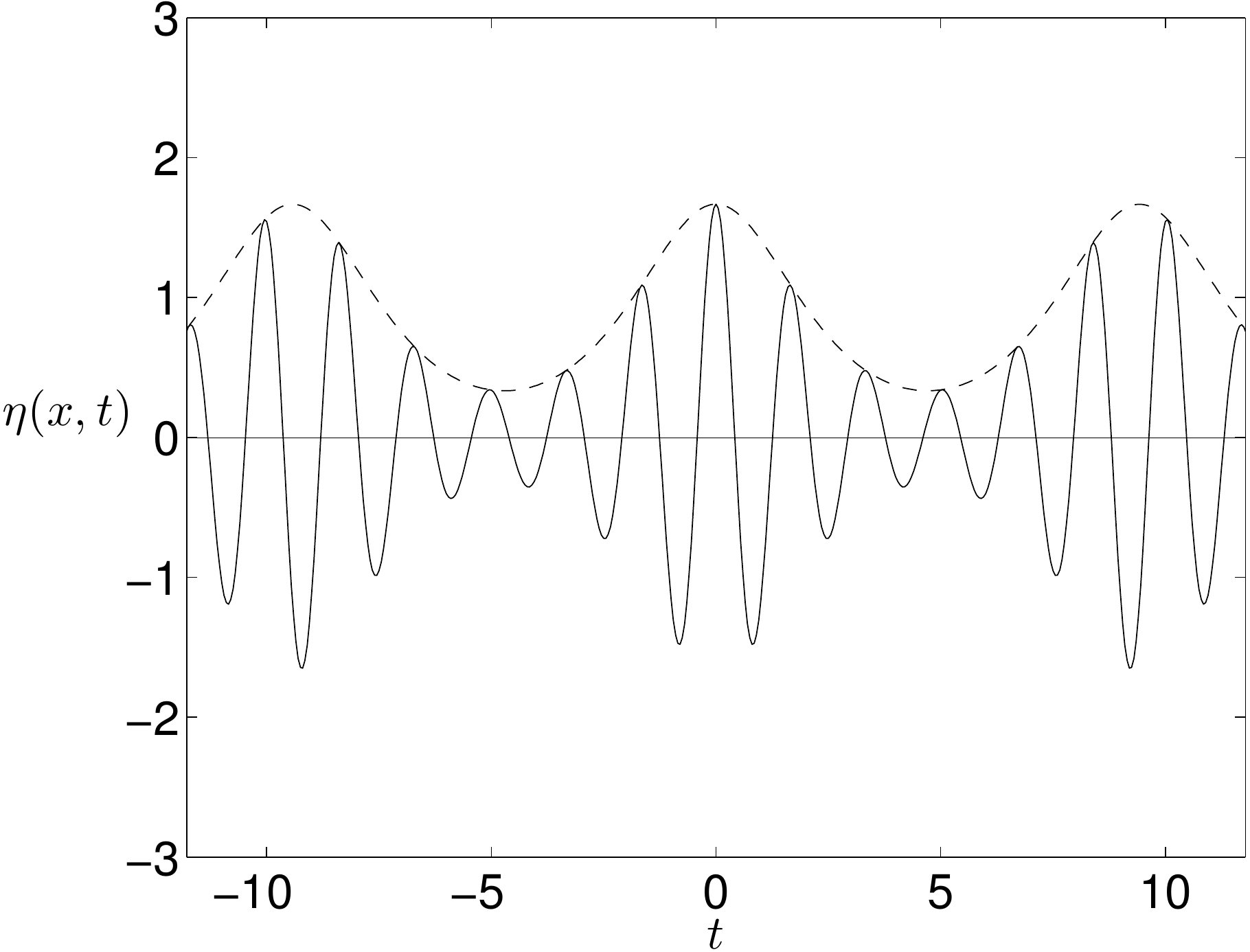}}
    \caption{The corresponding SFB wave signal (solid) and its envelope (dashed) for two different values of modulation frequency and at different positions. (a). $\tilde{\nu} = \sqrt{1/2}$ at $x = -10$; (b). $\tilde{\nu} = 4/3$ at $x = -20$; (c). $\tilde{\nu} = \sqrt{1/2}$ at $x = 0$ and (d). $\tilde{\nu} = 4/3$ at $x = 0$.} \label{SignalSFB}
  \end{center}
\end{figure}

A further illustration on the corresponding physical wave field of the SFB is presented in Figure \ref{SignalSFB}. SFB wave signals and their envelopes for two different modulation frequencies and different positions are depicted in this figure. We observe that the initial modulated wave signals at sufficiently far distance from $x = \xi = 0$ (Figure \ref{SignalSFB}(a) and \ref{SignalSFB}(b)), develop into relatively large and extreme wave signals at $x = 0$ (Figure \ref{SignalSFB}(c) and \ref{SignalSFB}(d)). We call these SFB wave signals at $x = 0$ the `extreme wave signals' and the position $x = 0$ the `extreme position'. Moreover, for modulation frequency $\tilde{\nu} = \sqrt{1/2}$, or for any value within the interval $0 < \tilde{\nu} \leq \sqrt{3/2}$, the envelope of the  wave signal vanishes at the extreme position and at two points within one modulation period. See Figure \ref{SignalSFB}(c). This is known as `phase singularity', which is closely related to wavefront dislocation we mentioned earlier. On the other hand, for $\tilde{\nu} = 4/3$, or for any value within the interval $\sqrt{3/2} < \tilde{\nu} < \sqrt{2}$, the envelope of the extreme wave signal remains positive for all time. See Figure \ref{SignalSFB}(d). Therefore we do not observe phase singularity or wavefront dislocation for the modulation frequency within that particular interval.

\subsubsection{Potential Energy Function}

\begin{figure}[h]
  \begin{center}
  \includegraphics[width=0.45\textwidth]{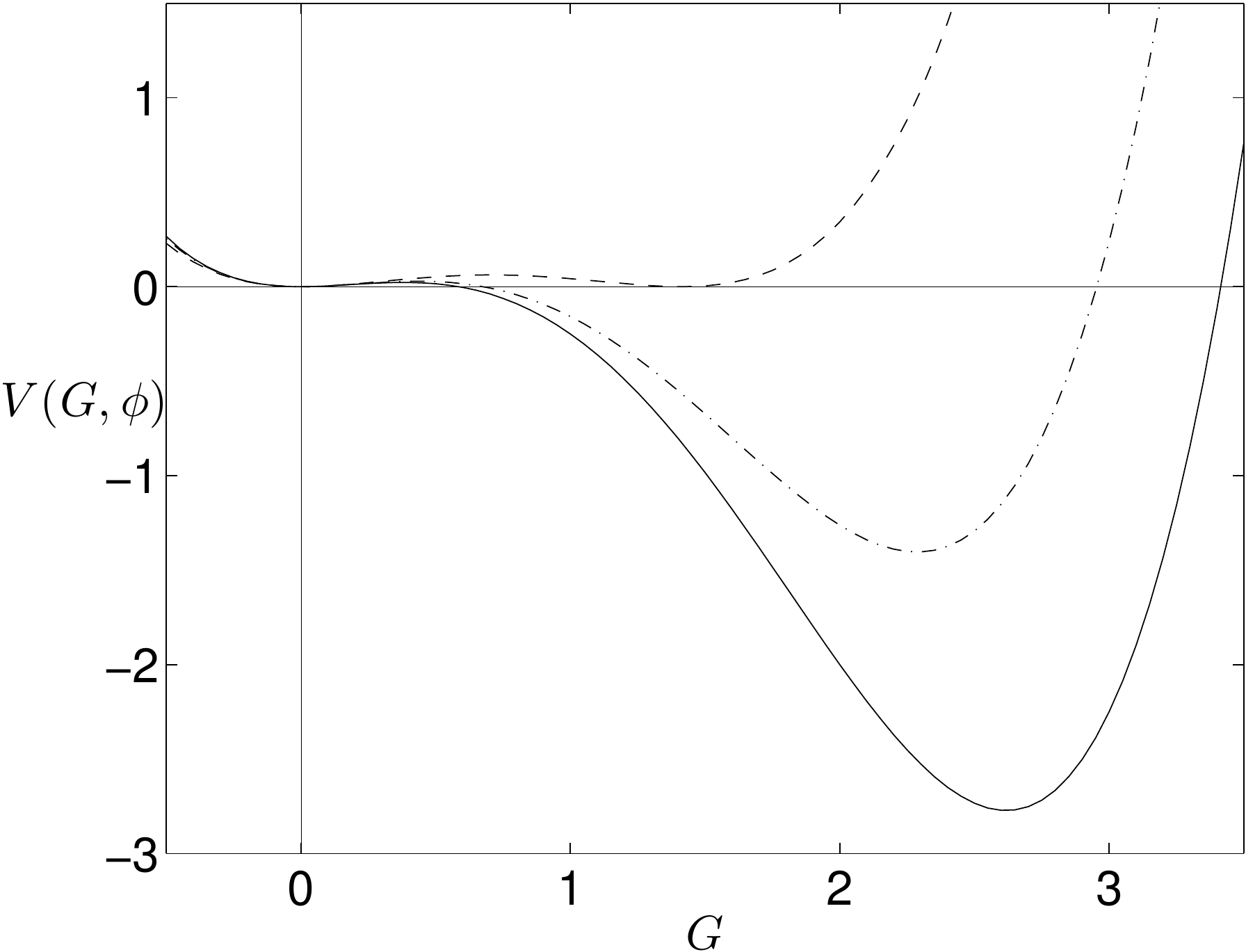}
  \caption{Sketch of the SFB potential energy function $V$ as a function of the displaced amplitude $G$ for $\tilde{\nu} = 1$ at different positions: $\xi = 0$ (solid), $\xi = \pm 1/2$ (dashed-dot) and $\xi \rightarrow \pm \infty$ (dashed).} \label{potenSFB}
  \end{center}
\end{figure}

Let us return again to our nonlinear oscillator equation \eqref{oscillator2} for the displaced amplitude $G$. For SFB, the corresponding potential energy function is given by \eqref{potential}, where the displaced phase satisfies the following expressions:
\begin{eqnarray}
  \cos \phi(\xi) &=& \frac{\tilde{\nu}^{2} \cosh (\sigma \xi)} {\sqrt{2\tilde{\nu}^{2} \cosh^{2}(\sigma \xi) - \tilde{\sigma}^{2}}}\\
  \textmd{and} \quad \phi'(\xi) &=& \frac{-\gamma r_{0}^{2} \tilde{\nu}^{2} \tilde{\sigma}^{2}}{2\tilde{\nu}^{2} \cosh^{2}(\sigma \xi) - \tilde{\sigma}^{2}}.
\end{eqnarray}
At $\xi = 0$, $\cos \phi(0) = 1$ and $\phi'(0) = -\gamma r_{0}^{2}(2 - \tilde{\nu}^{2})$, and thus the potential energy function becomes
\begin{equation}
  V(G,\phi(0)) = -\frac{1}{2} \gamma r_{0}^{2} (2 - \tilde{\nu}^{2}) G^{2} + \frac{1}{4} \gamma r_{0}^{2} G^{2} (G - 2)^{2}.
\end{equation}
For $\xi \rightarrow \pm \infty$, $\lim_{\xi \rightarrow \pm \infty} \cos \phi(\xi) = \tilde{\nu}/\sqrt{2}$ and $\lim_{\xi \rightarrow \pm \infty} \phi'(\xi) = 0$, and the potential energy function has asymptotic values
\begin{equation}
  \lim_{\xi \rightarrow \pm \infty} V(G,\phi(\xi)) = \frac{1}{4} \gamma r_{0}^{2} G^{2} (G - \tilde{\nu}\sqrt{2})^{2}.
\end{equation}
Figure \ref{potenSFB} shows the plot of the potential energy function $V$ as a function of the displaced amplitude $G$ for $\tilde{\nu} = 1$ at several positions. We observe that for $\xi \rightarrow \pm \infty$, the minimum value of the potential energy is zero and is reached when $G = \sqrt{2}$. The potential energy reaches its lowest minimum value at $\xi = 0$ when $G = \frac{1}{2}(3 + \sqrt{5}) \approx 2.618$.

\subsubsection{Asymptotic Behaviour}

It has been mentioned in the previous section that the SFB is a nonlinear extension of the Benjamin-Feir modulational instability. Now we would like to confirm that the asymptotic behaviour of the SFB for $\xi \rightarrow \pm \infty$ is indeed given by the Benjamin-Feir linear instability of a modulated wave signal. For $\xi \rightarrow \pm \infty$, the SFB \eqref{SFB} simply becomes the plane-wave solution with an additional phase $\phi_{0}$. So, the asymptotic behaviour of the SFB at the lowest order is given by:
\begin{equation*}
  \lim_{\xi \rightarrow \pm \infty} A(\xi,\tau) = A_{0}(\xi) e^{\mp i\phi_{0}},
\end{equation*}
where $e^{ i\phi_{0}} = \tilde{\nu}^{2} - 1 + i \tilde{\sigma}$. Thus, there is a phase shift of $2\phi_{0}$ during the evolution from $\xi = -\infty$ to $\xi = \infty$. However, in order to confirm the Benjamin-Feir initial signal, we also would like to know the linear term, namely the term that contains the sideband frequencies $\omega_{0} \pm \nu$.

The derivation given in this paper is only for $\xi \rightarrow -\infty$. One can apply a similar procedure to obtain the asymptotic behaviour for $\xi \rightarrow \infty$. Introduce a new variable $y = e^{\sigma \xi}$ to the SFB \eqref{SFB} without the plane-wave contribution since the plane-wave contributes a fast oscillation and distract the analysis. Denoting this complex-valued function as $F$, it reads:
\begin{equation}
  F(y,\tau) = \left(\frac{(\tilde{\nu}^{2} - i \tilde{\sigma})y^2 +
  \tilde{\nu}^{2} + i \tilde{\sigma}}{y^{2} + 1 - \sqrt{4 - 2\tilde{\nu}^{2}}\, y \,\cos (\nu \tau)} - 1\right).%
\end{equation}
Since for $\xi \rightarrow -\infty$, $y \rightarrow 0$, the function $F(y,\tau)$ can be expanded into a Taylor series about $y = 0$ as follows:
\begin{equation}
  F(y,\tau) = F(0,\tau) + \partial_{y}F(0,\tau)y + \frac{1}{2} \partial_{y}^{2}F(0,\tau)y^{2} + \dots \, .
\end{equation}
After some simple manipulations we arrive at
\begin{eqnarray}
  F(0,\tau) &=& \tilde{\nu}^{2} - 1 + i \tilde{\sigma}, \\
  \partial_{y}F(0,\tau) &=& \sqrt{4 - 2\tilde{\nu}^{2}}(\tilde{\nu}^{2} + i\tilde{\sigma})\cos(\nu \tau).%
\end{eqnarray}
Therefore, the asymptotic behaviour for the SFB at $\xi
\rightarrow \mp \infty$ is given by\index{asymptotic
behaviour!SFB}
\begin{equation}
  A(\xi,\tau) \approx A_{0}(\xi) \, \left[e^{\pm i\phi_{0}} +
  \sqrt{4 - 2\tilde{\nu}^{2}} e^{\pm i\phi_{1}} e^{\pm \sigma \xi} \cos (\nu \tau) \right],
  \label{SFB_asymp}
\end{equation}
where
\begin{equation}
  \tan \phi_{0} = \frac{\tilde{\sigma}}{\tilde{\nu}^{2} - 1}
  \qquad \textmd{and} \qquad
  \tan \phi_{1} = \frac{\tilde{\sigma}}{\tilde{\nu}^{2}}.
  \label{phaseshift}
\end{equation}
We observe that the phases of the sideband frequencies are equal and therefore in the previous section we concluded that $\phi_{1} = \phi_{2}$ even though initially we were not aware of this fact. The relationship between modulational instability and the asymptotic behaviour of the SFB was described already before \cite{Akhmediev86}.

\subsection{Ma Solution}

For the choice $\zeta(\tau) = \cosh(\mu \tau)$, where $\mu =
r_{0} \sqrt{\frac{\gamma}{\beta}} \tilde{\mu}$, the corresponding family of waves on finite background is known as the Ma solution \cite{Ma79}. The differential
equation for $G$ becomes:
\begin{equation}
  \partial_{\tau}^{2}G = \mu^{2} G - 3\mu^{2}\frac{Q}{P} G^{2} - 2\mu^{2} \frac{1 - Q^{2}}{P^{2}}
  G^{3}. \nonumber
\end{equation}
Comparing again with (\ref{oscillator}), the solution
(\ref{AnsatzG}) is obtained with $P(\phi) = -\tilde{\mu}^{2}
Q(\phi)/\cos \phi$, $Q^{2}(\phi) = 2 \cos^{2}\phi/(2 \cos^{2}\phi
+ \tilde{\mu}^{2})$ and the displaced phase satisfies
\begin{equation}
  \tan \phi(\xi) = - \frac{\tilde{\rho}}{\tilde{\mu}^{2}} \tan(\rho \xi) \label{disphiMa}
\end{equation}
where $\tilde{\rho} = \tilde{\mu} \sqrt{2 + \tilde{\mu}^{2}}$ and $\rho = \gamma r_{0}^{2} \tilde{\rho}$. The displaced amplitude $G$ reads:
\begin{eqnarray}
  G(\phi(\xi),\tau) &=& -\frac{\tilde{\mu}^{2} \sqrt{2}}{\sqrt{2} \cos \phi \mp \sqrt{2 \cos^{2}\phi + \tilde{\mu}^{2}} \cosh (\mu \tau)} \\
  &=& \frac{\tilde{\mu} \sqrt{2} \sqrt{\tilde{\rho}^{2} - 2\tilde{\mu}^{2} \cos^{2}(\rho \xi)}}{\tilde{\mu} \sqrt{2} \cos(\rho \xi) \pm \tilde{\rho} \cosh(\mu \tau)}. \label{disampMa}
\end{eqnarray}
\begin{figure}[h]
  \begin{center}
  \subfigure[]{\includegraphics[width=0.45\textwidth]{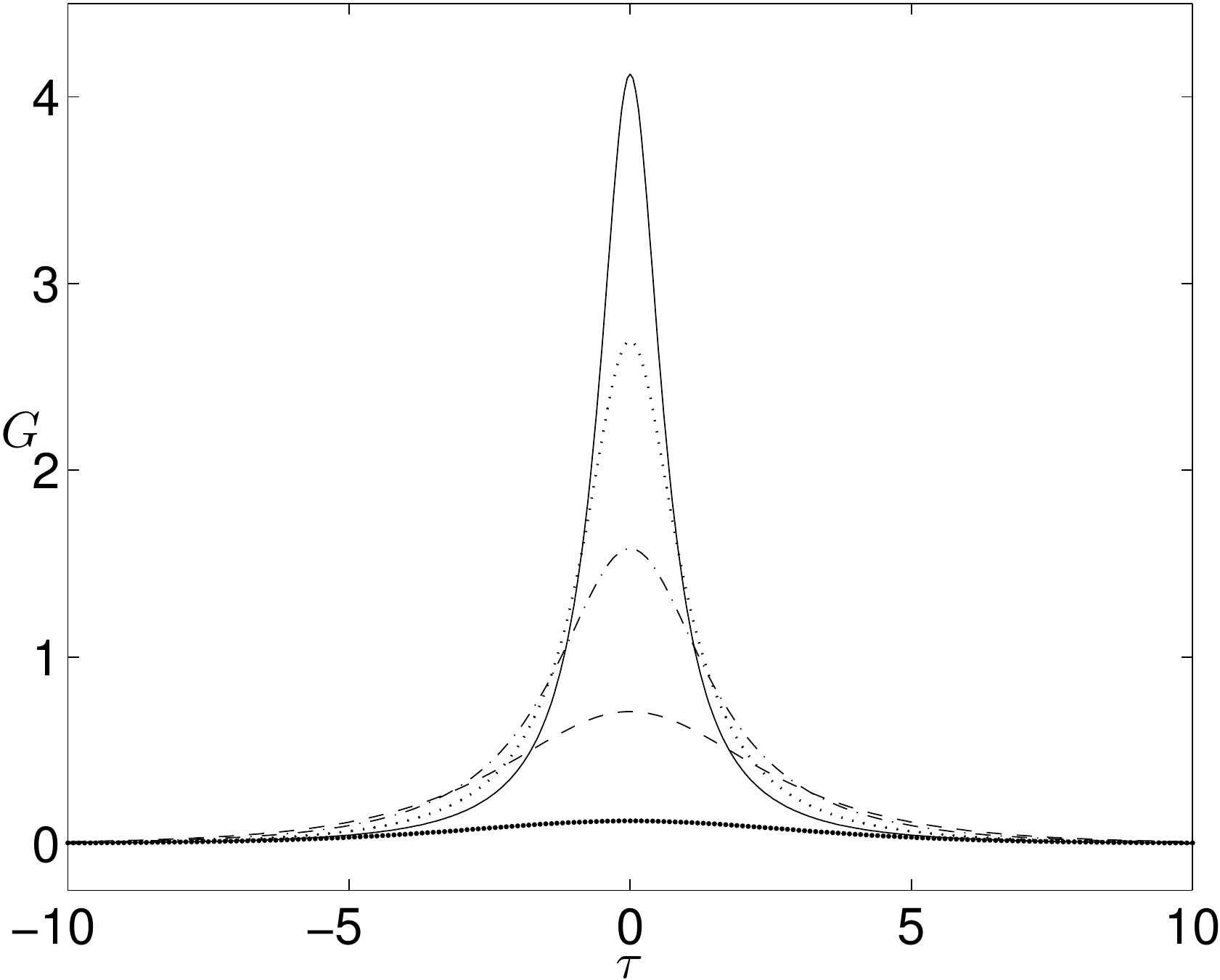}}
  \subfigure[]{\includegraphics[width=0.45\textwidth]{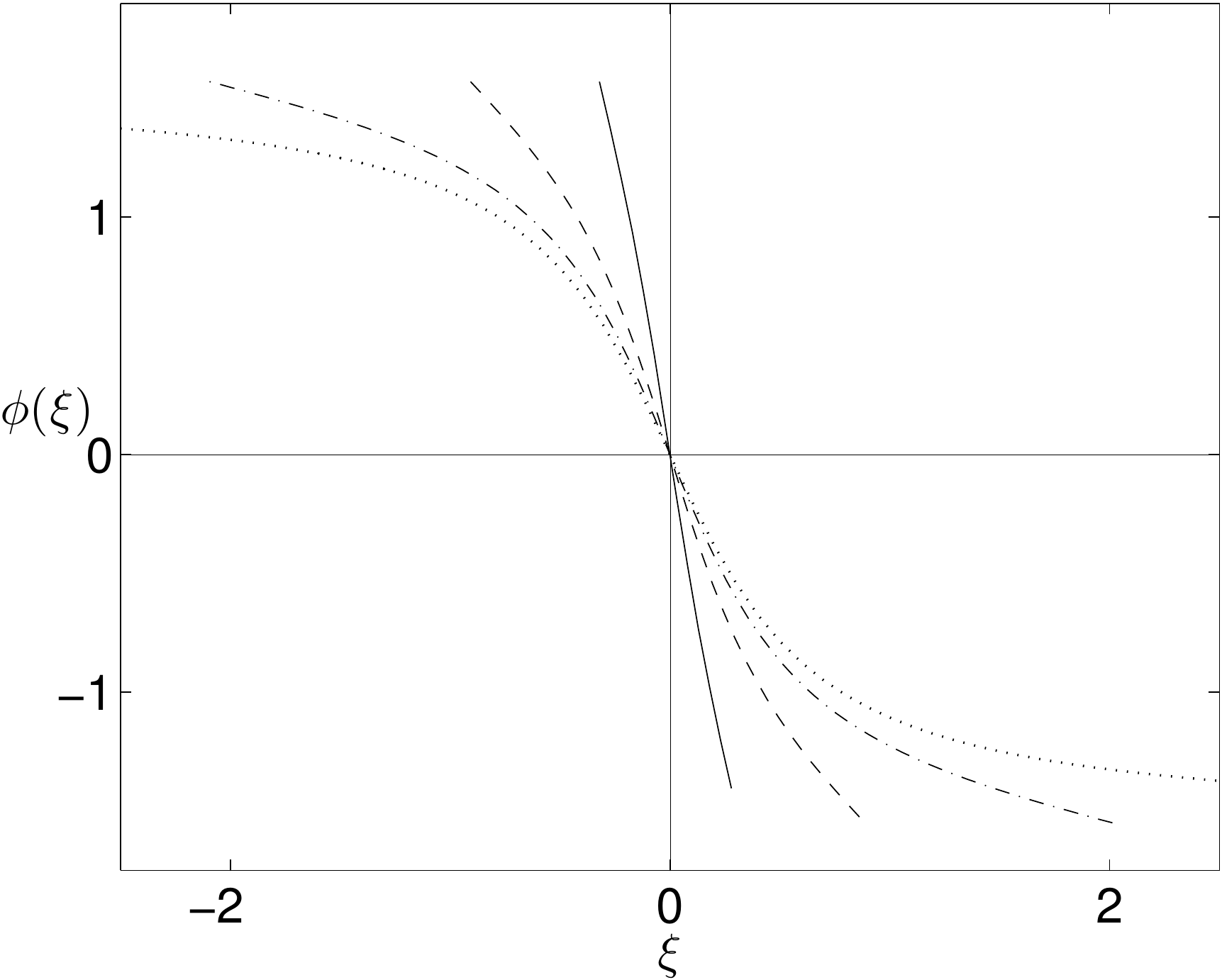}}
  \caption[]{(a). Sketch of the corresponding displaced amplitude of the Ma solution $G$ for $\tilde{\mu} = 1/2$ at different positions: $\xi = 0$ (heavy dotted), $\xi = \pi/(2\rho)$ (dashed), $\xi = 3\pi/(4\rho)$ (dashed-dot), $\xi = 7\pi/(8\rho)$ (light dotted) and $\xi = \pi/\rho$ (solid).
  (b). Sketch of the displaced phase $\phi$ for different values of $\tilde{\mu}$: $\tilde{\mu} = 4$ (solid), $\tilde{\mu} = 1$ (dashed), $\tilde{\mu} = 1/2$ (dashed-dot) and $\tilde{\mu} \rightarrow 0$ (dotted).} \label{disamphiMa}
  \end{center}
\end{figure}

Figure \ref{disamphiMa} shows plots of the corresponding displaced amplitude \eqref{disampMa} for $\tilde{\mu} = 1/2$ and the displaced phase \eqref{disphiMa} for several values of $\tilde{\mu}$ of the Ma solution. The displaced amplitude shows a periodic behaviour in $\xi$ for a spatial period of $2\pi/\rho$. It is almost flat at $\xi = 2n\pi/\rho$, $n \in \mathbb{Z}$ and it reaches its maximum at $\xi = (2n + 1)\pi/\rho$, $n \in \mathbb{Z}$. The displaced phase plots shown in Figure \ref{disamphiMa}(b) are only sketched for one spatial period. We observe that the smaller the value of $\tilde{\mu}$, the larger the period will be. For $\tilde{\mu} \rightarrow 0$, the period becomes infinite. The displaced phase for all $\tilde{\mu} > 0$ vanishes at $\xi = 0$, and thus the corresponding displaced amplitude at $\xi = 0$ becomes a real-valued function.

\subsubsection{Physical Characteristics}

The corresponding solution of the NLS equation (\ref{NLSsolution}) can be written after some manipulations as:
\begin{equation}
  A(\xi,\tau) = A_{0}(\xi) \left(\frac{-\tilde{\mu}^{2} \cos(\rho \xi) + i \tilde{\rho} \sin(\rho \xi)}{\cos(\rho \xi) \pm \sqrt{1 + \frac{1}{2} \tilde{\mu}^{2}} \cosh(\mu \tau)} - 1 \right).
\end{equation}
This is an explicit expression of the Ma solution. It has maxima at $(\xi,\tau) = (2n\pi/\rho,0)$, $n \in \mathbb{Z}$. It is periodic in space $\xi$ with spatial period depending on $\rho$ and has a soliton-like shape in $\tau$. For $\tau \rightarrow \pm \infty$, the profile goes to the plane-wave solution as a finite background. The smaller $\mu$ the smaller $\rho$, and the larger the spatial period. For $\mu \rightarrow 0$, similar as the SFB, the Ma solution also reduces to the rational solution that will be discussed in Subsection \ref{subsecrational}.
\begin{figure}[h]
  \begin{center}
    \subfigure[]{\includegraphics[width=0.45\textwidth]{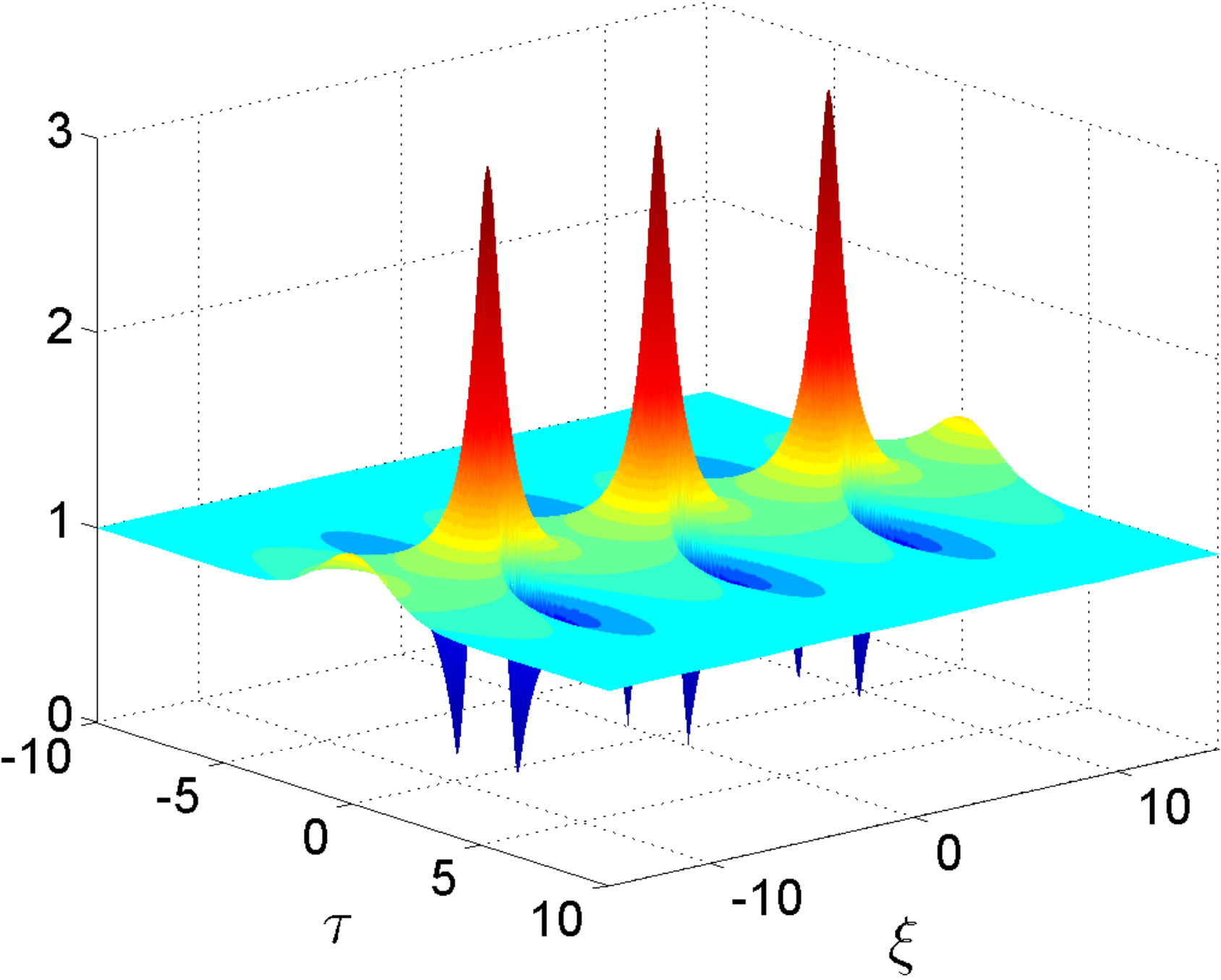}}         \hspace{1.0cm}
    \subfigure[]{\includegraphics[width=0.4\textwidth]{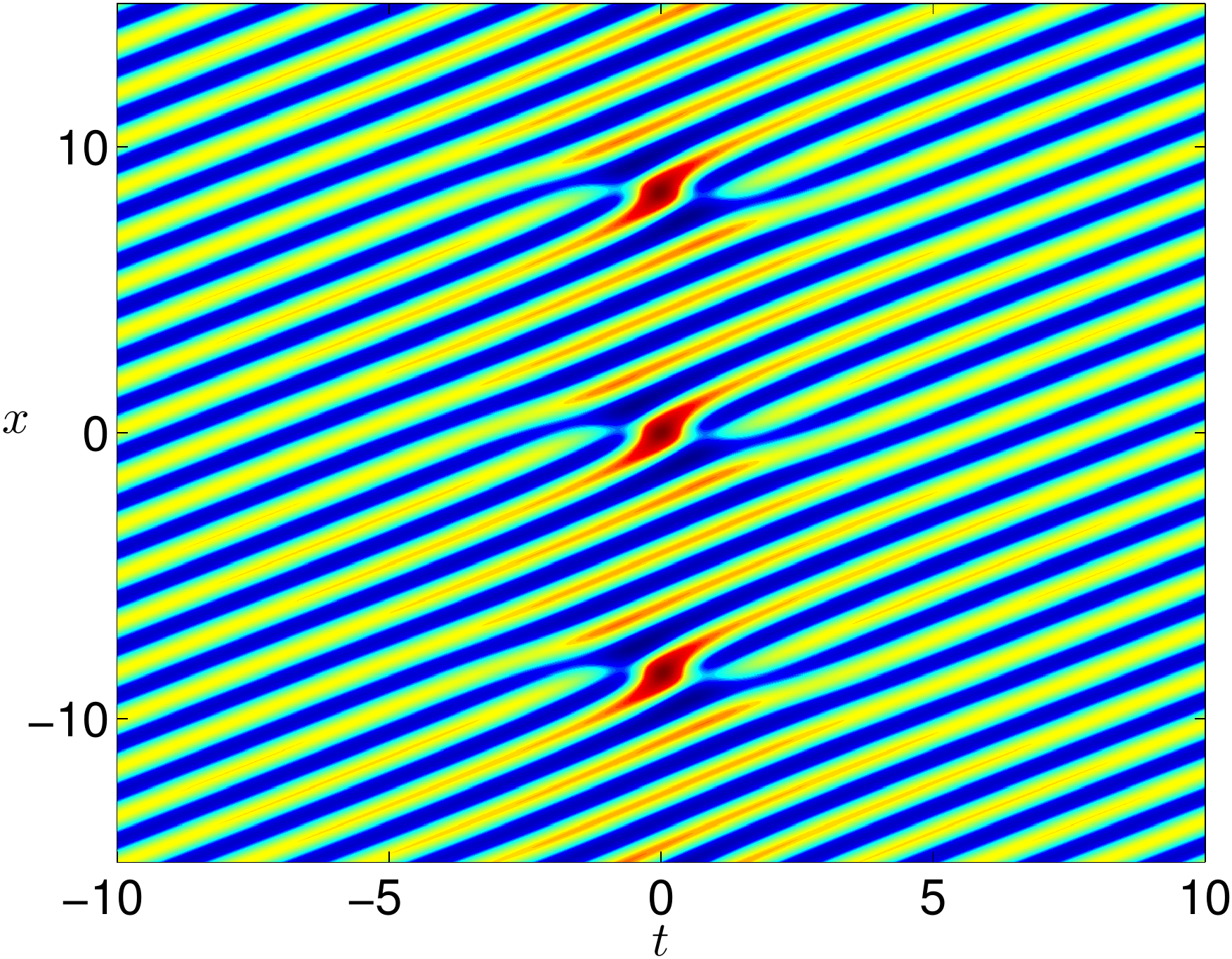}}
    \caption{(a). A three-dimensional plot of the absolute value of the Ma solution for $\tilde{\mu} = 1/2$, the dispersion and the nonlinear coefficients of the NLS equation are taken to be 1, i.e. $\beta = 1 = \gamma$. (b). A density plot of the corresponding physical wave field.} \label{PlotMa}
  \end{center}
\end{figure}
But for $\mu \rightarrow \infty$, the Ma solution becomes the single-soliton or bright-soliton solution of the NLS equation, given explicitly as follows \cite{Akhmediev97}:
\begin{equation}
  A(\xi,\tau) = A_{0}(\xi)\, \sqrt{2} \sech \left(r_{0} \sqrt{\frac{\gamma}{\beta}} \tau\right).
\end{equation}
As an illustration, Figure \ref{PlotMa} shows a three-dimensional plot of the absolute value of the Ma solution for $\tilde{\mu} = 1/2$ and its corresponding physical wave field shown in a moving frame of reference with a suitably chosen velocity. A similar phenomenon of wavefront dislocation is observed for all values of $\mu$ as the waves travel along the $x$-axis.

\begin{figure}[h]
  \begin{center}
    \subfigure[]{\includegraphics[width=0.45\textwidth]{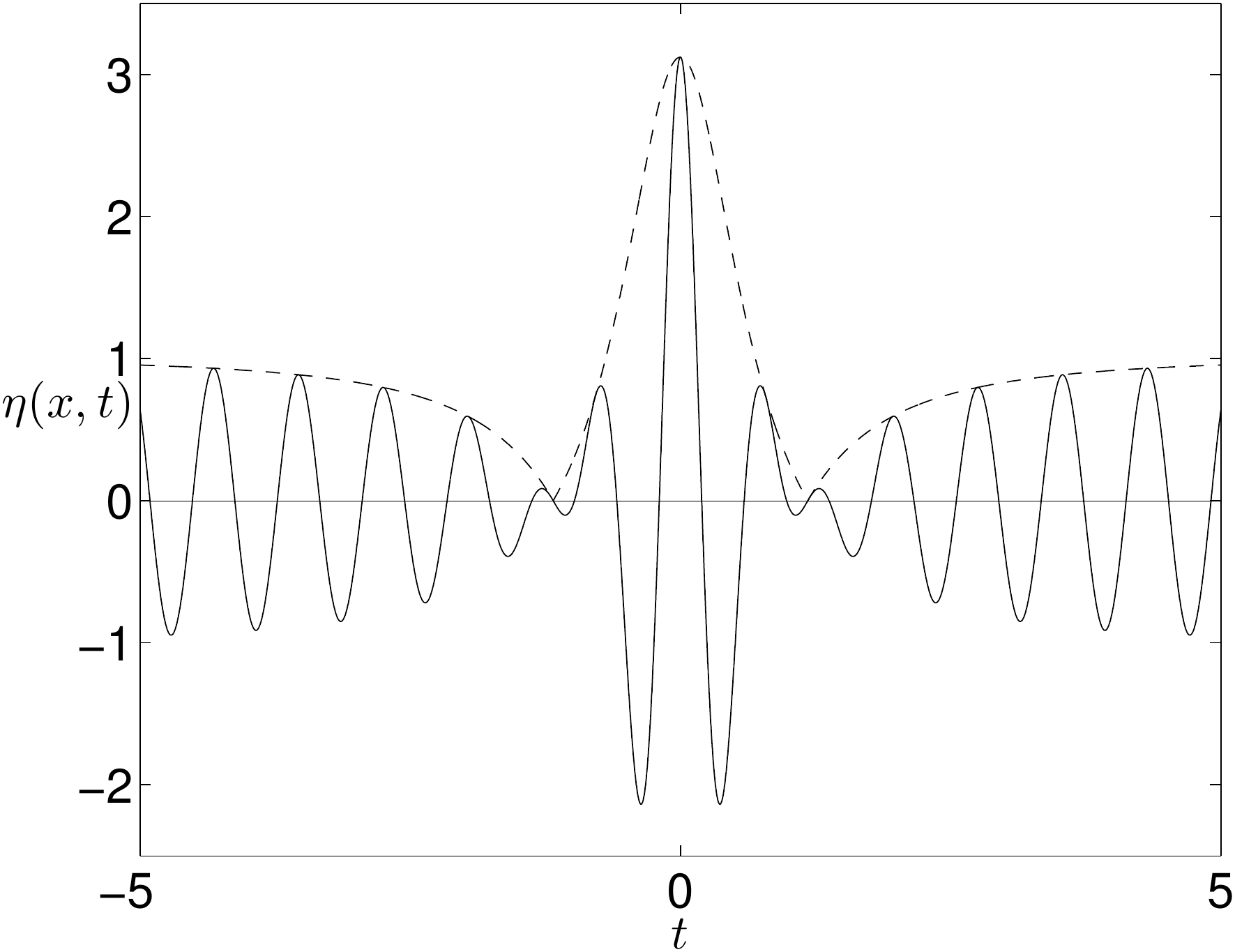}}    \hspace{1.0cm}
    \subfigure[]{\includegraphics[width=0.45\textwidth]{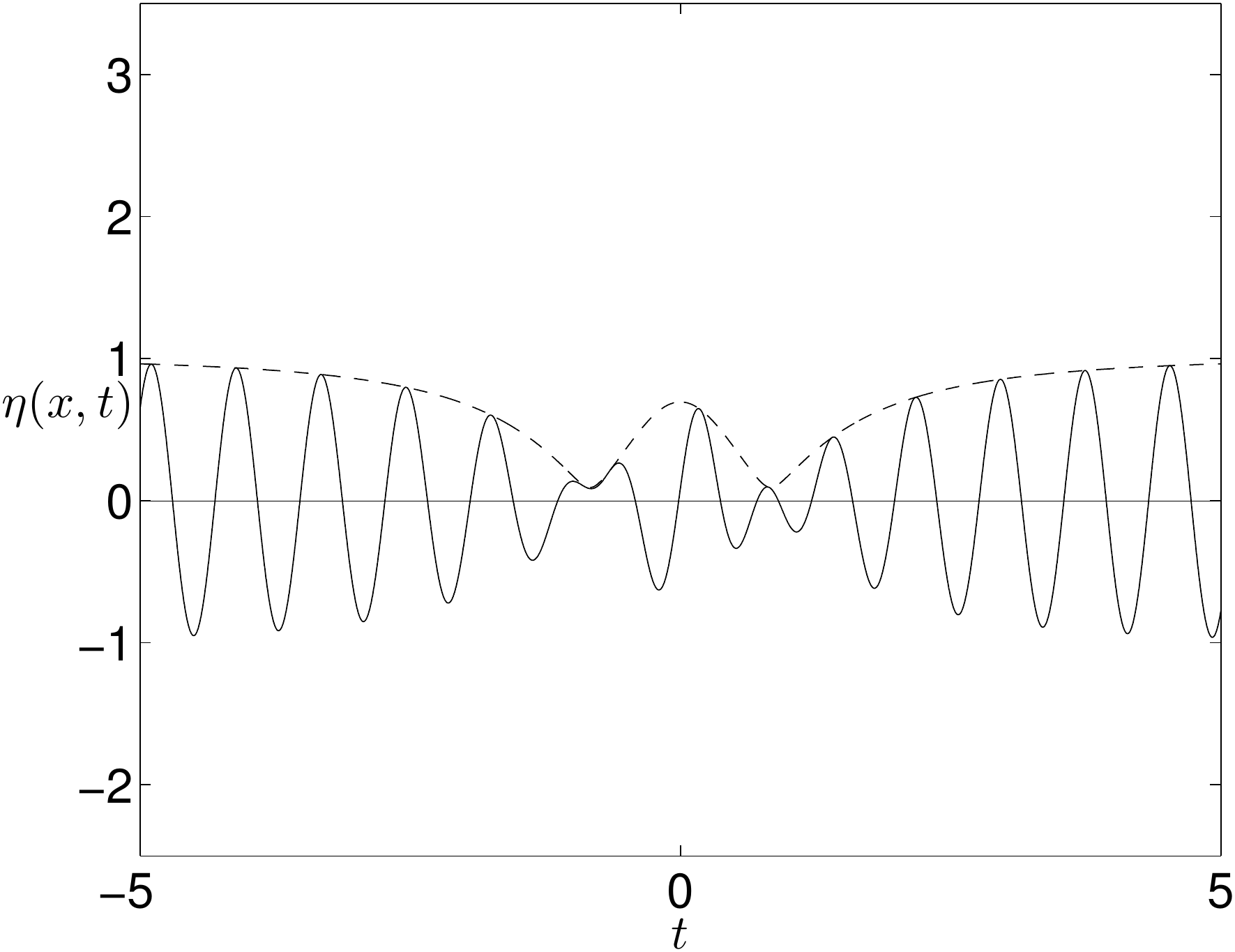}}   \vspace*{0.25cm} \\
    \subfigure[]{\includegraphics[width=0.45\textwidth]{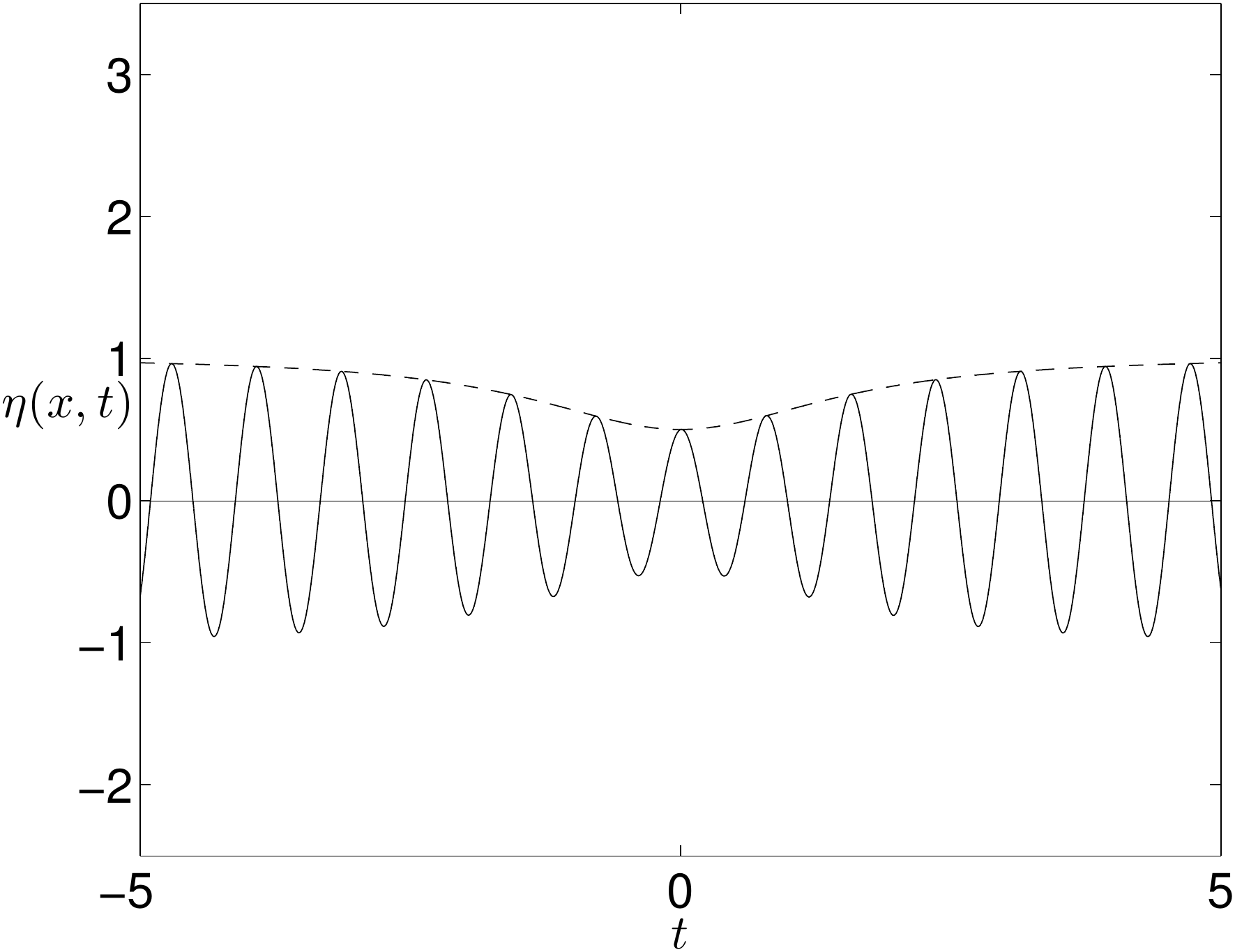}}   \hspace{1.0cm}
    \subfigure[]{\includegraphics[width=0.45\textwidth]{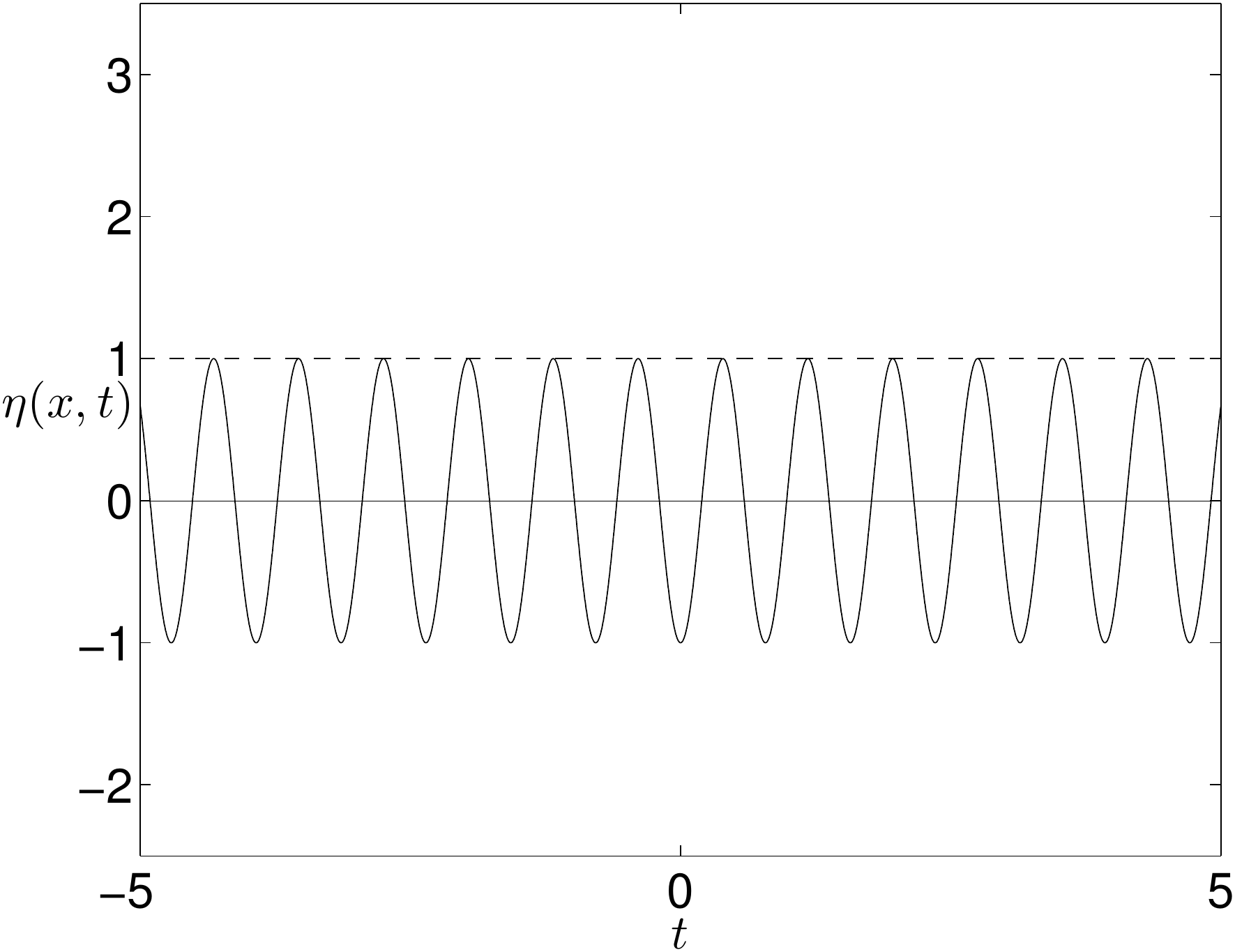}}
    \caption{The corresponding Ma wave signals (solid) and their envelopes (dashed) for $\tilde{\mu} = 1/2$ at different positions: (a). $x = 0$; (b). $x = \pi/(8\rho)$; (c). $x = \pi/(4\rho)$ and (d). $x = \pi/(2\rho)$.} \label{SignalMa}
  \end{center}
\end{figure}

Figure \ref{SignalMa} shows the corresponding Ma wave signals and their envelopes for $\tilde{\mu} = 1/2$ at different positions. We observe that at $x = 0$ the wave signal reaches its highest maximum, making this one as an extreme wave signal for the Ma solution. As the position progresses, the middle part of the signal decreases and increases to the state of the plane-wave solution as a finite background at $x = \pi/(2\rho)$. Furthermore, from $x = \pi/(2\rho)$ until $x = 3\pi/(2\rho)$, there is a slight increase of the middle part of the signal but it does not really give any significant contribution. From $x = 3\pi/(2\rho)$ until $x = 2\pi/\rho$, the reverse process of the one from $x = 0$ till $x = \pi/(2\rho)$ occurs. Within this interval, the middle part of the signal also decreases and increases until it finally reaches its highest maximum at $x = 2\pi/\rho$. A similar process occurs periodically for a distance of $2\pi/\rho$ in both positive and negative $x$ directions.

\subsubsection{Potential Energy Function}

\begin{figure}[h!]
  \begin{center}
  \includegraphics[width=0.45\textwidth]{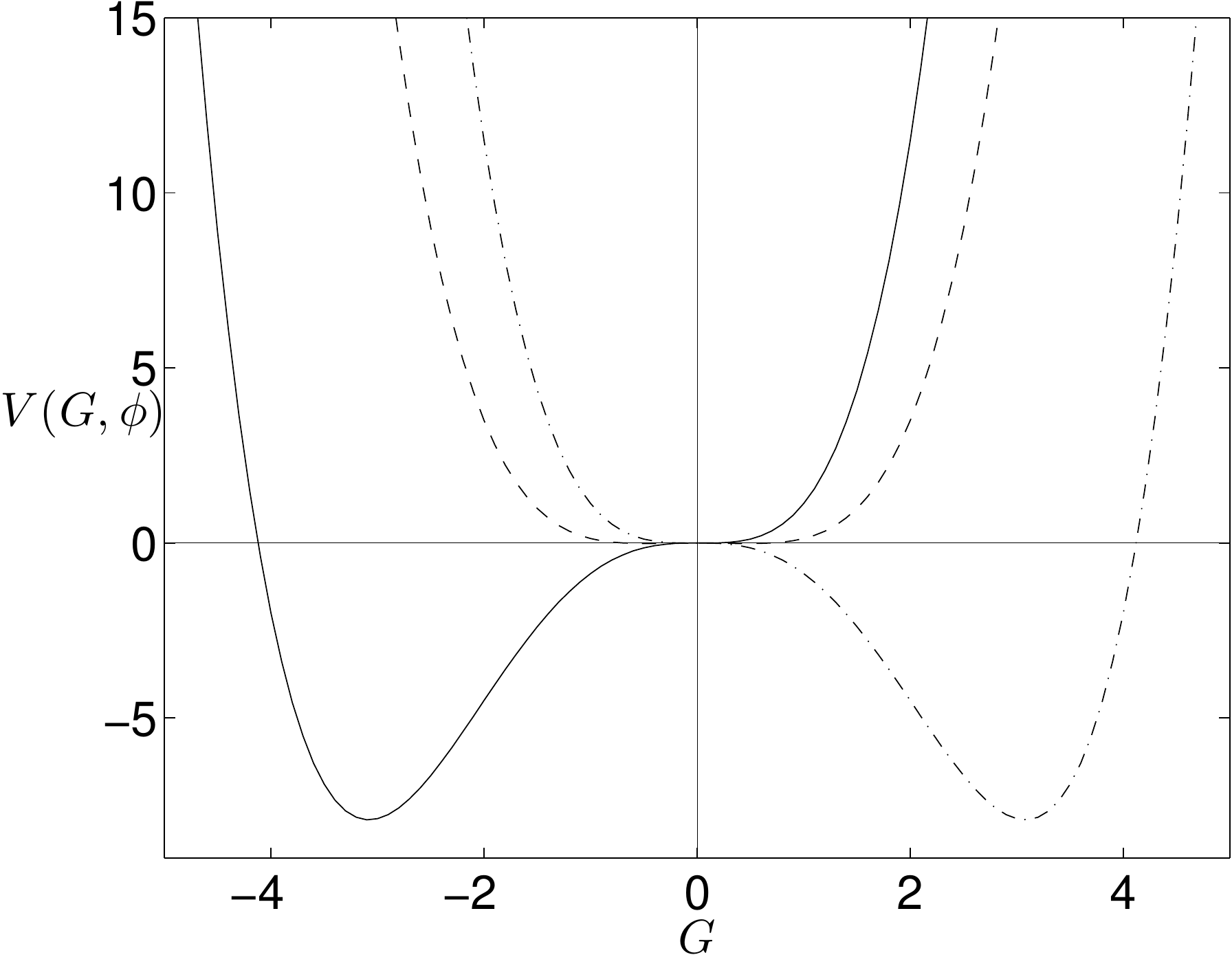}
  \caption{Sketch of the Ma solution potential energy function $V$ as a function of the displaced amplitude $G$ for $\tilde{\mu} = 1/2$ at different positions: $\xi = 2n\pi/\rho$ (solid), $\xi = n\pi/\rho$ (dashed-dot) and $\xi = (n + 1/2)\pi/\rho$ (dashed), $n \in \mathbb{Z}$.} \label{potenMa}
  \end{center}
\end{figure}

The corresponding potential energy function is also given by \eqref{potential}, where the displaced phase is given by the following conditions:
\begin{eqnarray}
  \cos \phi(\xi) &=& \frac{-\tilde{\mu}^{2} \cos (\rho \xi)} {\sqrt{\tilde{\rho}^{2} - 2\tilde{\mu}^{2} \cos^{2}(\rho \xi)}}\\
  \textmd{and} \quad \phi'(\xi) &=& \frac{-\gamma r_{0}^{2} \tilde{\mu}^{2} \tilde{\rho}^{2}}{\tilde{\rho}^{2} - 2 \tilde{\mu}^{2} \cos^{2} (\rho \xi)}.
\end{eqnarray}
At $\xi = (n + \frac{1}{2})\pi/\rho$, $n \in \mathbb{Z}$, $\cos \phi = 0$ and $\phi'(\xi) = -\gamma r_{0}^{2}\tilde{\mu}^{2}$, and thus the potential energy function becomes
\begin{equation}
  V(G,\phi) = -\frac{1}{2} \gamma r_{0}^{2} \tilde{\mu}^{2} G^{2} + \frac{1}{4} \gamma r_{0}^{2} G^{4}.
\end{equation}
For $\xi = n\pi/\rho$ and $\xi = 2n\pi/\rho$, $n \in \mathbb{Z}$, $\cos \phi = \pm 1$, and $\phi'(\xi) = -\gamma r_{0}^{2}(\tilde{\mu}^{2} + 2)$, and thus the potential energy function reads
\begin{equation}
 V(G,\phi) = -\frac{1}{2} \gamma r_{0}^{2} (\tilde{\mu}^{2} + 2) G^{2} + \frac{1}{4} \gamma r_{0}^{2} G^{2} (G \mp 1)^{2}
\end{equation}
respectively. Figure \ref{potenMa} shows the plot of the potential energy function $V$ as a function of the displaced amplitude $G$ for $\tilde{\mu} = 1/2$ at several positions. We observe that for $\xi = (n + 1/2)\pi/\rho$, $n \in \mathbb{Z}$, the curve is symmetric with respect to the vertical axis, i.e. $G = 0$, and it has a minimum value at the origin. The curve at $\xi = 2n\pi/\rho$ is simply a mirror of the curve at $\xi = n\pi/\rho$, $n \in \mathbb{Z}$, with respect to the vertical axis. The lowest minimum values are also attained at these positions.

\subsection{Rational Solution} \label{subsecrational}

Finally, for the choice of $\zeta(\tau) = 1 - \frac{1}{2} \nu^{2} \tau^{2}$, we obtain the waves on finite background known as the rational solution \cite{Peregrine83}.
Similarly, the same solution can also be obtained by substituting
$\zeta(\tau) = 1 + \frac{1}{2} \mu^{2} \tau^{2}$. As a
consequence, the differential equation for $G$ now becomes:
\begin{equation}
  \partial_{\tau}^{2}G = \frac{3\nu^{2}}{P} G^{2} - 4\nu^{2} \frac{Q - 1}{P^{2}} G^{3}. \nonumber%
\end{equation}
Comparing again with (\ref{oscillator}), we have $P(\phi) =
\tilde{\nu}^{2}/\cos \phi$, $Q(\phi) = 1 + P^{2}/(4\tilde{\nu}^{2})$ and the displaced phase satisfies $\tan \phi(\xi) = - 2\gamma r_{0}^{2} \xi$.
Substituting these into the Ansatz (\ref{AnsatzG}), the displaced amplitude $G$ reads:
\begin{equation}
  G(\phi(\xi),\tau) = \frac{4 \cos \phi}{1 + 2 \frac{\gamma}{\beta} r_{0}^{2} \cos^{2} \tau^2} = \frac{4\sqrt{1 + 4 (\gamma r_{0}^{2} \xi)^{2}}}{1 + 4 (\gamma r_{0}^{2} \xi)^{2} + 2 \frac{\gamma}{\beta} r_{0}^{2} \tau^{2}}
\end{equation}
and the rational solution is given by:
\begin{equation}
  A(\xi,\tau) = A_{0}(\xi) \left(\frac{4 (1 - 2 i \gamma r_{0}^{2} \xi)}
                {1 + 4 (\gamma r_{0}^{2} \xi)^{2} + 2 \frac{\gamma}{\beta} r_{0}^{2} \tau^{2}} - 1 \right).
\end{equation}

\begin{figure}[h]
  \begin{center}
    \subfigure[]{\includegraphics[width=0.45\textwidth]{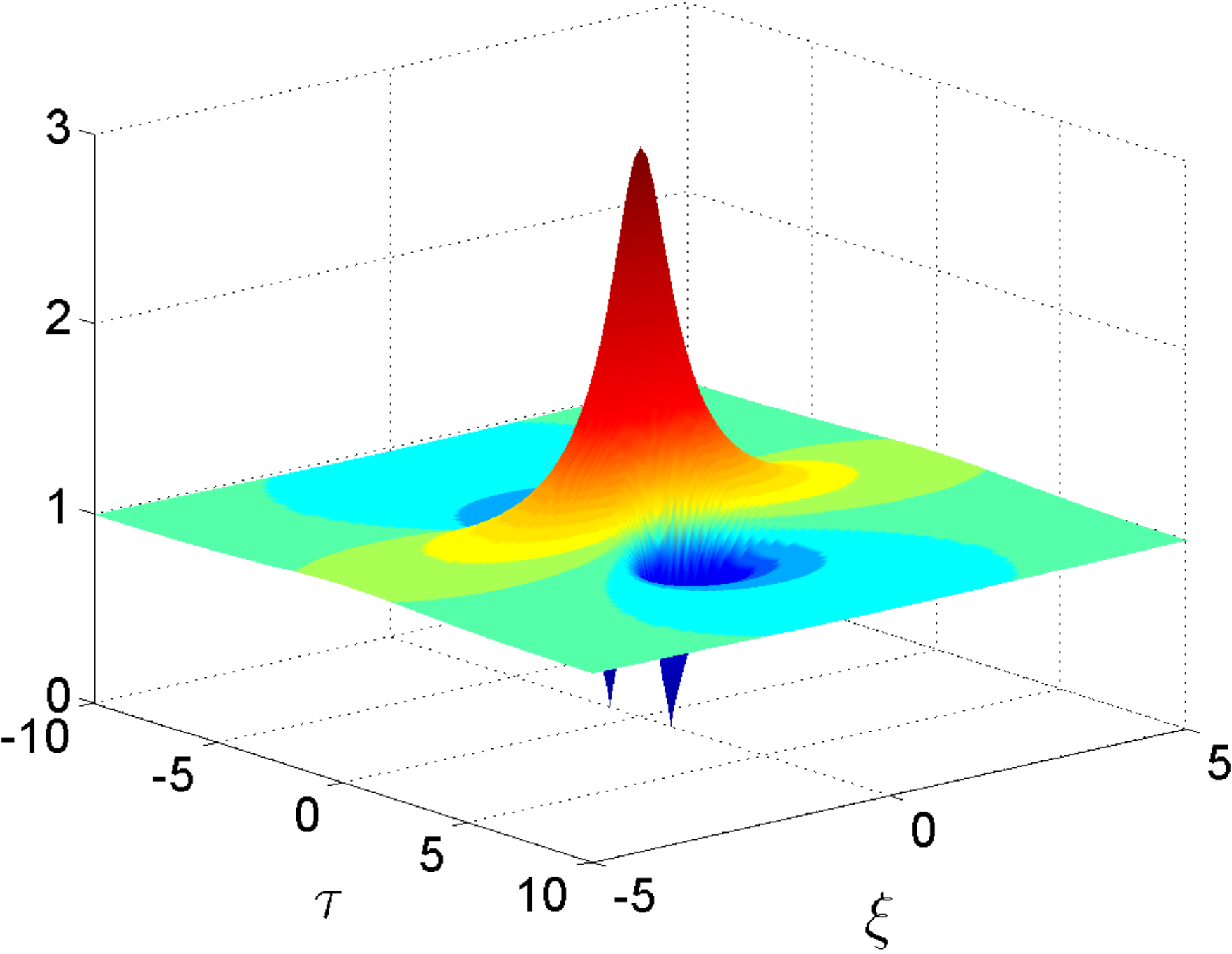}}     \hspace{1.0cm}
    \subfigure[]{\includegraphics[width=0.4\textwidth]{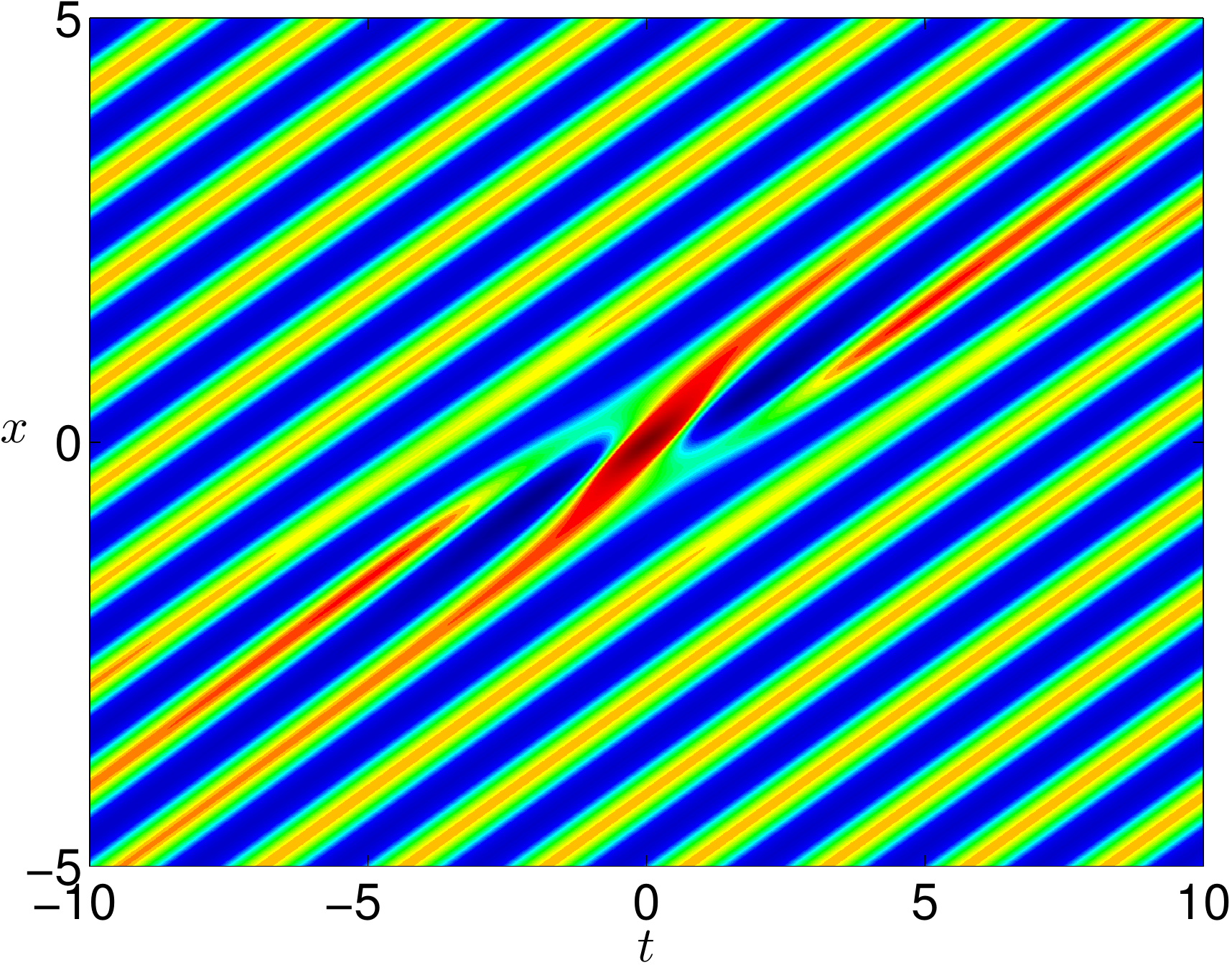}}
    \caption{(a). A three-dimensional plot of the absolute value of the rational solution. The dispersion and the nonlinear coefficients of the NLS equation are taken to be 1, i.e. $\beta = 1 = \gamma$. (b). A density plot of the corresponding physical wave field.} \label{PlotRa}
  \end{center}
\end{figure}

The rational solution can also be obtained by taking the limit of the modulation frequency $\nu \rightarrow 0$ from the SFB or taking the limit of the parameter $\mu \rightarrow 0$ from the Ma solution. Several names have been given for this solution, among others are the `Peregrine solution', to acknowledge the person who derived this solution \cite{Dysthe99}; the `isolated Ma solution', since it is confined in both space and time \cite{Henderson99}; and an `explode-decay solitary wave' solution, because of its soliton-like feature in space \cite{Nakamura85}. Figure \ref{PlotRa} shows a three-dimensional plot of the absolute value of the rational solution and the density plot of its corresponding physical wave field. We observe that indeed there is only a single and isolated extreme part that reach maximum at $(\xi,\tau) = (0,0)$. This is understandable since for $\nu \rightarrow 0$, the modulation period becomes infinity and thus the next neighbouring extreme part is also at the infinity and this leaves only one single extreme part mentioned earlier. Wavefront dislocation and phase singularity are also observed, with only one pair of them close to the extreme part.

\begin{figure}[h]
  \begin{center}
  \subfigure[]{\includegraphics[width=0.45\textwidth]{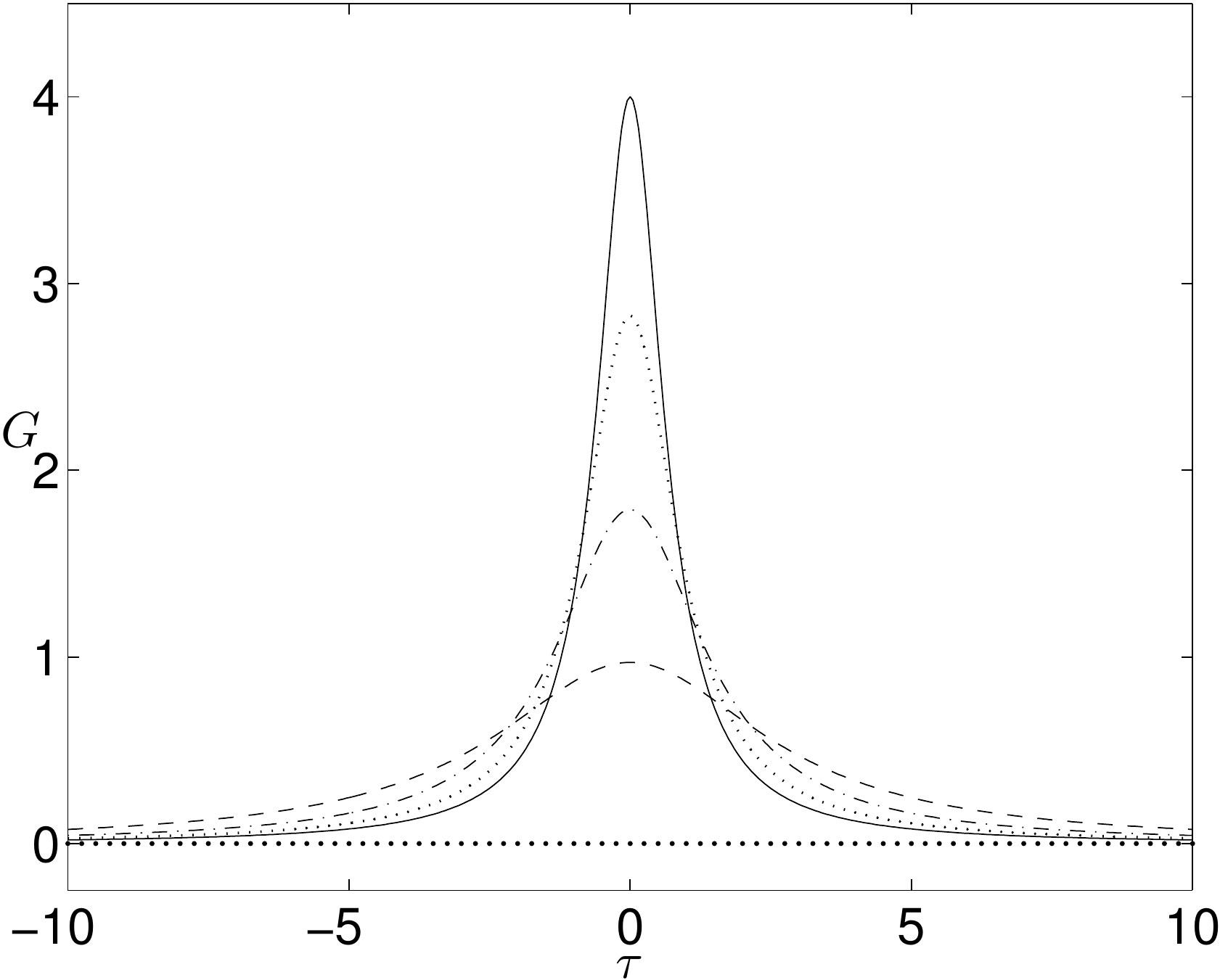}}
  \subfigure[]{\includegraphics[width=0.45\textwidth]{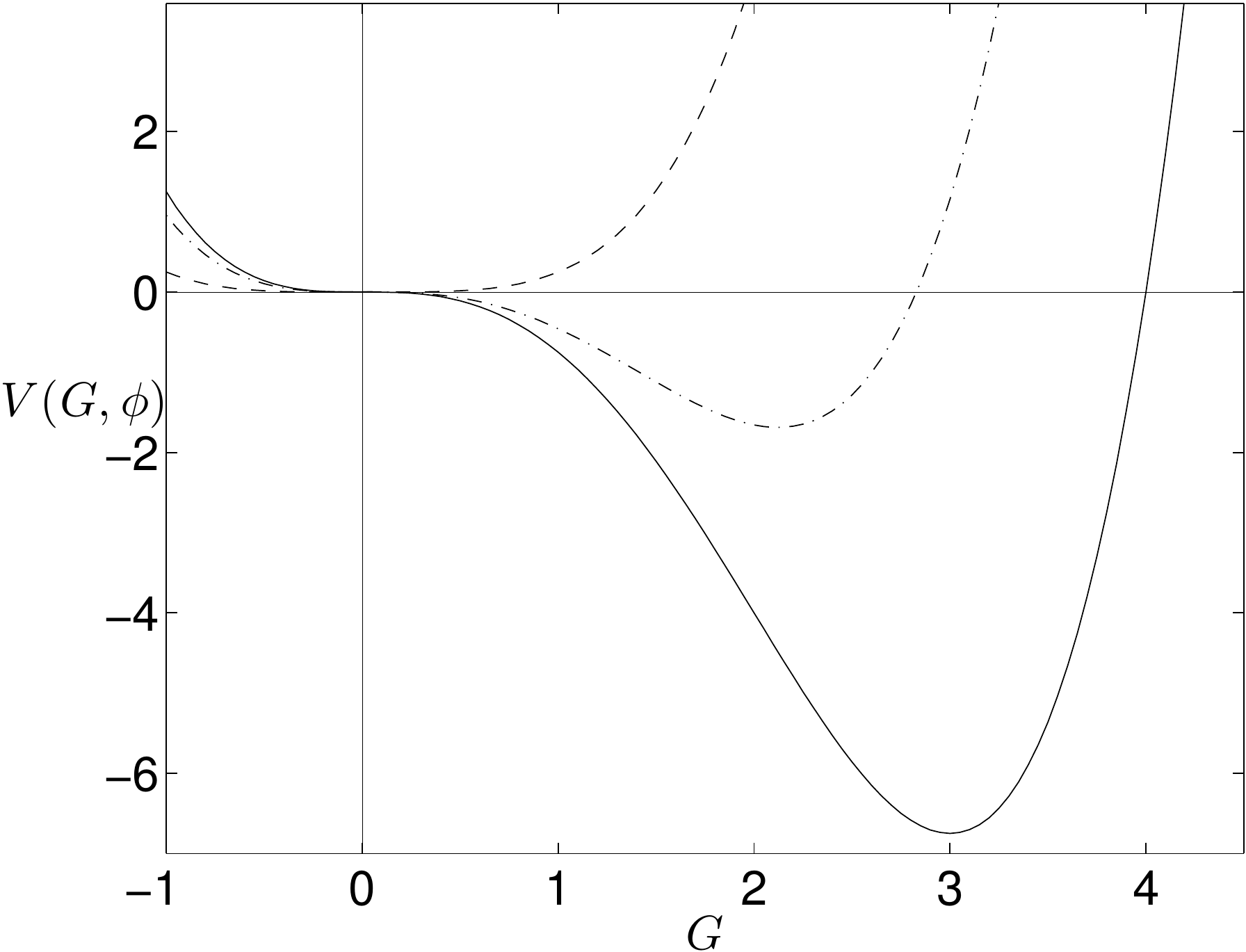}}
  \caption[]{(a). Sketch of the corresponding displaced amplitude of the rational solution at different positions: $\xi = 0$ (solid), $\xi = \pm 1/2$ (light dotted), $\xi = \pm 1$ (dashed-dot), $\xi = \pm 2$ (dashed) and $\xi \rightarrow \pm \infty$ (heavy dotted). (b). Sketch of the corresponding potential energy function $V$ for the rational solution as a function of the displaced amplitude $G$ at different positions: $\xi =0$ (solid), $\xi = \pm 1/2$ (dashed-dot) and $\xi \rightarrow \pm \infty$ (dashed).} \label{disampRa}
  \end{center}
\end{figure}

Figure~\ref{disampRa}(a) shows plots of the displaced amplitude $G$ at several positions. We observe that at $\xi = 0$, the displaced amplitude $G$ attains the largest maximum value of 4 which occurs for $\tau = 0$. The plot of the displaced phase $\phi$ is given by the dotted curve shown in Figure~\ref{disamphiSFB}(b) as well as in Figure~\ref{disamphiMa}(b). Similar to the corresponding displaced phases of the previous waves on finite background type of solutions, the displaced phase for the rational solution is monotonically decreasing as one travels along the positive $\xi$-direction. It also vanishes at $\xi = 0$, giving a consequence that the complex-valued amplitude reduces into a real-valued amplitude at this position. Since $\lim_{\xi \rightarrow \pm \infty} \phi(\xi) = \mp \pi/2$, the rational solution experiences the largest phase shift of $\pi$ compared to the other two solutions between $\xi \rightarrow -\infty$ and $\xi \rightarrow \infty$.

The corresponding potential energy function of the rational solution is also given by \eqref{potential}, where the displaced phase satisfies the following conditions:
\begin{eqnarray}
  \cos \phi(\xi) &=& \frac{1}{\sqrt{1 + 4 (\gamma r_{0}^{2} \xi)^{2}}}\\
  \textmd{and} \quad \phi'(\xi) &=& \frac{-2 \gamma r_{0}^{2}}{1 + 4 (\gamma r_{0}^{2} \xi)^{2}}.
\end{eqnarray}
At $\xi = 0$, $\cos \phi(0) = 1$ and $\phi'(0) = -\gamma r_{0}^{2}$, and thus the potential energy function becomes
\begin{equation}
  V(G,\phi(0)) = -\gamma r_{0}^{2} G^{2} + \frac{1}{4} \gamma r_{0}^{2} G^{2} (G - 2)^{2}.
\end{equation}
For $\xi \rightarrow \pm \infty$, $\lim_{\xi \rightarrow \pm \infty} \cos \phi(\xi) = 0$ and $\lim_{\xi \rightarrow \pm \infty} \phi'(\xi) = 0$, and thus the potential energy function has asymptotic values proportional to $G^{4}$:
\begin{equation}
  \lim_{\xi \rightarrow \pm \infty} V(G,\phi(\xi)) = \frac{1}{4} \gamma r_{0}^{2} G^{4}.
\end{equation}
Figure~\ref{disampRa}(b) shows plots of the potential energy function $V$ as a function of the displaced amplitude $G$ at several positions. We observe that for $\xi \rightarrow \pm \infty$, the potential energy function is asymptotically proportional to $G^{4}$. It is symmetric with respect to the vertical axis, i.e. $G = 0$, and has the highest minimum value of zero for $G = 0$. For $\xi = 0$, the potential energy reaches its lowest minimum value of $-6.75\gamma r_{0}^{2}$ for $G = 3$.

In the following subsection, we give an overview of one application of waves on finite background type of solution in the area of nonlinear hydrodynamics.

\subsection{Application in Nonlinear Hydrodynamics}

The waves on finite background type of solutions have been proposed as a mathematical model for the nonlinear dynamics of freak or extreme wave events which arise from the Benjamin-Feir modulational instability \cite{Dysthe99,Henderson99,Osborne00,Osborne01}. The authors of the latter two papers derive the waves on finite background solutions of the NLS equation using the nonlinear Fourier formulation of the periodic inverse scattering transform, a new method of mathematical physics developed by \citeasnoun{Gardner67}. \citeasnoun{Osborne00} present the dynamical behaviour of freak waves as solutions of the NLS equation using analytical and numerical results in one-dimensional and two-dimensional cases, respectively. Although it is mentioned that there is an infinite class of wave on finite background type of  solutions, only two explicit expressions of waves on finite background are presented in that paper. These ones correspond to the SFB for $\tilde{\nu} = 1$ and to the Ma solution for $\tilde{\mu} = \sqrt{2}$. In \cite{Osborne01}, the SFB for $\tilde{\nu} = 1$ is referred as the `rogue wave solution' of the spatial NLS equation and the Ma solution for $\tilde{\mu} = \sqrt{2}$ is called `another rogue wave spectral component'.

In particular, an implementation of the SFB wave signal as one type of waves on finite background for freak wave generation in hydrodynamic laboratories including some experimental results is discussed in \cite{Huijsmans05,Karjanto06}. The experiment on freak wave generation using the SFB as our theoretical model for wave signal propagation has been conducted at the high-speed basin of the Maritime Research Insitute Netherlands (MARIN). We choose the SFB instead of the Ma solution since the initial SFB wave signal is more reasonable to generate by the wavemaker than the initial Ma wave signal. The SFB wave signal has a sufficiently long distance to develop into an extreme wave signal while the Ma wave signal develops into its extreme signal within a much shorter distance. Considering the condition of the wave basin, we conclude that it is rather impossible to generate the Ma wave signal and therefore we decided to do experiments using the SFB wave signal.

After collecting experimental data and perform some analysis, we observe that all experimental results show amplitude increase according to the Benjamin-Feir modulational instability phenomenon corresponding to the SFB solution of the NLS equation. The carrier wave frequency and the modulation period are conserved during the wave propagation in the wave basin. A significant difference is that the experimental wave signal does not preserve the symmetry structure as possessed by the SFB wave signal. Nevertheless, the experimental signal shows a closely related phenomenon to wavefront dislocation, namely phase singularity. In one modulation period, the SFB wave signal has one pair of phase singularities but the experimental signal has only one singularity in this same time interval.

In the following section, we will investigate some interesting physical properties of higher order waves on finite background.

\section{Higher Order Waves on Finite Background}

Higher order waves on finite background refer to the higher order hierarchy of exact solutions of the NLS equation which have a characteristic finite background for $\xi \rightarrow \pm \infty$ or $\tau \rightarrow \pm \infty$. In this paper, we only discuss the SFB type of higher order solution, particularly the one we refer as SFB$_{2}$. An index 2 is now attached to distinguish from the previous SFB, which now we refer as SFB$_{1}$. These indices refer to the number of pairs of initial sideband in the spectral domain. Hence, in this context, SFB$_{1}$ has one pair of initial sidebands $\omega \pm \nu_{1}$, corresponding to the initial Benjamin-Feir spectrum. In order to ensure the validity of the solution, the sideband frequency $\nu_{1}$ must be inside the stability interval, or in normalized variable it is given by $0 < \tilde{\nu}_{1} < \sqrt{2}$.

However, if there is more than one pair of sidebands as the initial condition of modulation, and all these are inside the instability interval, then, due to the four-wave mixing process, all will be amplified. Since this process is a nonlinear superposition of the elementary processes of the Benjamin-Feir
instability with one pair of initial sidebands, the total effect can no longer be described by the SFB$_{1}$. As a consequence, two, three or more pair of initial sidebands have a correspondence with higher order SFBs. Accordingly, SFB$_{2}$ has two pairs of initial sidebands, $\omega \pm \nu_{2}$ and $\omega_{0} \pm 2 \nu_{2}$. Both pairs of sideband must be inside the stability interval to ensure the validity of the solution. So, in the normalized quantity, $0 < \tilde{\nu}_{2} < 1/\sqrt{2}$. Generally, SFB$_{n}$ has $n$-pairs of initial sidebands in the spectrum domain, $\omega_{0} + \nu_{n}$, $\omega_{0} + 2\nu_{n}, \dots$ and $\omega_{0} + n\nu_{n}$. All pairs of sidebands have to be inside the instability interval, so in the normalized quantity, $0 < \tilde{\nu}_{n} < \sqrt{2}/n$, $n \in \mathbb{N}$.

It is important to note that in the description above we assume the distance from one sideband to another is equal. Nonetheless, one can choose an arbitrary distance from one sideband to another but all sidebands are still inside the instability interval. As a result, there are infinitely many solutions within the same family of SFB$_{n}$, $n \in \mathbb{N}$. Explicit expressions for these solutions are not known in literature but an explicit expression for SFB$_{2}$ where the distance of the sidebands is the same has been discovered \cite{Akhmediev85,Ablowitz90,Calini02,Akhmediev97}. As a matter of fact, higher order solutions of the NLS equation can be constructed using a Darboux transformation \cite{Matveev91,Rogers02}. Another possibility is by implementing Hirota's method and associating the solution with the dark-hole soliton solutions of the defocusing NLS equation.
\begin{figure}[h]
  \begin{center}
  \includegraphics[width=0.45\textwidth]{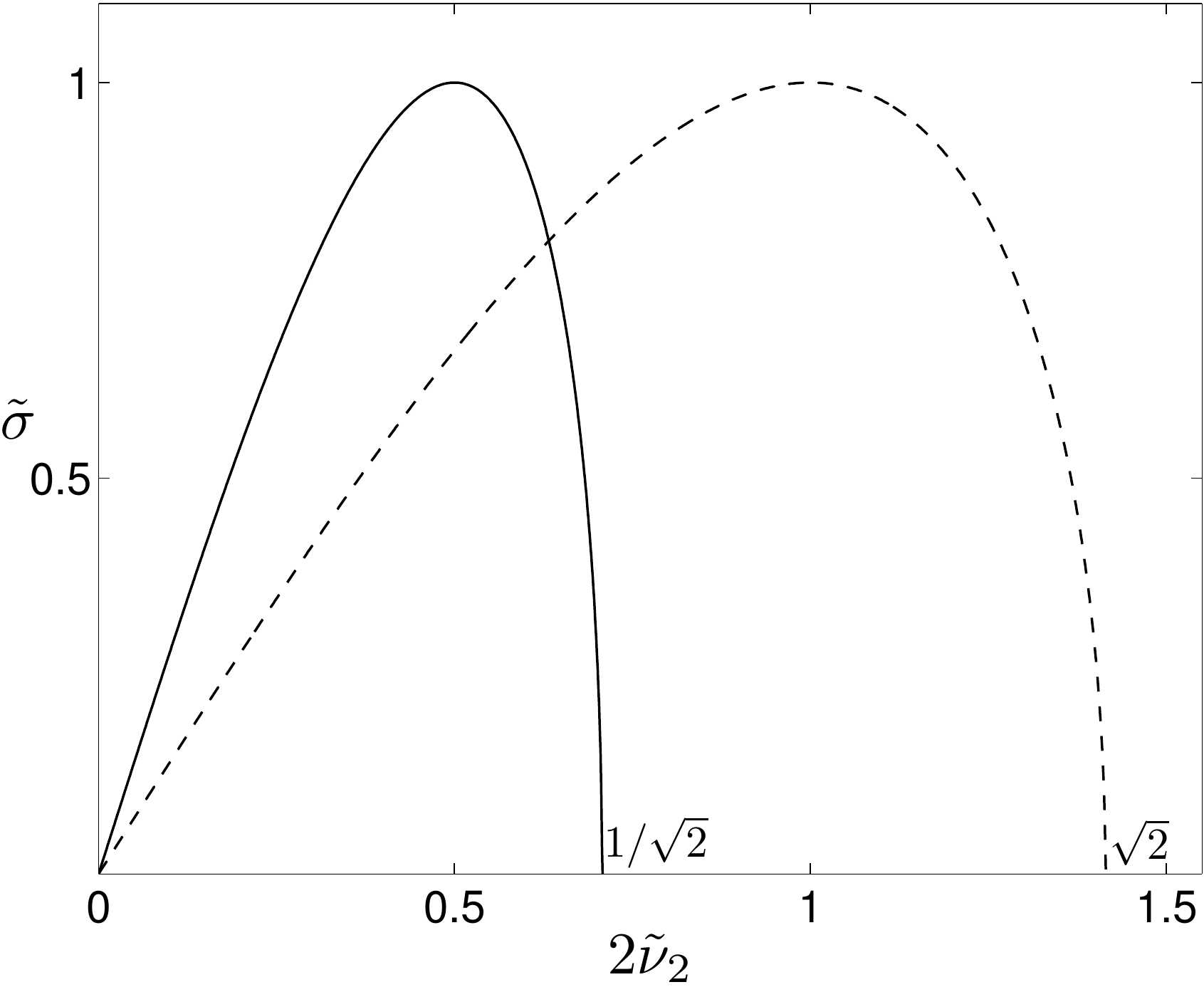}
  \caption{The plot of growth rates $\sigma_{1}$ (dashed) and $\sigma_{2}$ (solid) as a function of modulation frequency $\tilde{\nu}_{2}$. Observe that the range of modulation frequencies for $\sigma_{1}$ and $\sigma_{2}$ is $0 < 2\tilde{\nu}_{2} < \sqrt{2}$. The first pair of sidebands has $\sigma_{2}$ as its growth rate and the second pair of sidebands has $\sigma_{1}$ as its growth rate. The maximum growth rates for $\sigma_{1}$ and $\sigma_{2}$ are reached when $\tilde{\nu}_{2} = 1/2$ and both growth rates are equal for $\nu_{2} = \sqrt{2/5}$.}
  \end{center}
\end{figure}

\subsection{An Explicit Expression for SFB$_{2}$}

An expression of SFB$_{2}$ can be written in the simplest form as follows:
\begin{equation}
  A_{2}(\xi,\tau) = A_{0}(\xi) \left(1 - \frac{P_{2}(\xi,\tau) - i Q_{2}(\xi,\tau)}{H_{2}(\xi,\tau)} \right) \label{SFB2}
\end{equation}
where
\begin{align}
  P_{2}(\xi,\tau) = \tilde{\nu}_{2}^{2}
  \left[\frac{1}{2} \tilde{\sigma}_{2} \cosh \sigma_{1}(\xi - \xi_{11}) \cos 2\nu_{2}(\tau - \tau_{12}) \right. \nonumber \\
  \left. - \; 4\tilde{\sigma}_{1} \cosh \sigma_{2}(\xi - \xi_{02}) \cos \nu_{2} (\tau - \tau_{11}) \right] \nonumber \\
  + \; 3\sqrt{2}\tilde{\nu}_{2}^{3} \cosh \sigma_{1}(\xi - \xi_{11})\,\cosh \sigma_{2}(\xi - \xi_{12}) 
\end{align}
\begin{align}
  Q_{2}(\xi,\tau) = \frac{1}{2}\tilde{\sigma}_{1}\,\tilde{\sigma}_{2} \sinh \sigma_{1}(\xi - \xi_{11})\,\cos 2\nu_{2}(\tau - \tau_{12}) \nonumber \\
  - \; \tilde{\sigma}_{1}\,\tilde{\sigma}_{2} \sinh \sigma_{2}(\xi - \xi_{12})\,\cos \nu_{2}(\tau - \tau_{01}) \nonumber \\
  + \; \tilde{\nu}_{2}\sqrt{2}\left[\tilde{\sigma}_{2}\cosh \sigma_{1}(\xi - \xi_{11})\,\sinh \sigma_{2}(\xi - \xi_{12}) \right. \nonumber \\
  \left. - \; \tilde{\sigma}_{1}\sinh \sigma_{1}(\xi - \xi_{11})\,\cosh \sigma_{2}(\xi - \xi_{12}) \right] \\
  H_{2}(\xi,\tau) = \frac{1}{2} \tilde{\sigma}_{2} \cosh \sigma_{1}(\xi - \xi_{11})\,\cos 2\nu_{2}(\tau - \tau_{12}) \nonumber \\
  - \; \tilde{\sigma}_{1} \cosh \sigma_{2}(\xi -   \xi_{12})\,\cos \nu_{2}(\tau - \tau_{11}) \nonumber \\
  - \; \frac{2}{3\tilde{\nu}_{2}\sqrt{2}} \left[\tilde{\nu}_{2}^{2} (4\tilde{\nu}_{2}^{2} - 5)\cosh \sigma_{1}(\xi - \xi_{11})\,\cosh \sigma_{2}(\xi - \xi_{12}) \right. \nonumber \\
  \left. + \; \tilde{\sigma}_{1}\, \tilde{\sigma}_{2} \sinh \sigma_{1}(\xi - \xi_{11})\,\sinh \sigma_{2}(\xi - \xi_{12}) \right] \nonumber \\
  - \;\frac{3\tilde{\sigma}_{1}\,\tilde{\sigma}_{2}}{4\tilde{\nu}_{2}\sqrt{2}}
  \left[\cos \nu_{2}(\tau + \tau_{11} - 2\tau_{12})
  + \; \frac{1}{9}\cos \nu_{2}(3\tau - 2\tau_{12} - \tau_{11}) \right].
\end{align}

We now have two growth rates corresponding to the first and the second sideband modulation frequencies, given as $\sigma_{2}$ and $\sigma_{1}$, respectively. These growth rates are given by $\sigma_{1} = \sigma(\tilde{\nu}_{2}) = \gamma r_{0}^{2} \tilde{\sigma}_{1}$, where $\tilde{\sigma}_{1} = \tilde{\nu}_{2} \sqrt{2 - \tilde{\nu}_{2}^{2}}$ and $\sigma_{2} = \sigma(2\tilde{\nu}_{2}) = \gamma r_{0}^{2} \tilde{\sigma}_{2}$, where $\tilde{\sigma}_{2} = 2\tilde{\nu}_{2} \sqrt{2 - 4\tilde{\nu}_{2}^{2}}$. A set of real parameters $\xi_{11},\xi_{12}, \tau_{11}$ and $\tau_{12}$ determines when and where SFB$_{2}$ reaches its maxima. For nonzero $\xi_{11}$ and $\xi_{12}$, we observe that SFB$_{2}$ is composed by two SFB$_{1}$s. The larger the difference between these time parameters, the farther the distance between SFB$_{1}$s. On the other hand, the time parameters $\tau_{11}$ and $\tau_{12}$ only contribute in the time shifting of the individual SFB$_{1}$s, yet their periods remain the same. When $\xi_{11} = 0 = \xi_{12}$, both SFB$_{1}$s interact and form a completely different shape of the complex-valued amplitude, depending on the choice of time parameters $\tau_{11}$ and $\tau_{12}$. Interestingly, for a particular choice of parameters of $\xi_{11} = 0 = \xi_{12}$ and $\tau_{11} = 0 = \tau_{12}$, SFB$_{2}$ simply reduces to SFB$_{1}$ for modulation frequency $\tilde{\nu}_{2} \rightarrow \sqrt{1/2}$:
\begin{equation}
  \lim_{\tilde{\nu}_{2}\rightarrow \sqrt{1/2}}
  A_{2}(\xi,\tau;\tilde{\nu}_{2}) = A_{1}\left(\xi,\tau;\tilde{\nu}_{1} = \sqrt{1/2} \right),
\end{equation}
where $A_{1}$ denotes SFB$_{1}$.\\

\begin{remark*}
In the previous section we have seen that an explicit expression of SFB$_{1}$ can be written in the displaced phase-amplitude representation. However, we are not able to write SFB$_{2}$ in a single displaced phase-amplitude representation since the displaced phase fails to be expressed as a space-dependent function only. By observing the terms of SFB$_{2}$ and three-dimensional plots of the absolute value of its complex-valued amplitude, we conclude that SFB$_{2}$ is an interaction of two SFB$_{1}$s where one modulation frequency is twice of the other. Let $A_{11}$ and $A_{12}$ be two SFB$_{1}$ with modulation frequencies $\nu_{2}$ and $2\nu_{2}$, respectively. The maxima of $A_{11}$ and $A_{12}$ occur at $(\xi_{11}, \tau_{11} + 2n\pi/\nu_{1})$ and $(\xi_{12}, \tau_{12} + 2n\pi/\nu_{1})$, $n \in \mathbb{Z}$, respectively. Explicitly they are given as follows:
\begin{eqnarray}
  A_{11}(\xi,\tau) &=& A_{0}(\xi) F_{11}(\xi,\tau) \nonumber \\
  &=& A_{0}(\xi) \left(\frac{\tilde{\nu}_{2}^{2} \cosh \sigma_{1} (\xi - \xi_{11}) - i \tilde{\sigma}_{1} \sinh \sigma_{1} (\xi - \xi_{11})}{\cosh \sigma_{1} (\xi - \xi_{11}) - \tilde{\sigma}_{1}/(\tilde{\nu}_{2} \sqrt{2}) \cos \nu_{2} (\tau - \tau_{11})} - 1 \right)
\end{eqnarray}
and
\begin{eqnarray}
  A_{12}(\xi,\tau) &=& A_{0}(\xi) F_{12}(\xi,\tau) \nonumber \\
  &=& A_{0}(\xi) \left(\frac{4\tilde{\nu}_{2}^{2} \cosh \sigma_{2} (\xi - \xi_{12}) - i \tilde{\sigma}_{2} \sinh \sigma_{2} (\xi - \xi_{12})}{\cosh \sigma_{2} (\xi - \xi_{12}) - \tilde{\sigma}_{2}/(2\tilde{\nu}_{2} \sqrt{2}) \cos 2\nu_{2} (\tau - \tau_{12})} - 1 \right),
\end{eqnarray}
where $\sigma_{1}$ and $\sigma_{2}$ are the Benjamin-Feir instability growth rates corresponding to modulation frequencies $\nu_{2}$ and $2\nu_{2}$, respectively. The interaction of these two SFB$_{1}$ can be written as the multiplication of non oscillating parts $F_{11}$ and $F_{12}$, i.e., $\tilde{A}_{2}(\xi,\tau) = A_{0}(\xi) F_{11}(\xi,\tau) F_{12}(\xi,\tau)$. The terms of this product also occur in the terms of SFB$_{2}$. However, SFB$_{2}$ is not simply the product of two SFB$_{1}$s even though it is an interaction of the twos. It can be checked that $\tilde{A}_{2}$ is generally not a solution of the NLS equation since after substituting it into \eqref{spatialNLS}, it produces additional nonzero terms of $\partial_{\tau} F_{11} \partial_{\tau} F_{12}$ and $F_{11}F_{12}(|F_{11}|^{2} - 1)(|F_{12}|^{2} - 1)$.
\end{remark*}

\subsection{Asymptotic Behaviour}

Similar to SFB$_{1}$, for $\xi \rightarrow \pm \infty$ SFB$_{2}$ \eqref{SFB2} also reduces to the plane-wave solution with a phase factor of $\phi_{20}$. The asymptotic behaviour of SFB$_{2}$ at the lowest order can be written as:
\begin{equation}
  \lim_{\xi \rightarrow \pm \infty} A(\xi,\tau) = A_{0}(\xi) e^{\mp i \phi_{20}}
\end{equation}
where the phase $\phi_{20}$ satisfies
\begin{equation}
  \tan(\phi_{20}) = \frac{3 \tilde{\nu}_{2}^{2} (\tilde{\sigma}_{2} - \tilde{\sigma}_{1})}{\tilde{\nu}_{2}^{2} (13 \tilde{\nu}_{2}^{2} - 5) + \tilde{\sigma}_{1} \tilde{\sigma}_{2}}.
\end{equation}
For SFB$_{2}$, the corresponding initial wave signal has two pairs of sidebands in the spectral domain, i.e. $\omega_{0} \pm \nu_{2}$ and $\omega_{0} \pm 2 \nu_{2}$ as the first and the second pairs of sidebands frequency, respectively. Therefore, we are interested in finding the asymptotic behaviour which includes the terms $\cos \nu_{2}(\tau - \tau_{11})$ and $\cos 2\nu_{2}(\tau - \tau_{12})$. A similar procedure of Taylor series expansion as in the previous section can be applied in this regard.
Since for $\xi - \xi_{11} \rightarrow -\infty$ implies $\xi - \xi_{12} \rightarrow -\infty$ and vice versa for all $\xi_{11}$ and $\xi_{12}$ real numbers, then the asymptotic behaviour of $e^{\sigma_{1}(\xi - \xi_{11})}$ and $e^{\sigma_{2}(\xi - \xi_{12})}$ at $\xi \rightarrow -\infty$ will be similar.
Therefore, we introduce a new variable $y = e^{\sigma_{1}(\xi - \xi_{11})} = e^{\sigma_{2}(\xi - \xi_{12})}$ and substitute to SFB$_{2}$ \eqref{SFB2} after removing the plane-wave solution contribution. We denote this new function as $F_{2}(y,\tau)$ which reads as follows: 
\begin{equation}
  F_{2}(y,\tau) = 1 - \frac{P_{2}(y,\tau) - i Q_{2}(y,\tau)}{H_{2}(y,\tau)}
\end{equation}
where
\begin{eqnarray}
  P_{2}(y,\tau) &=& \frac{\tilde{\nu}_{2}^{2}}{2} \left[\tilde{\sigma}_{2} \cos 2\nu_{2}(\tau - \tau_{12}) - 8 \tilde{\sigma}_{1} \cos \nu_{2} (\tau - \tau_{11}) \right]  (y^{3} + y) \nonumber \\ && + \; 3 \tilde{\nu}_{2}^{3} \sqrt{2} (y^{2} + 1)^{2} \\
  Q_{2}(y,\tau) &=& \frac{1}{2} \tilde{\sigma}_{1}\,\tilde{\sigma}_{2} \left[\cos 2\nu_{2}(\tau - \tau_{12}) - 2 \cos \nu_{2} (\tau - \tau_{11}) \right](y^{3} - y) \nonumber \\ &&  + \; \tilde{\nu}_{2}\sqrt{2} (\tilde{\sigma}_{2} - \tilde{\sigma}_{1}) (y^{4} - 1) \qquad \\
  H_{2}(y,\tau) &=& \frac{1}{2} \left[\tilde{\sigma}_{2} \cos 2\nu_{2}(\tau - \tau_{12}) - 2\tilde{\sigma}_{1} \cos \nu_{2} (\tau - \tau_{11}) \right] (y^{3} + y) \nonumber \\
  && - \; \frac{2}{3\tilde{\nu}_{2}\sqrt{2}} \left[\tilde{\nu}_{2}^{2} (4\tilde{\nu}_{2}^{2} - 5) (y^{2} + 1)^{2} + \tilde{\sigma}_{1}\, \tilde{\sigma}_{2} (y^{2} - 1)^{2} \right] \nonumber \\ && - \; \frac{3\tilde{\sigma}_{1}\,\tilde{\sigma}_{2}} {4\tilde{\nu}_{2}\sqrt{2}} \left[\cos \nu_{2}(\tau + \tau_{11} - 2\tau_{12}) + \frac{1}{9}\cos \nu_{2} (3\tau - 2\tau_{12} - \tau_{11}) \right] y^{2}. \qquad
\end{eqnarray}
Expanding $F_{2}$ into a Taylor series about $y = 0$ gives
\begin{equation}
  F_{2}(y,\tau) = F(0,\tau) + \partial_{y}F(0,\tau) y + \frac{1}{2} \partial_{y}^{2}F(0,\tau) y^{2} + \dots
\end{equation}
where
\begin{eqnarray}
  F(0,\tau) &=& \frac{\tilde{\nu}_{2}^{2} (13 \tilde{\nu}_{2}^{2} - 5) + \tilde{\sigma}_{1} \tilde{\sigma}_{2} + 3 i \tilde{\nu}_{2}^{2} (\tilde{\sigma}_{2} - \tilde{\sigma}_{1})}{\tilde{\nu}_{2}^{2} (4 \tilde{\nu}_{2}^{2} - 5) + \tilde{\sigma}_{1} \tilde{\sigma}_{2}} \\
  \partial_{y}F(0,\tau) &=& \rho_{11} e^{i\phi_{11}} \cos(\nu_{2} \tau) + \rho_{12} e^{i\phi_{12}} \cos(2\nu_{2} \tau)
\end{eqnarray}
where
\begin{eqnarray}
  \rho_{11} e^{i\phi_{11}} &=& -\frac{3}{\sqrt{2}} \tilde{\nu}_{2} \tilde{\sigma}_{1} \left( \frac{\tilde{\nu}_{2}^{2} \left[\tilde{\nu}_{2}^{2}(16 \tilde{\nu}_{2}^{2} - 11) + 4 \tilde{\sigma}_{1} \tilde{\sigma}_{2} \right] + i \tilde{\sigma}_{1} \left[\tilde{\sigma}_{2} (2 \tilde{\sigma}_{1} + \tilde{\sigma}_{2}) - 3\tilde{\nu}_{2}^{2} \right]}{\left[\tilde{\nu}_{2}^{2}(4 \tilde{\nu}_{2}^{2} - 5) + \tilde{\sigma}_{1} \tilde{\sigma}_{2} \right]^{2}} \right) \qquad \\
  \rho_{12} e^{i\phi_{12}} &=& \frac{3}{4} \tilde{\nu}_{2} \tilde{\sigma}_{2} \sqrt{2} \left( \frac{\tilde{\nu}_{2}^{2} \left[4\tilde{\nu}_{2}^{2}(\tilde{\nu}_{2}^{2} + 1) + \tilde{\sigma}_{1} \tilde{\sigma}_{2} \right] + i \left[\tilde{\sigma}_{1}^{2} (\tilde{\sigma}_{2} - 4 \tilde{\sigma}_{1}) + 3\tilde{\nu}_{2}^{2} \tilde{\sigma}_{2}\right] }{\left[\tilde{\nu}_{2}^{2}(4 \tilde{\nu}_{2}^{2} - 5) + \tilde{\sigma}_{1} \tilde{\sigma}_{2} \right]^{2}} \right).
\end{eqnarray}
Therefore, the asymptotic behaviour for SFB$_{2}$ at $\xi \rightarrow \mp \infty$ is given by
\begin{eqnarray}
  A_{2}(\xi,\tau) &\approx& A_{0}(\xi) \, \left[e^{\pm i\phi_{20}} +
  \rho_{11} e^{\pm i\phi_{21}} e^{\pm \sigma_{2} (\xi - \xi_{11})} \cos \nu_{2} (\tau - \tau_{11}) \right. \nonumber \\
  && \left. + \; \rho_{12} e^{\pm i \phi_{22}} e^{\pm \sigma_{1} (\xi - \xi_{12})} \cos 2 \nu_{2} (\tau - \tau_{12}) \right].
\end{eqnarray}
We observe that each sideband carries different phase information as well as the growth rates. It is important again to emphasize that the first pair of sidebands $\omega_{0} \pm \nu_{2}$ corresponds to growth rate $\sigma_{2}$ and the second pair of sidebands $\omega_{0} \pm 2\nu_{2}$, which is equivalent to $\omega_{0} \pm \nu_{1}$ in the context of SFB$_{1}$, corresponds to growth rate $\sigma_{1}$.

\subsection{Physical Characteristics}

In this subsection, we will discuss some particular physical characteristics of SFB$_{2}$. It is even more interesting to compare these characteristics with the ones of SFB$_{1}$; we will consider the amplitude amplification factor, physical wave field and wave signal evolution.

\subsubsection{Amplitude Amplification Factor}

\begin{figure}[h]
  \begin{center}
  \includegraphics[width=0.45\textwidth]{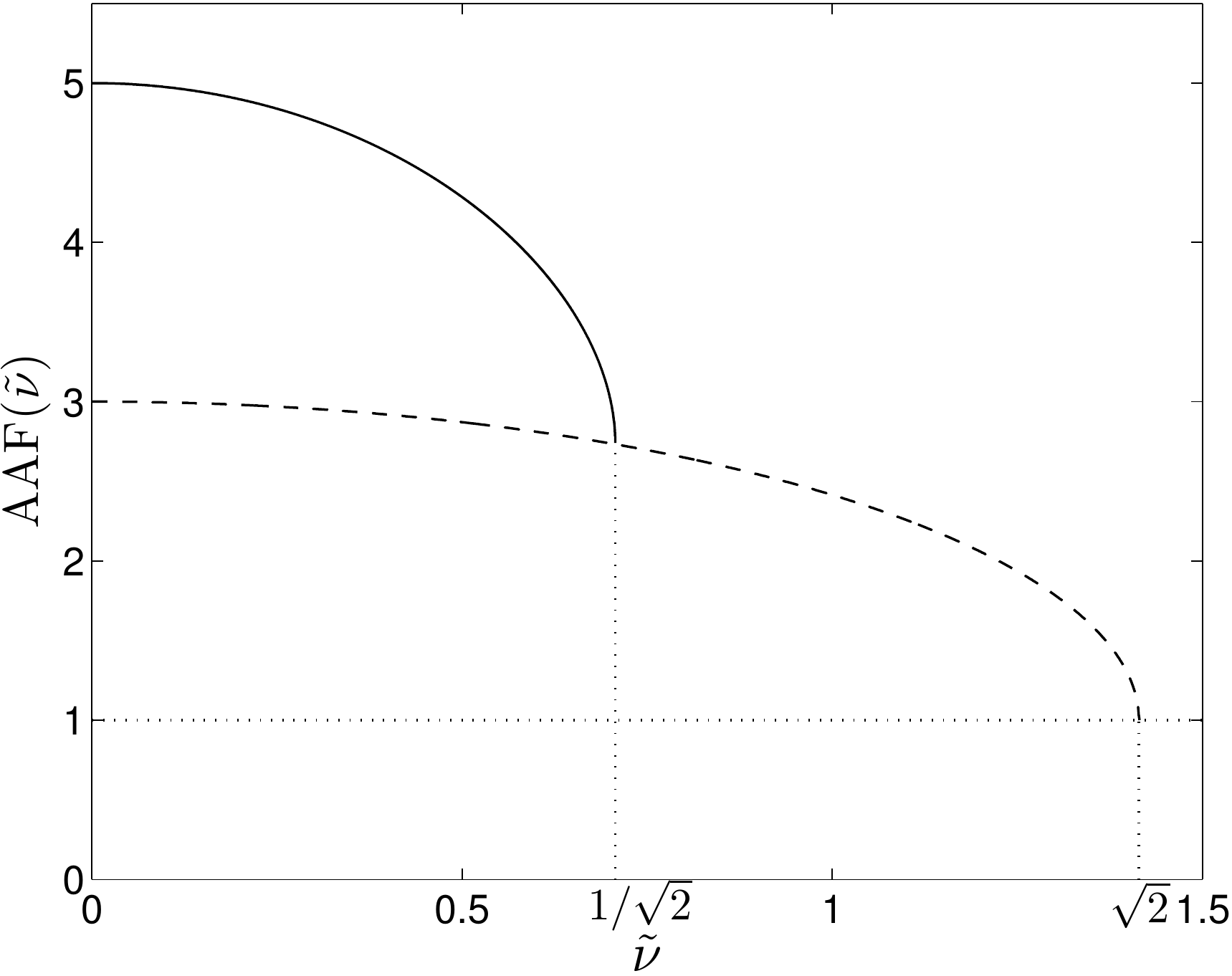}
  \caption{Plots of the amplitude amplification factors corresponding to SFB$_{1}$ (dashed) and SFB$_{2}$ (solid) as a function of modulation frequency. Both curves are monotonically decreasing for increasing modulation frequency. When $\tilde{\nu}_{2} \rightarrow 1/\sqrt{2} = \tilde{\nu}_{2}$, the amplification factor reaches the same value of $1 + \sqrt{3}$.} \label{AAF}
  \end{center}
\end{figure}
The amplitude amplification factor (AAF) of waves on finite background is defined as the ratio of the maximum amplitude and the value of its background. This value depends on a parameter of the waves on finite background. In the case of SFB$_{1}$ and SFB$_{2}$, the amplitude amplification factor depends on the modulation frequency. For SFB$_{1}$, the corresponding amplitude amplification factor is given by
\begin{equation}
  \textmd{AAF}_{1}(\tilde{\nu}_{1}) = 1 + \sqrt{4 - 2\tilde{\nu}_{1}}, \qquad \textmd{for} \quad 0 < \tilde{\nu}_{1} < 2.
\end{equation}
It can reach up to a maximum factor of 3 for a very long modulation when $\nu_{1} \rightarrow 0$. For SFB$_{2}$ with the choice of parameters of $\xi_{11} = 0 = \xi_{12}$ and $\tau_{11} = 0 = \tau_{12}$, the corresponding amplitude amplification factor is  obtained by choosing the set of parameters to be zero, i.e. $\xi_{11} = 0 = \xi_{12}$ and $\tau_{11} = 0 = \tau_{12}$. Explicitly it is given by:
\begin{equation}
  \textmd{AAF}_{2}(\tilde{\nu}_{2}) = 1 + \sqrt{4 - 2\tilde{\nu}_{2}^{2}} +
  2\sqrt{1 - 2\tilde{\nu}_{2}^{2}}, \qquad \textmd{for} \quad 0 < \tilde{\nu}_{2} < 1/\sqrt{2}.
\end{equation}
It is interesting to note that there exists a relationship between the amplitude amplification factors of SFB$_{1}$ and SFB$_{2}$, which can be written explicitly as a simple linear relationship:
\begin{equation}
  \textmd{AAF}_{2}(\tilde{\nu}_{2}) = 1 + \left(\textmd{AAF}_{1}(\tilde{\nu}_{2}) - 1 \right) +
  \left(\textmd{AAF}_{1}(2\tilde{\nu}_{2}) - 1 \right).
\end{equation}
This quantity can reach up to a maximum factor of 5 for a very long modulation when $\nu_{2} \rightarrow 0$. Figure \ref{AAF} shows the plot of the amplitude amplification factors of SFB$_{1}$ and SFB$_{2}$.

Other interesting physical characteristics of SFB$_{2}$ are three-dimensional plots of the absolute value of the complex-valued amplitude, the corresponding physical wave fields and the evolution of the corresponding wave signals. Figures \ref{3D1}--\ref{SignalSFB2nu12Xi0} show these descriptions for several different combinations of the parameters $\xi_{11}, \xi_{12}, \tau_{11}$ and $\tau_{12}$.

\subsubsection{Wavefront Dislocations around a Single Position}

\begin{figure}[h]
  \begin{center}
    \subfigure[]{\includegraphics[width=0.45\textwidth]{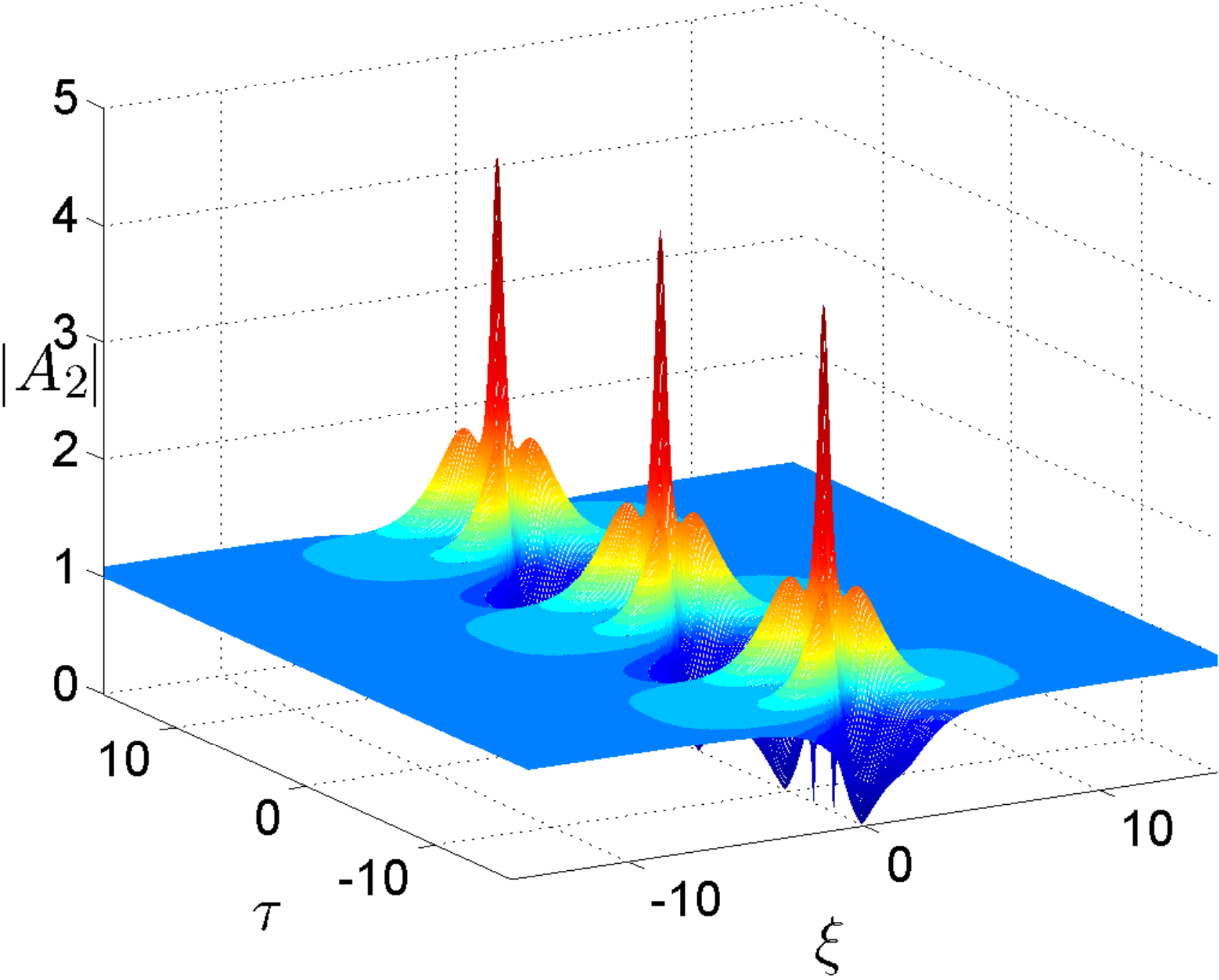}}    \hspace{0.5cm}
    \subfigure[]{\includegraphics[width=0.45\textwidth]{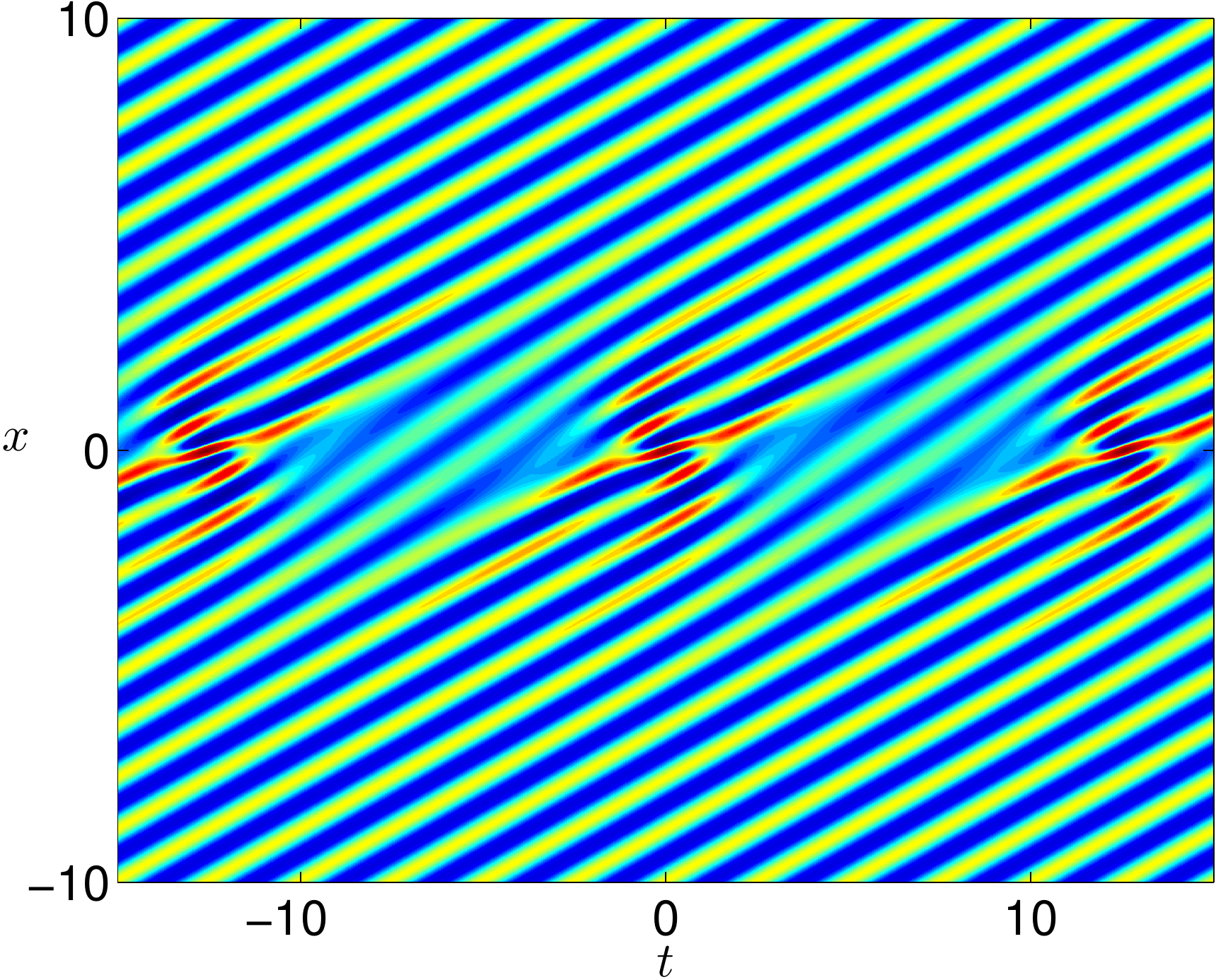}}
    \caption{(a). A three-dimensional plot of the absolute value of SFB$_{2}$ for the modulation frequency $\tilde{\nu}_{2} = 1/2$, $\xi_{11} = 0 = \xi_{12}$ and $\tau_{11} = 0 = \tau_{12}$. (b). A density plot of the corresponding physical wave field.} \label{3D2}
  \end{center}
\end{figure}

\begin{figure}[h]
  \begin{center}
    \includegraphics[width=0.75\textwidth,angle=-90]{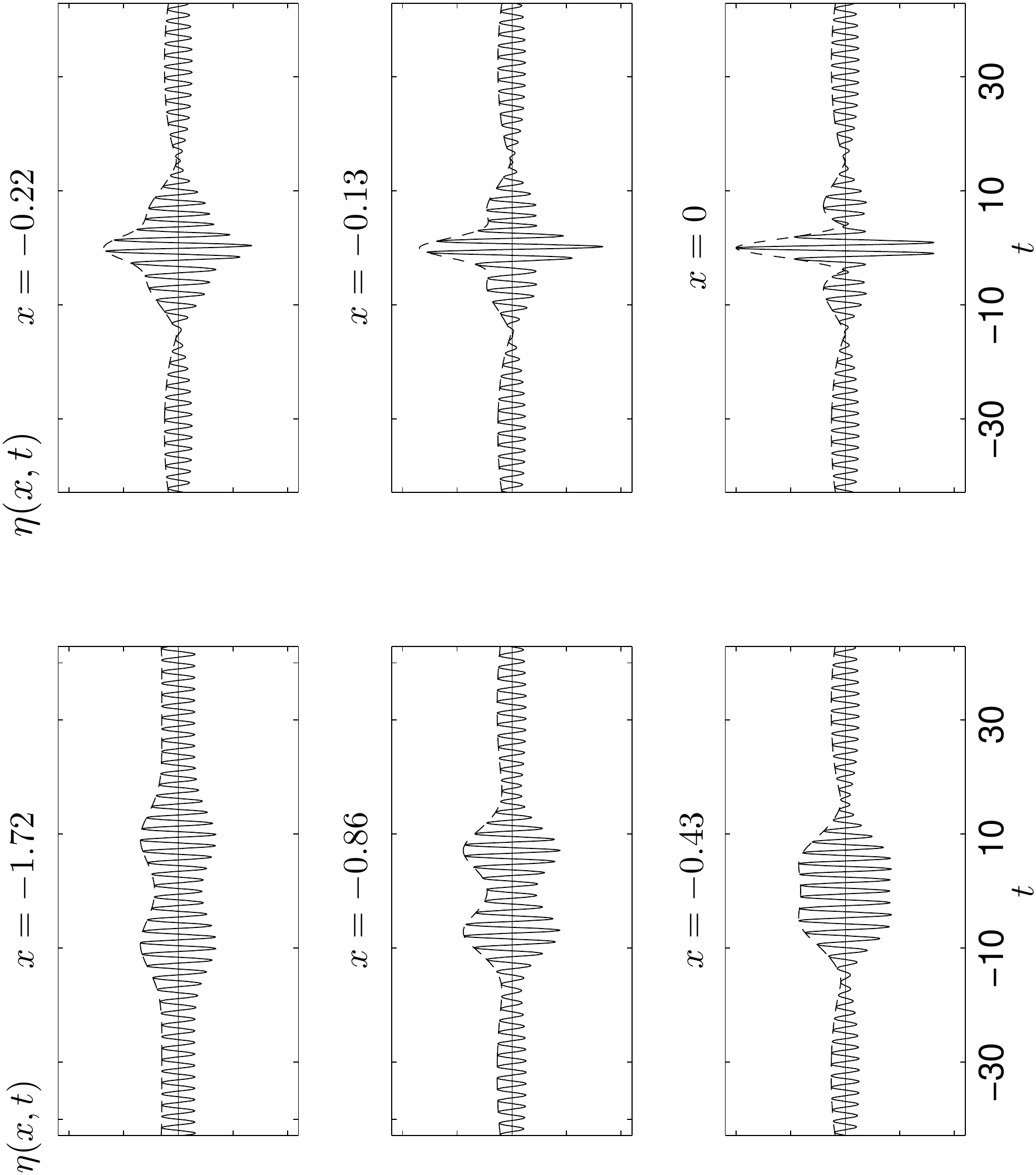}
    \caption{Plots of SFB$_{2}$ wave signal evolution for the modulation frequency $\tilde{\nu}_{2} = 1/2$ at several positions, where the all the parameters are taken to be zero: $\xi_{11} = 0 = \xi_{12}$ and $\tau_{11} = 0 = \tau_{12}$.} \label{SignalSFB2nu12Xi0Tau0}
  \end{center}
\end{figure}

This particular characteristic is obtained by choosing the set of parameters all to be zero, i.e. $\xi_{11} = 0 = \xi_{12}$ and $\tau_{11} = 0 = \tau_{12}$. Figure \ref{3D2} and Figure \ref{SignalSFB2nu12Xi0Tau0} show a three-dimensional plot of the absolute value of SFB$_{2}$, a density plot of the corresponding physical wave field and the corresponding wave signal evolution. For this choice of parameters, we observe that SFB$_{2}$ is the result of an interaction of two SFB$_{1}$s with two different modulation frequencies. See Figure \ref{3D2}(a). This interaction can be obtained by allowing both $\xi_{11} \rightarrow 0$ and $\xi_{12} \rightarrow 0$ of SFB$_{2}$ in Figure \ref{3D1}(a). As a consequence, we now have only one extreme part at $\xi = 0$. It remains periodic in $\tau$ with the modulation period of $2\pi/\nu_{2}$. Since the interaction is constructive, now the extreme part of SFB$_{2}$ reaches higher maximum point than each individuals of SFB$_{1}$s. Recall that the AAF can reach up to a maximum factor of 5 for $\nu_{2} \rightarrow 0$. As a matter of fact, the highest possible amplitude amplification factor for SFB$_{2}$ is obtained by this choice of parameters.

The corresponding physical wave field of this type of SFB$_{2}$ also shows wavefront dislocation, but with a rather different pattern as is observed in the physical wave field of SFB$_{1}$. See Figure \ref{3D2}(b). The waves merge and split two times in one modulation period as we move along in the spatial direction. Thus, we now have two pairs of wavefront dislocations, as confirmed by the wave signal evolution that shows two pairs of phase singularities in one modulation period at $x = 0$, see Figure \ref{SignalSFB2nu12Xi0Tau0}. From this figure, we can observe that an initially modulated wave signal with two different modulation frequencies interacts among each other and develops into an extreme wave signal as it propagates through space. It is important to note that not all values of modulation frequency in the interval $0 < \tilde{\nu}_{2} < \sqrt{1/2}$ will give two pairs of phase singularities. We will only observe two pairs of wavefront dislocations and phase singularities in one modulation period for modulation frequency in the interval $0 < \tilde{\nu}_{2} \leq \frac{1}{6}\sqrt{8\sqrt{10} - 10}$. The numerical value of this upper boundary is approximately $0.652$. Furthermore, for $\frac{1}{6}\sqrt{8\sqrt{10} - 10} < \tilde{\nu}_{2} < \sqrt{1/2}$, we will only observe one pair of singularities within one modulation period. Thus, there is only a tiny range of modulation frequency of approximately $0.055$,  merely less than 8\% of the interval, where this case occurs. The pair of singularities which remains is the one that is closer to the extreme wave signal. It is interesting to note that different with SFB$_{1}$ where it can be that there is no wavefront dislocation and phase singularity, this type of SFB$_{2}$ always shows the occurrence of the phenomena, whether one pair or two pairs of them.

\subsubsection{Wavefront Dislocations at Two Different Positions}

\begin{figure}[h]
  \begin{center}
    \subfigure[]{\includegraphics[width=0.45\textwidth]{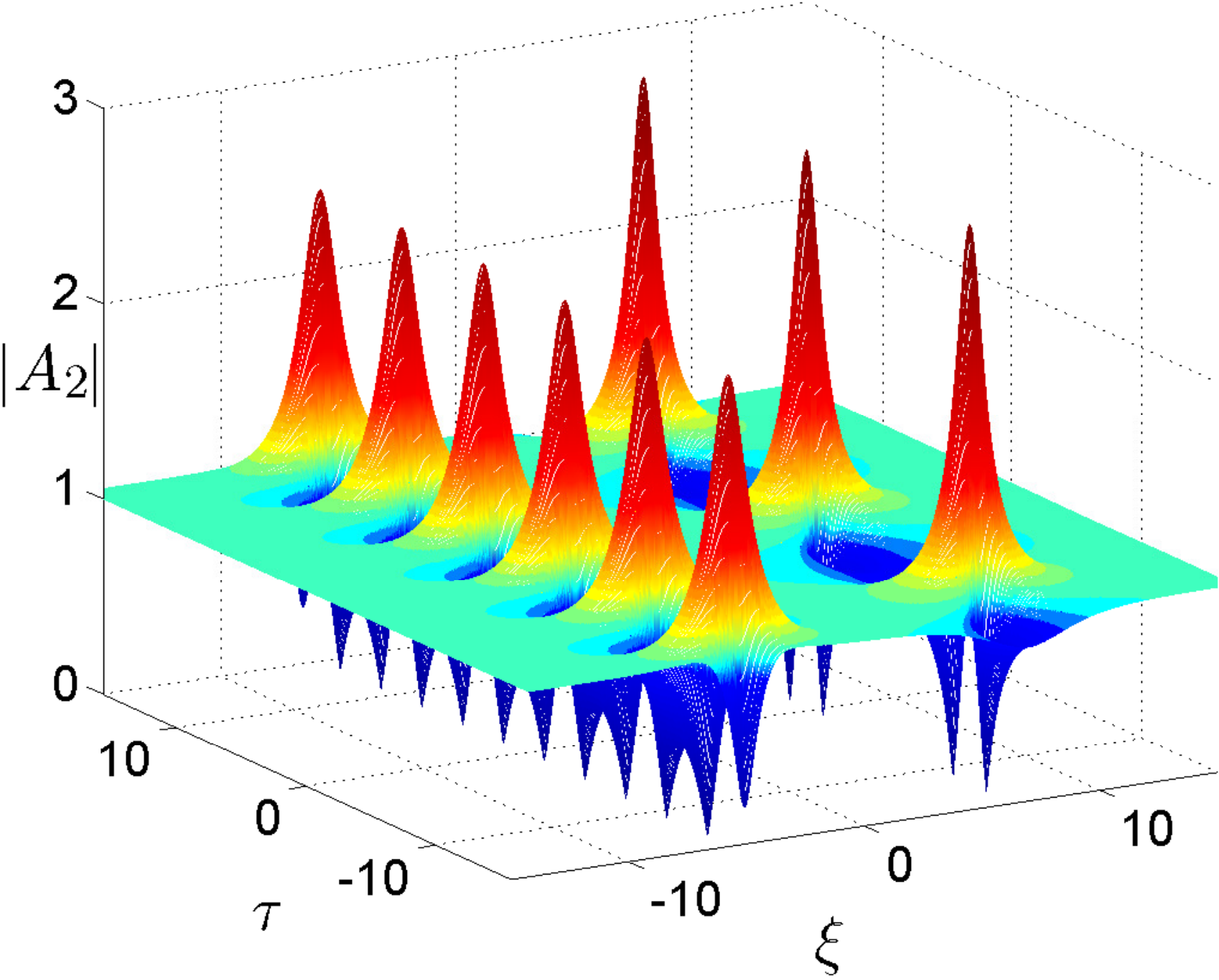}}     \hspace{0.5cm}
    \subfigure[]{\includegraphics[width=0.45\textwidth]{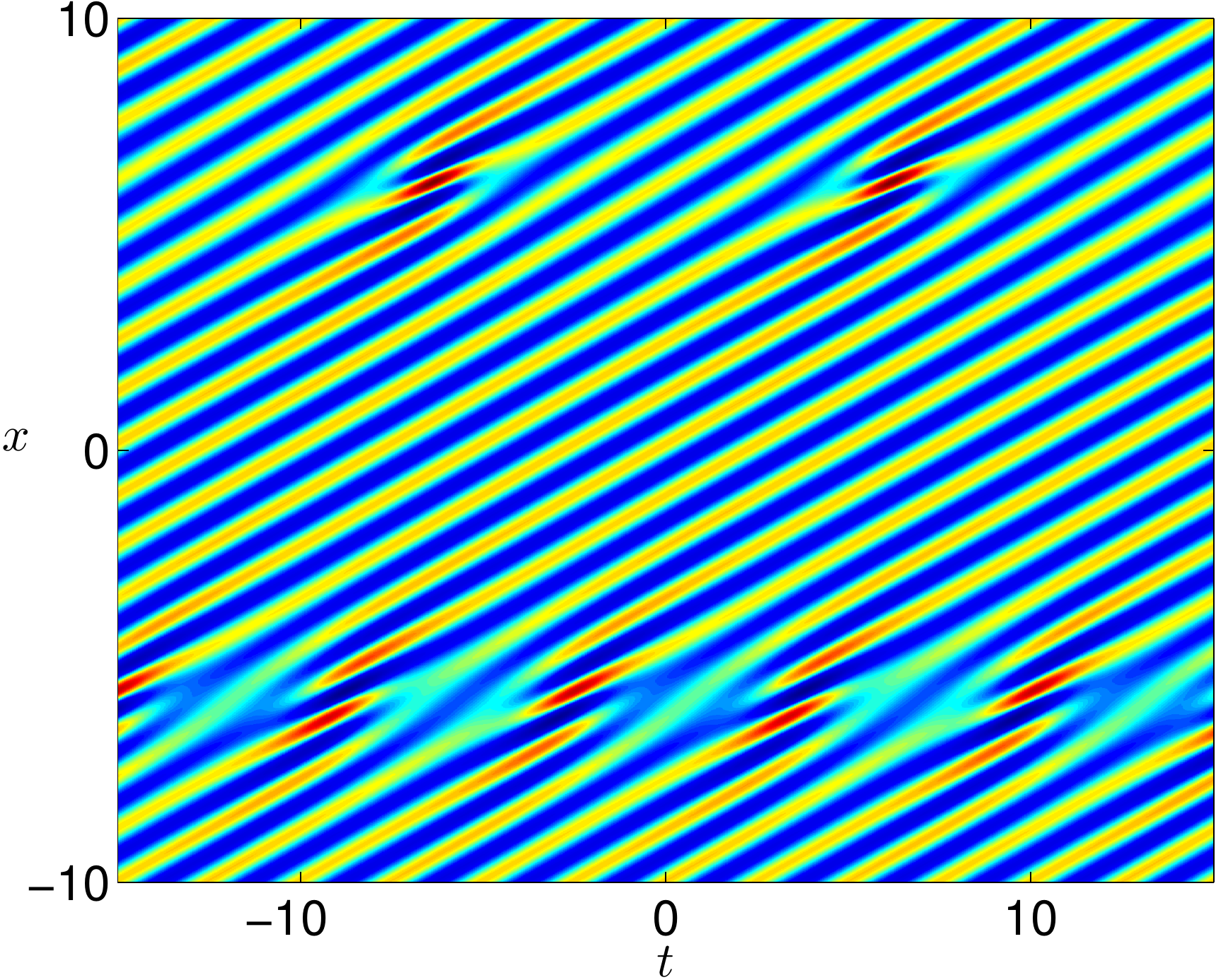}}
    \caption{(a). A three-dimensional plot of the absolute value of SFB$_{2}$ for the modulation frequency $\tilde{\nu}_{2} = 1/2$, $\xi_{11} = 5$, $\xi_{12} = -5$ and $\tau_{11} = 0 = \tau_{12}$. (b). A density plot of the corresponding physical wave field.} \label{3D1}
  \end{center}
\end{figure}
\begin{figure}[h]
  \begin{center}
    \includegraphics[width=0.75\textwidth,angle=-90]{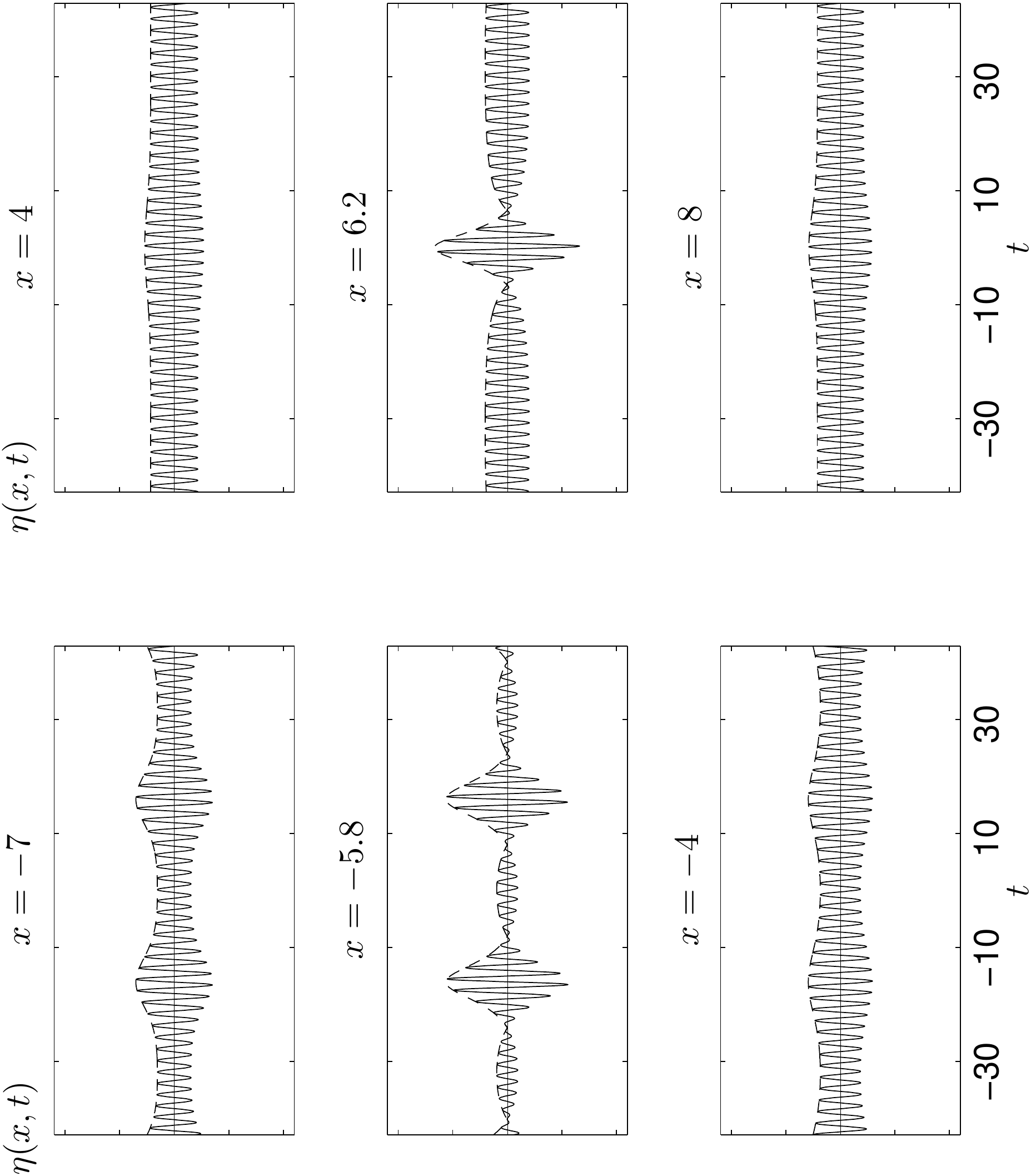}
    \caption{Plots of SFB$_{2}$ wave signal evolution for the modulation frequency $\tilde{\nu}_{2} = 1/2$ at different locations. The set parameters are chosen as follows: $\xi_{11} = 5$, $\xi_{12} = -5$ and $\tau_{11} = 0 = \tau_{12}$.} \label{SignalSFB2nu12Tau0}
  \end{center}
\end{figure}

To show the wavefront dislocation at two places, we take the time parameters to be zero and the space parameters nonzero. The plots of the absolute value of the complex-value amplitude, the corresponding physical wave field and the wave signal evolution for $\xi_{11} = 5$, $\xi_{12} = -5$, and $\tau_{11} = 0 = \tau_{12}$ are shown in Figure \ref{3D1} and Figure \ref{SignalSFB2nu12Tau0}.
We observe that for nonzero $\xi_{11}$ and $\xi_{12}$, the absolute value of the complex-valued amplitude of SFB$_{2}$ now has extreme parts at two different positions. This SFB$_{2}$ consists of two SFB$_{1}$s with modulation frequencies $\tilde{\nu}_{2} = 1/2$ and $2\tilde{\nu}_{2} = 1$. By imposing a shift of $\xi_{11}$, one part of SFB$_{2}$ which is SFB$_{1}$ with modulation frequency $\nu_{2}$ will shift in the positive $\xi$ direction for $\xi_{11} > 0$ and in the negative $\xi$ direction for $\xi_{11} < 0$. Likewise, imposing a shift of $\xi_{12}$, part of SFB$_{2}$ which is SFB$_{1}$ with modulation frequency $2\nu_{2}$ will also shift according to the sign of $\xi_{12}$, in the positive $\xi$ direction for $\xi_{12} > 0$ and in the negative $\xi$ direction for $\xi_{12} < 0$. In our example, the SFB$_{1}$ on the right corresponds to the shift $\xi_{11} = 5$ and the SFB$_{1}$ on the left corresponds to the shift $\xi_{12} = -5$. That is why the right part has a larger modulation period than on the one on the left. See Figure \ref{3D1}(a). The larger the difference between $\xi_{11}$ and $\xi_{12}$, the farther are the two SFB$_{1}$s split apart.

Since SFB$_{2}$ has extreme parts at two different locations, we observe that wavefront dislocations occur likewise. In our example, wavefront dislocations occur at $x < 0$ and $x > 0$. For $x = 0$, it is simply the plane-wave solution, which is similar as for $x \rightarrow \pm \infty$, see Figure \ref{3D1}(b). The corresponding wave signal evolution also confirms that phase singularities occur at two different locations, as can be seen in Figure \ref{SignalSFB2nu12Tau0}. We observe that an initially modulated wave signal develops into an extreme signal at $x = -5.8$ with modulation period of $2\pi/(2\nu_{2})$. The signal then returns to its initial state and at $x = 0$ it is simply the plane-wave solution. Another modulated wave signal appears and develops into another extreme wave signal at $x = 6.2$ with modulation period of $2\pi/\nu_{2}$. This signal also eventually returns to the previous state of modulated wave signal.

\subsubsection{Wavefront Dislocations at Three Different Positions}

\begin{figure}[h]
  \begin{center}
    \subfigure[]{\includegraphics[width=0.45\textwidth]{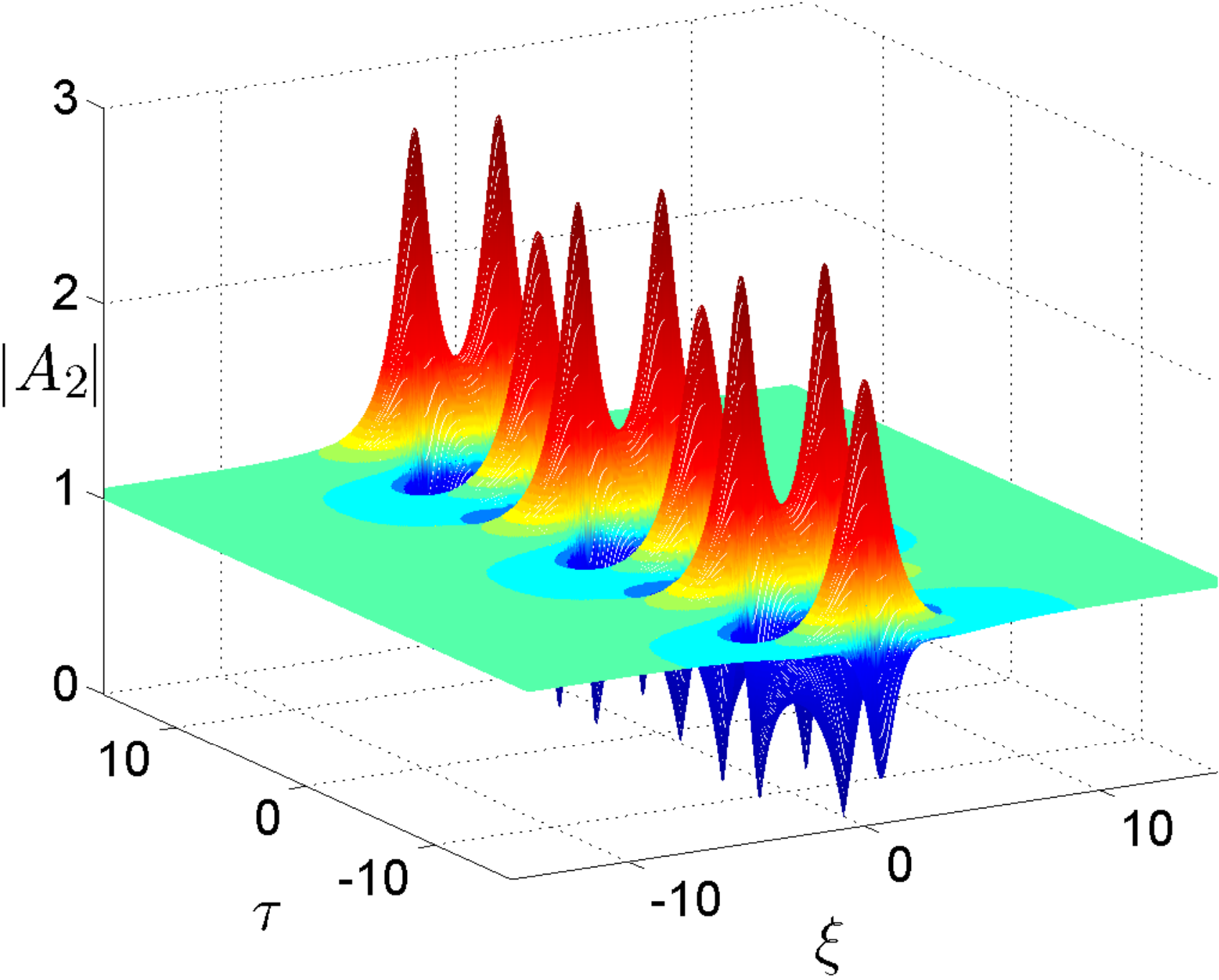}}    \hspace{0.5cm}
    \subfigure[]{\includegraphics[width=0.45\textwidth]{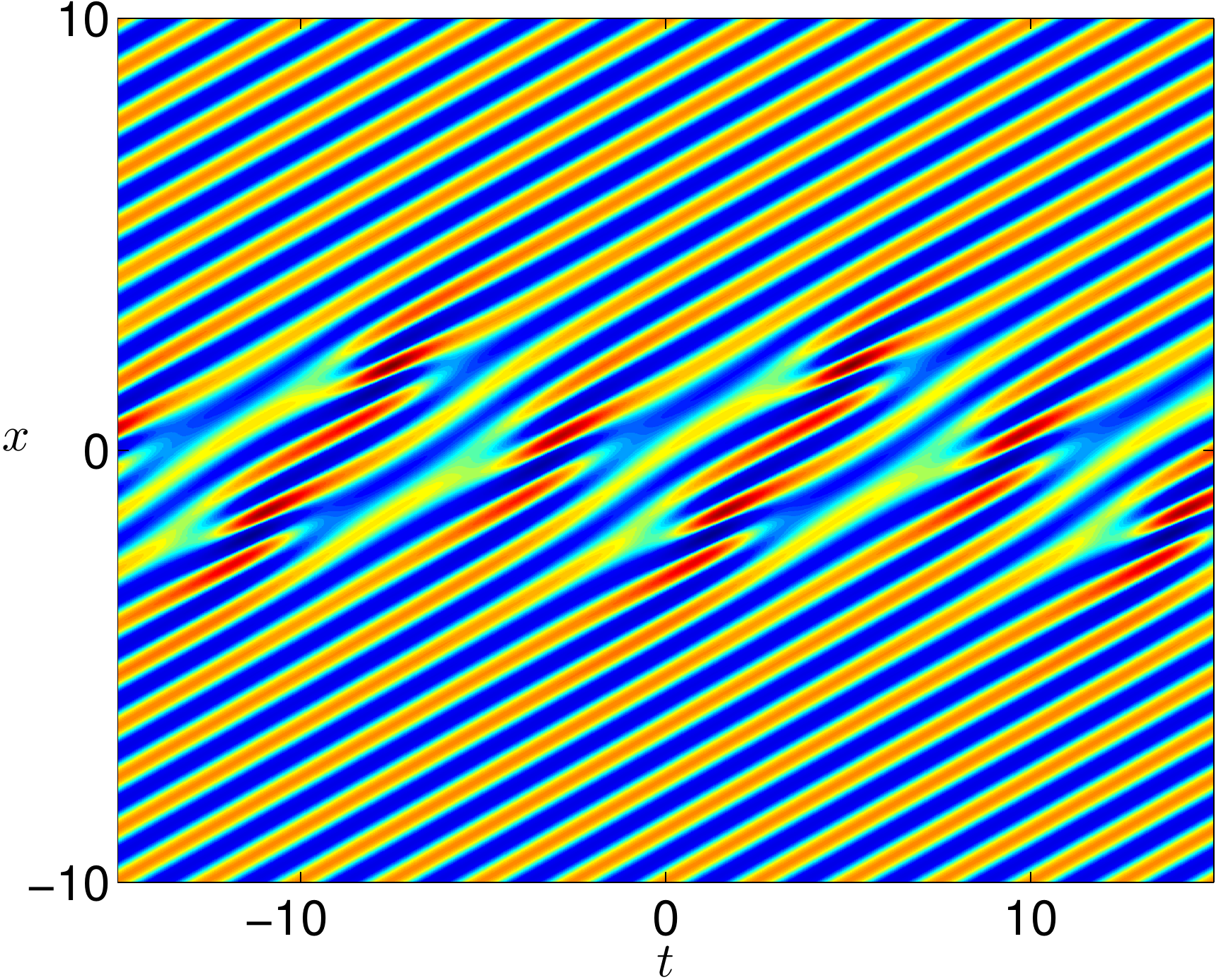}}
    \caption{(a). A three-dimensional plot of the absolute value of SFB$_{2}$ for the modulation frequency $\tilde{\nu}_{2} = 1/2$, $\xi_{11} = 0 = \xi_{12}$, $\tau_{11} = \pi/(2\nu_{2})$ and $\tau_{12} = 0$. (b). A density plot of the corresponding physical wave field.} \label{3D3}
  \end{center}
\end{figure}

Now we consider a different set of parameters, i.e. $\xi_{11} = 0 = \xi_{12}$, $\tau_{11} = \pi/(2\nu_{2})$ and $\tau_{12} = 0$. We observe that this set of parameters gives a different pattern for SFB$_{2}$, as compared to the two previous cases. Figure \ref{3D3} shows for a three-dimensional plot of the absolute value of the complex-valued amplitude and a density plot of the corresponding physical wave field for the SFB$_{2}$ with this choice of parameters. The interaction results to SFB$_{2}$ with single and double extreme parts arranged alternatingly in the temporal direction. The modulation period between each extreme part, whether it is single or double, is $2\pi/\nu_{2}$. The double extreme part has larger maximum values than the single part. While the single part reaches its maxima at $\xi = 0$, the double extreme part reaches its maxima at two different positions, $\xi < 0$ and $\xi > 0$, symmetric with respect to $\xi = 0$, see Figure \ref{3D3}(a).

\begin{figure}[h]
  \begin{center}
    \includegraphics[width=0.75\textwidth,angle=-90]{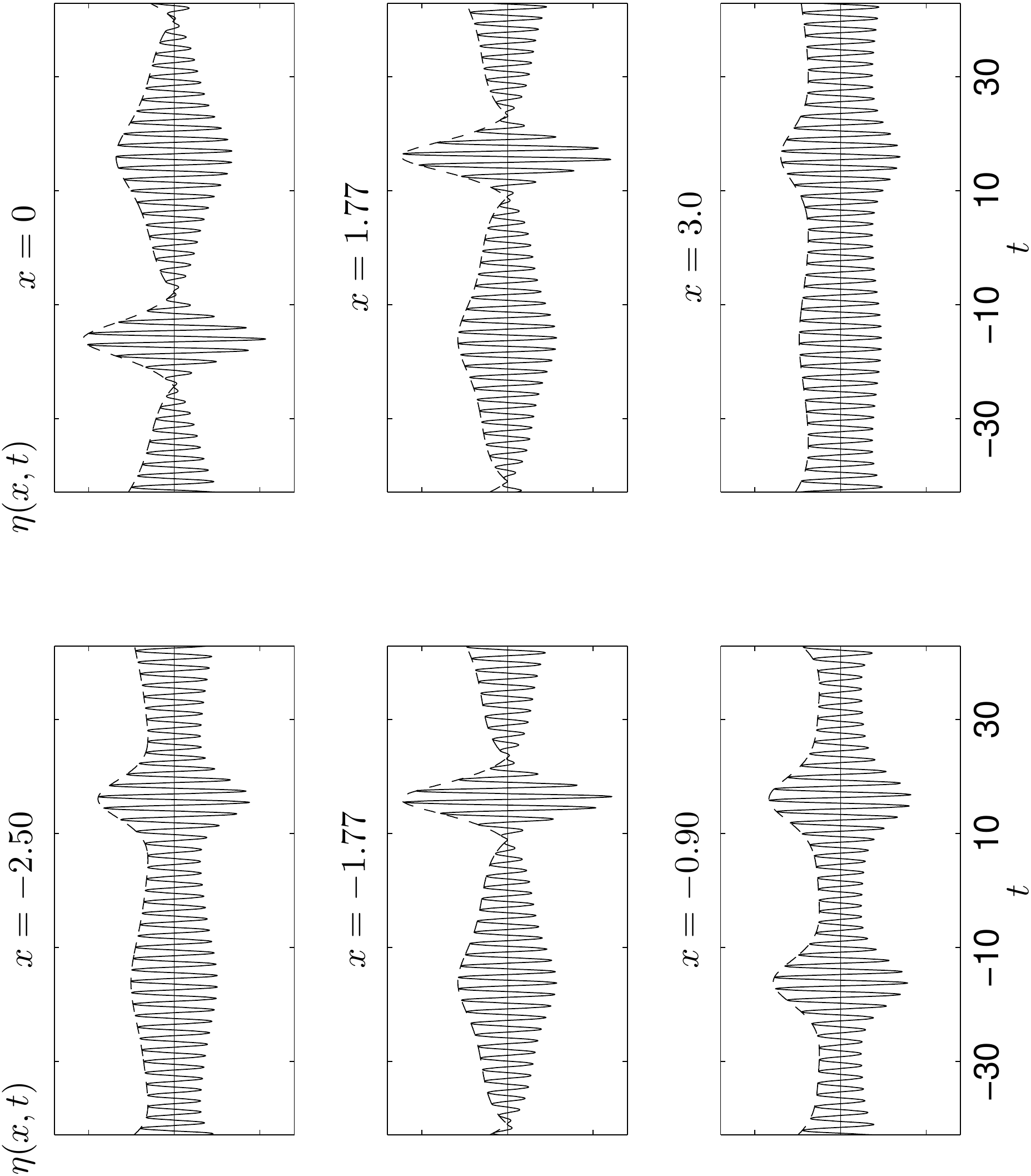}
    \caption{Plots of SFB$_{2}$ wave signal for the modulation frequency $\tilde{\nu}_{2} = 1/2$ at several positions. The chosen parameters for SFB$_{2}$ are given by $\xi_{11} = 0 = \xi_{12}$ and $\tau_{11} = \pi/2$ and $\tau_{12} = 0$.} \label{SignalSFB2nu12Xi0}
  \end{center}
\end{figure}

The corresponding physical field also shows wavefront dislocation which is periodic in time. Now it occurs at three different positions, $x < 0$, $x = 0$ and $x > 0$. See Figure \ref{3D3}(b). Each time we follow a wave crest or a wave trough along its propagation, splitting waves will be followed by merging waves, and vice versa. By observing the propagation of the corresponding wave signal, we observe that phase singularity occurs at three different positions too. Figure \ref{SignalSFB2nu12Xi0} shows plots of the SFB$_{2}$ wave signal evolution with this choice of parameters. In this example phase singularity occurs at $x = -1.77$, $x = 0$ and $x = 1.77$. Notice that the wave signals at $x = \pm 1.77$ have a similar pattern since the whole profile is symmetric with respect to $x = 0$. Note also that there is always one pair of phase singularities within one modulation period. We observe that a modulated wave signal develops into an extreme wave signal at $x = -1.77$, but after slightly decaying, it rises again to another extreme wave signal at $x = 0$, which has smaller maximum value than the one at $x = \pm 1.77$. This wave signal also decays a little bit, but not until becoming the plane-wave solution, then it rises again to reach a similar extreme wave signal as previously at $x = 1.77$. After this position, the wave signal returns to its initial condition and eventually becomes the plane-wave solution at the far field.
Regarding applications, the SFB$_{2}$ family of solutions of the NLS equation can be implemented for freak wave generation in hydrodynamic laboratory, as is the case for SFB$_{1}$.

\section{Conclusion}

In this article, we have discussed several characteristics of waves on finite background type of solutions of the NLS equation. These exact solutions are derived using displaced phase-amplitude variables. Although the exact solutions themselves are complex-valued functions, both the displaced phase and the displaced amplitude are real-valued functions. It is interesting to note that the displaced amplitude satisfies a Riccati-like equation as well as a nonlinear oscillator equation. The corresponding potential equation of this nonlinear oscillator equation depends on the displaced phase and a quartic function of the displaced amplitude. The total energy for the displaced phase-amplitude representation is zero for all waves on finite background type of solutions discussed in this article.

The waves on finite background type of solutions that we derived are known as the SFB, the Ma solution and the rational solution. The SFB is periodic in time but at $\tau = 2\pi/\nu$ has a soliton-like profile which goes to a finite background for $\xi \rightarrow \pm \infty$. The Ma solution is periodic in space and at $\xi = 2\pi/\rho$ has a soliton-like form for $\tau \rightarrow \pm \infty$. The rational solution is on the other hand neither space-periodic  nor time-periodic. Instead, it is isolated around $(\xi,\tau) = (0,0)$. The rational solution can be obtained by taking the modulation frequency $\nu \rightarrow 0$ in the SFB or the parameter $\mu \rightarrow 0$ in the Ma solution. All these three exact solutions have been proposed as mathematical models for freak wave events both in the open oceans as well as in the hydrodynamic laboratories.

In particular, we have chosen a family of SFB solution, or SFB$_{1}$ in the context of higher order waves on finite background, as a good candidate for generating freak wave events in the wave basin of a hydrodynamic laboratory. This is due to the fact that the SFB has an initial modulated wave signal which is much easier to be generated by the wavemaker compared to the Ma wave signal, which even could have a rather high amplitude at or close to the wavemaker. Compared to the rational solution wave signal that has infinite modulation period, the SFB wave signal has a finite modulation period. Besides, experiments on the rational wave solution have been discussed by \citeasnoun{Henderson99}. Although there are some limitations on the theoretical model based on the SFB, several important robust phenomena are still observed in the experimental results, amongst others are an amplitude increase based on the Benjamin-Feir modulational instability, phase singularity and the preservation of the carrier wave frequency as well as the modulation period.

In order to get the best possible match between the theoretical model and the experimental result, improving the model for the wave propagation might become a necessary step for a particular research direction. A number of experts in the theory of water waves have suggested implementing the modified NLS equation that contains higher order terms \cite{Dysthe79,Trulsen96}. Another effort to improve the model equation  has been done by one of us (EvG) which leads to nonlinear dispersive higher order KdV-type of equations for waves above finite or infinite depth \cite{vanGroesen07}.

There are no tangible experimental results for the freak wave generation using SFB$_{2}$ family of solutions yet. Nonetheless, we are convinced that several aspects of our theoretical model will support the experimental finding for a particular case of SFB$_{2}$ freak wave generation. We remark that experiments on freak wave generation using SFB$_{2}$ is interesting for future research. Since we only consider unidirectional waves on finite background type of solutions, an extension to bidirectional or multidirectional waves on finite background type of solutions for freak wave generation is another interesting possible future research direction.

\section*{Acknowledgement}
The research is supported by the project TWI.5374 of the Netherlands Organization of Scientific Research NWO, subdivision Applied Sciences STW.
\addcontentsline{toc}{section}{Acknowledgment} The authors would like to acknowledge the Chair of Applied Analysis and Mathematical Physics at University of Twente and Faculty of Engineering at The University of Nottingham Malaysia Campus. Many fruitful discussions with Dr. Andonowati (Bandung Institute of Technology and LabMath Indonesia), Gert Klopman (Albatross Flow Research and University of Twente), Dr. Miguel Onorato (Universit\'{a} di Torino), Dr. Panayotis Kevrekidis (University of Massachusetts at Amherst), Professor Nail Akhmediev (The Australian National University) and Professor Karsten Trulsen (University of Oslo)  are also appreciated.

\end{document}